\documentclass{egpubl}
\usepackage{pg2025}
 
\SpecialIssueSubmission    

\CGFStandardLicense

\usepackage[T1]{fontenc}
\usepackage{dfadobe}
\usepackage{amsmath}
\usepackage{amssymb}
\usepackage{algorithm}
\usepackage{algpseudocode}
\usepackage[percent]{overpic}
\usepackage{cite}  
\usepackage{tikz}
\usepackage{lineno}

\newcommand{\diff}{\mathop{}\!\mathrm{d}}
\newcommand\rev[1]{#1}

\BibtexOrBiblatex

\electronicVersion
\PrintedOrElectronic

\usepackage{egweblnk}
\usepackage{hyperref}

\title[Geometric Integration for Neural Control Variates]%
      {Geometric Integration for Neural Control Variates}

\author[D. Meister \& T. Harada]
{\parbox{\textwidth}{\centering D. Meister\orcid{0000-0002-3149-1442} and T. Harada\orcid{0000-0001-5158-8455}}\\
{\parbox{\textwidth}{\centering Advanced Micro Devices, Inc.}}}

\begin{document}

\newcommand{\imgheight}{0.22}

\teaser{
\centering
\setlength{\tabcolsep}{1pt}
\begin{tabular}{cccccc}
\includegraphics[height=\imgheight\linewidth]{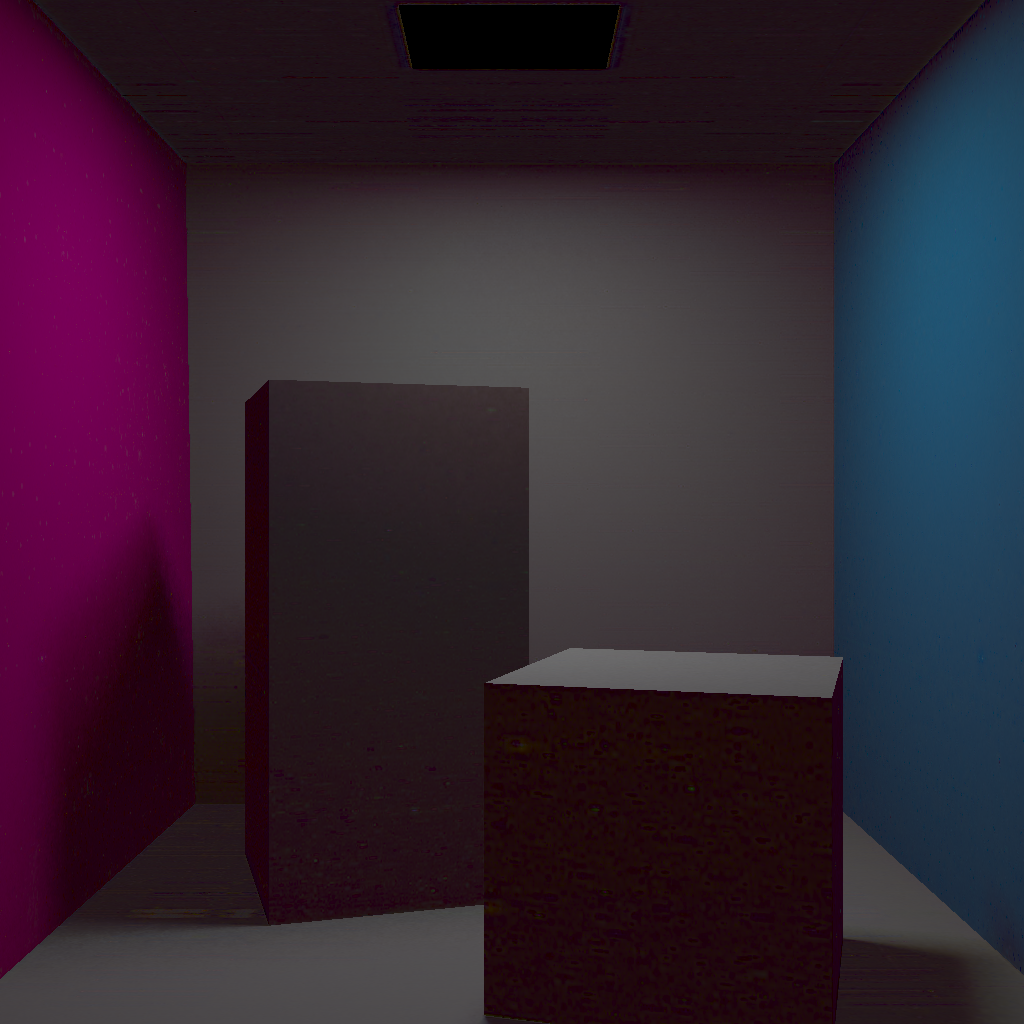} &&
\includegraphics[height=\imgheight\linewidth]{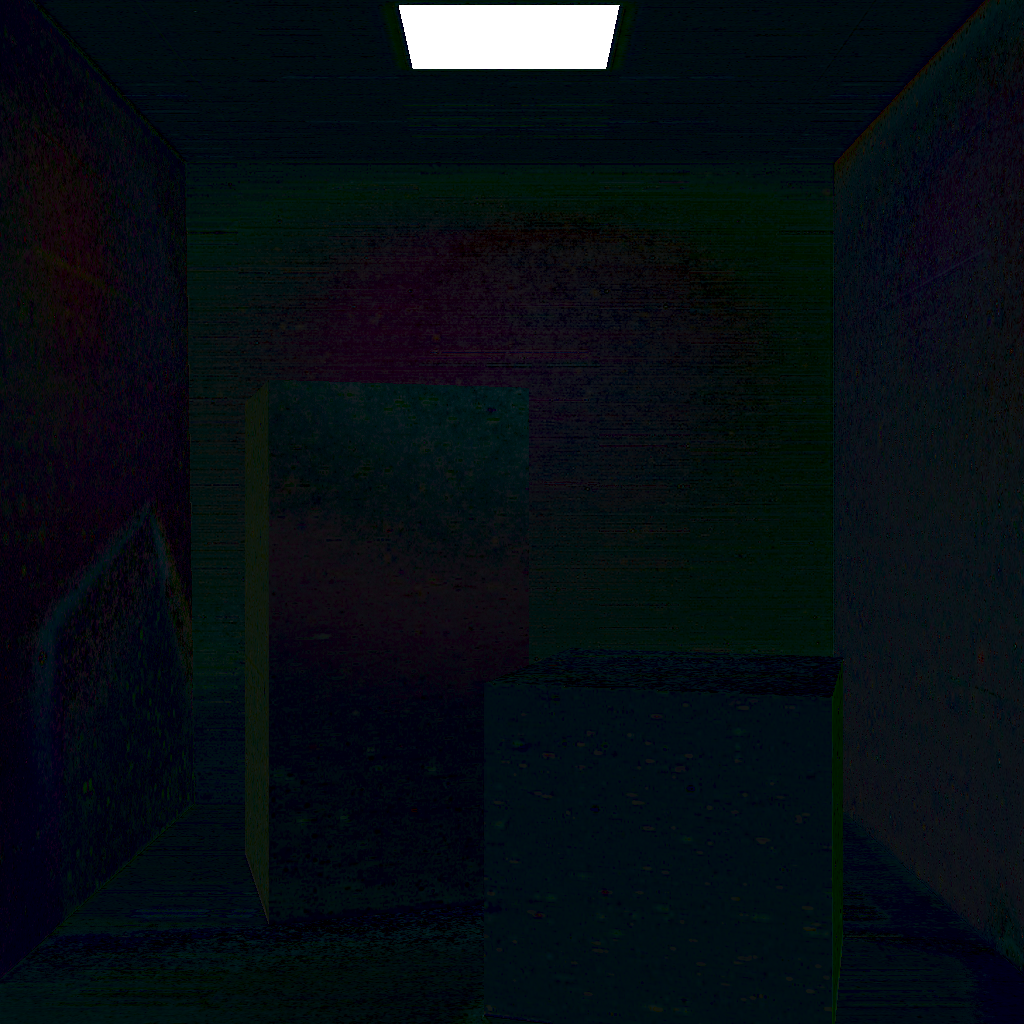} &&
\includegraphics[height=\imgheight\linewidth]{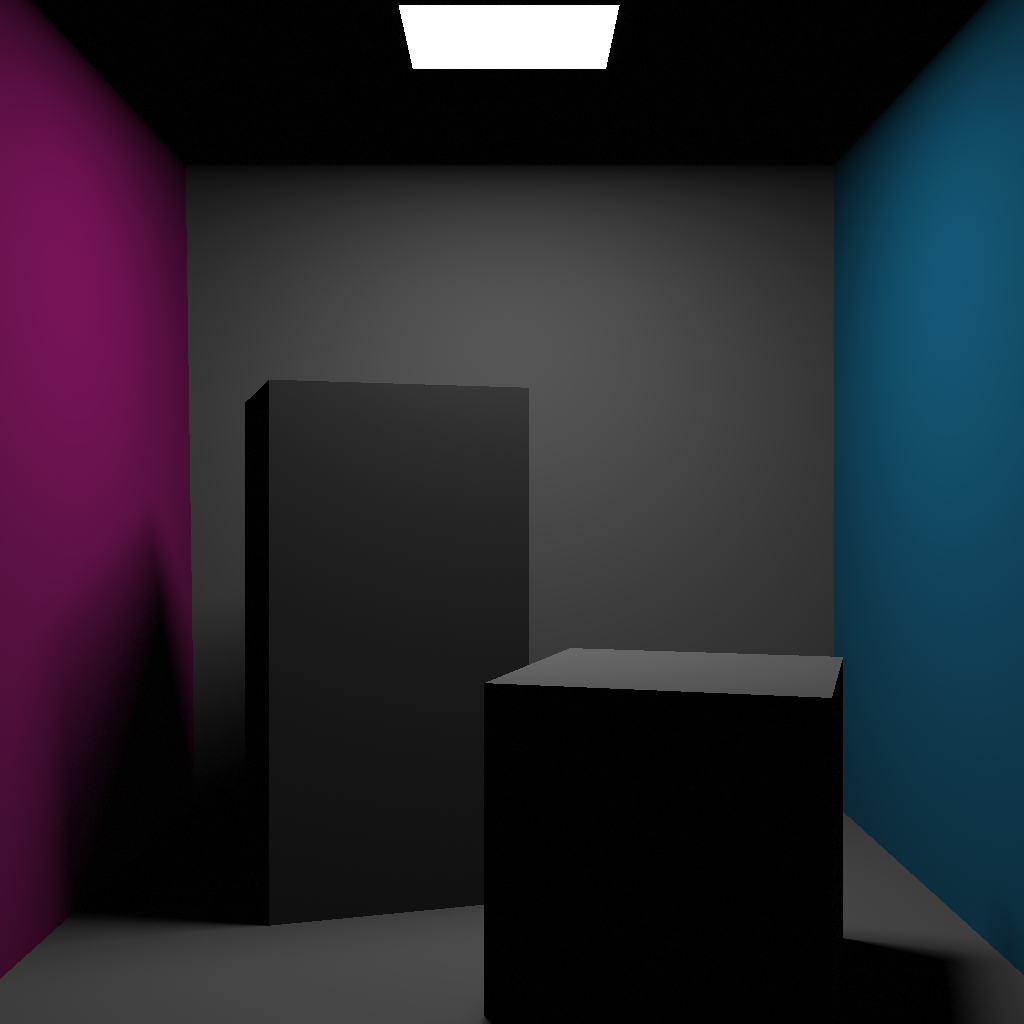} &
\includegraphics[height=\imgheight\linewidth]{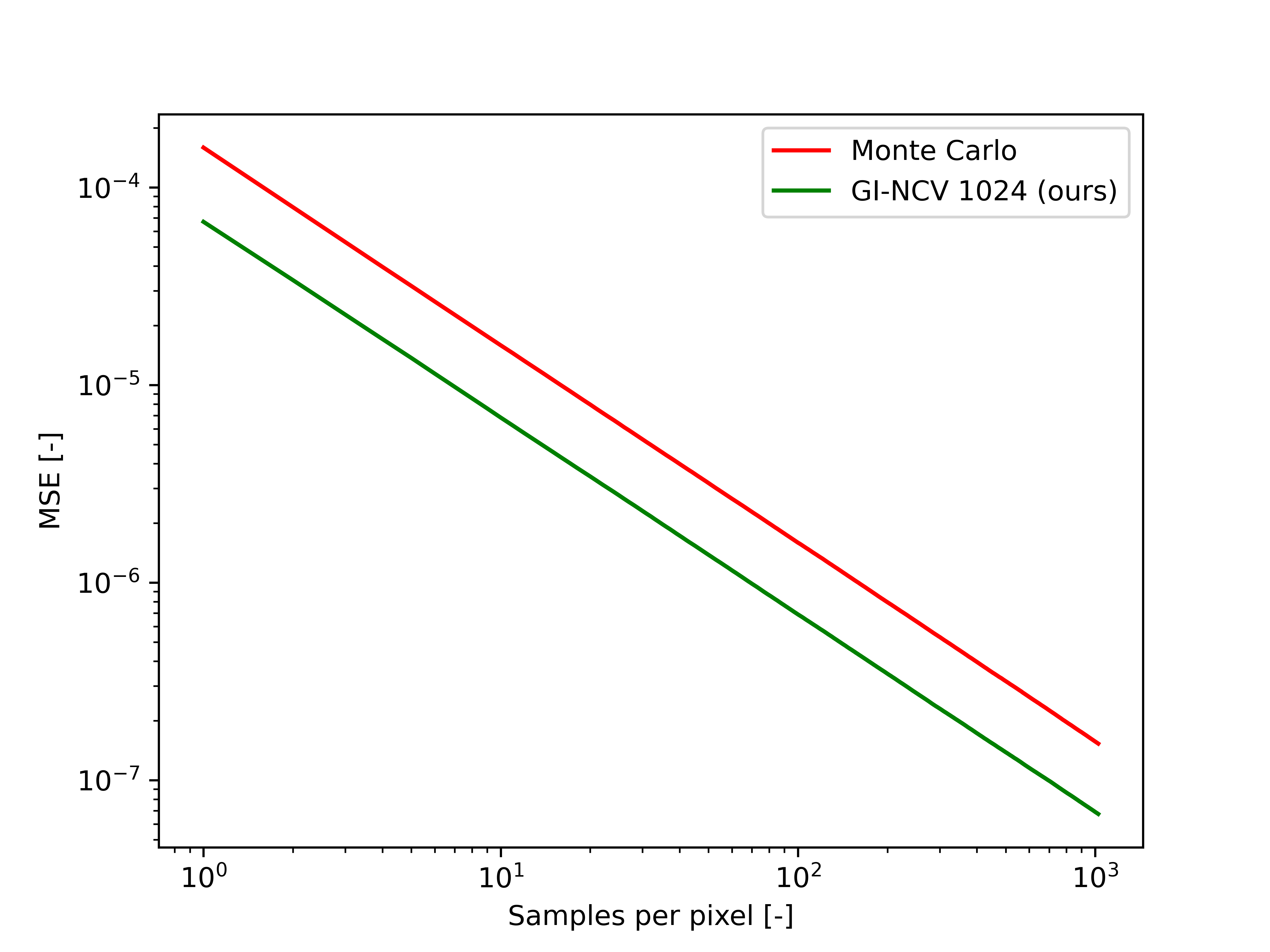}\\
$\alpha G$ & $+$ & $\int_\mathcal{D} f(x) - \alpha g(x) \diff x$ & $=$ & $\int_\mathcal{D} f(x) \diff x$ & $G = \int_\mathcal{D} g(x) \diff x$\\
\end{tabular}
\caption{We approximate the integrand by a small neural network (i.e., multilayered perceptron). Thanks to our geometric subdivision, we can integrate the neural network analytically, and use it as a control variate for Monte Carlo integration. The integral of the approximation provides a biased estimate (left), which is corrected by Monte Carlo integration of the residual integrand (center left), obtaining the final unbiased estimate (center right), which can achieve a lower error than vanilla Monte Carlo (right).}
\label{Fig:Teaser}
}

\maketitle

\begin{abstract}
Control variates are a variance-reduction technique for Monte Carlo integration. The principle involves approximating the integrand by a function that can be analytically integrated, and integrating using the Monte Carlo method only the residual difference between the integrand and the approximation, to obtain an unbiased estimate. Neural networks are universal approximators that could potentially be used as a control variate. However, the challenge lies in the analytic integration, which is not possible in general. In this manuscript, we study one of the simplest neural network models, the multilayered perceptron (MLP) with continuous piecewise linear activation functions, and its possible analytic integration. We propose an integration method based on integration domain subdivision\rev{, employing techniques from computational geometry to solve this problem in 2D}. We demonstrate that an MLP can be used as a control variate in combination with our integration method, showing applications in the light transport simulation.
\end{abstract}

\section{Introduction}
To synthesize photorealistic images, we need to solve notoriously complex integrals that model the underlying light transport. In general, these integrals do not have an analytic solution, and thus we employ tools of numerical integration to solve them. Among those, Monte Carlo integration is prominent, providing a general and robust solution, efficiently dealing with, for example, high dimensions or discontinuities that other numerical methods may struggle with. Monte Carlo converges to a correct solution with an increasing number of samples; however, it may require a large number of samples to suppress variance, that otherwise exhibits as high-frequency noise in the rendered images, under an acceptable threshold.

Thus, various variance-reduction techniques have been proposed. Widely studied importance sampling minimizes the variance by sampling from a probability distribution that matches the underlying integrand. In physically-based rendering, the goal is to sample directions from which incoming radiance is high. Such techniques include sampling according to the bidirectional reflectance distribution function (BRDF), sampling of directly visible light sources, or more general path guiding methods accounting for complex global illumination phenomena. 

Control variates are another variance-reduction technique that received significantly less attention in computer graphics than importance sampling. The principle is to approximate the integrand by a simpler function that can be analytically integrated. To get an unbiased estimate the analytic integration is corrected by integrating the difference between the integrand and the approximation, thanks to the following identity:
\begin{equation}
F = \int_\mathcal{D} f(x) \diff x = \alpha G + \int_\mathcal{D} f(x) - \alpha g(x) \diff x,
\label{Eq:ControlVariates}
\end{equation}
where $f(x)$ is the integrand, $\mathcal{D}$ is an integration domain, $g(x)$ is a function approximating $f(x)$ such that $G = \int_\mathcal{D} g(x) \diff x$, and $\alpha$ is a scale parameter, controlling influence of the control variates. The variance reduction depends on how $g(x)$ matches $f(x)$.

Neural networks are universal approximators, making them excellent candidates for control variates. However, integrating a general neural network could be as difficult as integrating the original integrand. Recently, it has been showed that the multilayer perceptron (MLP), the simplest neural network model, with sophisticated input encoding~\cite{Fujieda2023b} is capable to represent credibly a high variety of functions~\cite{Muller2021, Takikawa2021, Fujieda2023a}. Compared to complex architectures, the MLP is small and can be online-trained efficiently on contemporary GPUs~\cite{Muller2021}.

In this manuscript, we focus on the MLP with continuous piecewise linear activation functions. The key observation is that the integration domain of such function can be split into convex subdomains, where each subdomain corresponds to a single affine function \cite{Liu2024}. Thanks to the linearity of the integrals, we can formulate the integration as a sum of the integrals of the individual subdomains. We utilize some of the algorithms and data structures from computational geometry~\cite{Berg2008} to efficiently compute the convex regions and the corresponding affine functions \rev{in 2D}. In particular, our contributions are as follows:
\begin{itemize}
    \item We formulate an analytic integration of the MLP as a geometric problem, offering a novel perspective on the subject.
    \item We propose an algorithm employing algorithms and data structures from computational geometry to solve this problem \rev{in 2D}.
    \item We employ the proposed integration in combination with control variates, demonstrating the application in the light transport simulation.
\end{itemize}


\section{Related Work}

Control variates have been successfully adapted in various area such as operations research~\cite{Hesterberg1998} or financial mathematics~\cite{Kemna1990}. However, their application in computer graphics remains relatively limited. Rousselle et al.~\cite{Rousselle2016} applied image-space control variates for re-rendering and the Poisson reconstruction in the gradient-domain rendering. Xu et al.~\cite{Xu2024} proposed importance sampling for sampling the residual integral in the problem of re-rendering. Crespo et al.~\cite{Crespo2021} proposed using piecewise polynomials as a control variate in the primary sample space. Similarly, Sala\"{u}n et al.~\cite{Salaun2022} use the polynomial regression, additionally showing the connection between the least-square regression and Monte Carlo integration. Kondapaneni et al.~\cite{Kondapaneni2019} showed that the optimal weights for multiple importance sampling can be interpreted as an optimal control variate. Nicolet et al.~\cite{Nicolet2023} propose recursive control variates for differentiable rendering.

Since neural networks can approximate complex distributions better than traditional methods, it is tempting to use them as control variates. The challenging part is the integration of a neural network ($G$ in Equation~\ref{Eq:ControlVariates}). M\"{u}ller et al.~\cite{Muller2020} proposed to use the normalizing flows scaled by an additional neural network to fit a given distribution, where the scale is trivially the integration of the approximation. The architecture of normalizing flows is quite complex due to the requirement for the invertibility of the mapping, consisting of a chain of coupling layers. Normalizing flows excel in high dimensions, but it is not straightforward to use them in lower dimensions (e.g., 1D or 2D). Note that the invertibility is a stronger requirement than integrability, limiting the expressiveness of the model. The parameters of each coupling layer are provided by a neural network, resulting in a model with a large number of parameters. 

Lindell et al.~\cite{Lindell2021} proposed an automatic integration of an arbitrary neural network based on the fundamental theorem of calculus. The idea is to represent an antiderivative (that can be used for the integration) as a neural network (\emph{integral network}) such that the output is differentiated (using an autodiff) before feeding to a loss function. The authors proposed to construct another neural network (\emph{grad network}) that represents the derivative of the integral network such that both networks share the same parameters, and thus the grad network can be trained directly avoiding excessive use of the autodiff. Li et al.~\cite{Li2024} proposed to use the automatic integration in combination with control variates, demonstrating the variance reduction in estimation of the Poisson and Laplace equations using the walk-on-spheres algorithm. Subr~\cite{Subr2021} proposed a Q-NET integration algorithm, exploiting the fact that an antiderivative of a sum of sigmoids can be computed as a weighted sum of antiderivatives of the individual sigmoids.

There are several caveats about the automatic integration. For a $d$-dimensional integration domain, we need to compute $d$-th order derivatives, which might be very expensive for higher dimensions. Even if we can afford to compute higher order derivatives, the derivatives must be non-zero to be able to fit the function. For example, the second-order derivatives of an MLP with linear layers and rectified linear units (ReLUs) results in zero (i.e., applying the product rule on the result of the chain rule). Another caveat is that the architecture of the grad network cannot be controlled directly (the size of the grad network is actually quadratic with the respect of the size of the integral network~\cite{Lindell2021}), making the training more difficult.

M\"{u}ller et al.~\cite{Muller2021} and Hadadan et al.~\cite{Hadadan2021} showed that even a simple MLP can represent high-frequency radiance field. Furthermore, such a small neural network can be fit into an on-chip-memory, and thus can be trained/infer efficiently on modern GPUs~\cite{Muller2021}. An input encoding is an essential component to learn high-frequency signals from lower dimensional inputs. M\"{u}ller et al.~\cite{Muller2022} proposed multi-resolution hashgrid with learnable parameters (i.e., latent feature vectors) for encoding world-space positions. In this case, the MLP can be relatively shallow, serving as a decoder of the latent feature vectors. Our goal is to use such a small MLP with continuous piecewise linear activation functions~\cite{Humayun2023} as a control variate. \rev{Liu~\cite{Liu2024} showed that an MLP with the ReLU activation functions represents a piecewise affine function, and thus the integration can be formulated as a sum of integrals, where each integral is an affine function defined on a convex subdomain. However, the author does not show how to find the subdomains.} Dereviannykh et al.~\cite{Dereviannykh2025} proposed to use the fully-fused MLP to cache incidence radiance in combination with multi-level Monte Carlo integration, replacing the analytical integration in control variates by another Monte Carlo integration.


\section{Geometric Integration}
In order to use an MLP as a control variate, we need to be able to analytically integrate it (see Equation~\ref{Eq:ControlVariates}). In this section, we describe our integration method based on the integration domain subdivision. The key observation is that an MLP with the continuous piecewise linear activation functions, such as the step function or the (leaky) rectified linear unit (ReLU), itself represents a piecewise linear (affine) function~\cite{Humayun2023}. The integration domain can be split into disjoint subdomains, where we can solve the integration separately on each subdomain. Assuming that function $g(x)$ is represented by such MLP, we can obtain $G$ by summing up $n$ integrals on the $n$ corresponding subdomains:
\begin{equation}
G = \int_\mathcal{D} g(x) \diff x = \sum_{i=1}^n \int_{\mathcal{D}_i} g(x) \diff x = \sum_{i=1}^n \int_{\mathcal{D}_i} g_i(x) \diff x,
\label{Eq:Subdomains}
\end{equation}
such that $\bigcup_{i=1}^n \mathcal{D}_i = \mathcal{D}$, $\bigcap_{i=1}^n = \emptyset$, and $g_i(x)$ is an affine function defined on $\mathcal{D}_i$. Determining $\mathcal{D}_i$ and $g_i(x)$ presents a challenge.

\subsection{Integration Domain Subdivision}
\label{Sec:DomainSubdivision}
\rev{From now on, w}e restrict ourselves to integration in 2D (using variables $x$ and $y$). Without loss of generality, we can also assume that the integration domain is a unit square and the activation functions are ReLUs. The goal is to split the integration domain (i.e., the unit square) into subdomains such that each subdomain has an associated affine function that is identical to the function induced by the MLP.

We construct the subdomains and the corresponding affine functions incrementally, proceeding from the input layer to the output layer, based on whether previous neurons are activated or not. It is important to keep in mind that the elements we are working with are affine functions (i.e., an affine combination of $x$ and $y$ in the form of $ax+by+c$) instead of numbers, relying on the fact that the affine functions are closed under composition. 

\begin{figure*}
\centering
\includegraphics[width=0.95\textwidth]{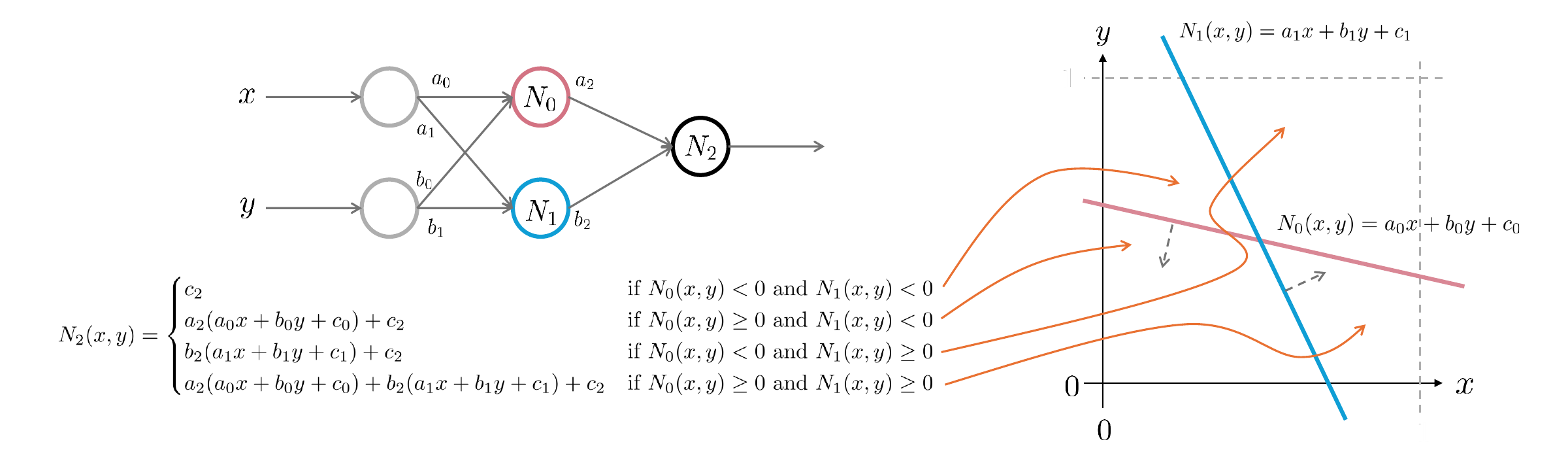}
\caption{An example of the integration domain subdivision for a simple MLP with one hidden layer containing two neurons (the weights are denoted by $a_i$, $b_i$, and biases $c_i$), using the ReLU activation function. The input to each hidden neuron is an affine combination of $x$ and $y$, which is subsequently fed to the ReLU. The ReLU function is activated based on the sign of the input affine function, such that we can split the 2D plane by a line into positive (indicated by the dashed arrows) and negative half-planes, where the line itself corresponds to $x$ and $y$, at which the function is zero. Two hidden neurons correspond to two lines, resulting in four subdomains defined by the intersection of either positive or negative half-planes.}
\label{Fig:Overview}
\end{figure*}

The ReLU functions in neurons pass positive values and clamp negative ones \rev{to zero}. Since, in our case, the input is an affine function, we need to decide for which $(x,y)$, the function is positive and negative. This geometrically means that a line (corresponding to where the function is zero) splits the 2D plane into negative and positive half-planes. In the case of single hidden layer, each neuron defines such a line, and a subdomain is an intersection of \rev{the corresponding} half-planes; either positive or negative half-plane of the neuron participating in the intersection. An example of such subdivision is illustrated in Figure~\ref{Fig:Overview}.

For multiple hidden layers, we process each subdomain from the previous layer recursively. The inputs for the current layer are outputs of the previous layer transformed by the corresponding weights and biases, resulting in new affine functions, which define the lines of the current layer. Notice that the lines (and the corresponding affine functions) inside each subdomain are different, as different neurons of the previous layer have been activated (providing different inputs for the current layer) on different subdomains.

We repeat this process until we reach the output layer, where the outputs of the last hidden layer are transformed by the output layer weights and biases, obtaining the affine function for a bottom-most subdomain. If we use the ReLU function on the output, we need to perform one more level of subdivision. We can also use other continuous piecewise linear activation functions other than ReLU such as the leaky ReLU or the step function, as long as the output is an affine function. Otherwise, a non-linear output would result in a non-linear input for the following layers, resulting in non-linear boundaries (\rev{i.e., general curves instead of straight lines}) between subdomains.

\subsection{Line Arrangement}
\label{Sec:LineArrangement}
In the previous section, we described how to find subdomains $\mathcal{D}_i$ and the associated functions $g_i$ that can be used for the integration (see Equation~\ref{Eq:Subdomains}). However, so far, it is not clear how to algorithmically compute the intersections of half-planes. 

\begin{figure*}
\centering
\includegraphics[width=0.95\textwidth]{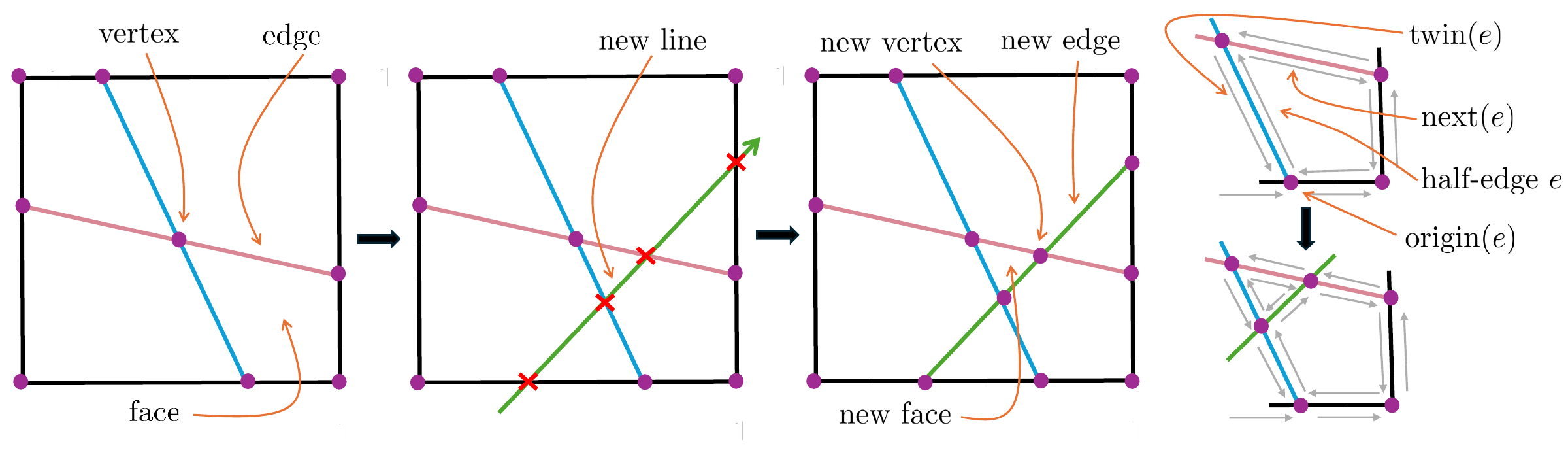}
\caption{An example illustrating a planar subdivision (left), the process of inserting a new line (middle), and a doubly-connected edge list (DCEL) (right). Inserting a new line requires finding the intersection with the existing edges, which correspond to new vertices. New edges are added between the consecutive intersections along the line. The planar subdivision is stored in the DCEL, that allows us to traverse through the planar subdivision, splitting edges and faces, thanks to the topological information \rev{provided by the DCEL}.}
\label{Fig:CGE}
\end{figure*}

There is a problem known as \emph{line arrangement}, and it is well-studied in computational geometry~\cite{Berg2008}. Given a set of lines, the goal is to find a planar subdivision defined by vertices (intersections of the lines), edges (line segments between intersections of one line with other lines), and faces (convex regions containing no lines), where the faces correspond to the subdomains we want to find. The lines are typically cropped by a bounding box to avoid unbounded edges. In our case, we can naturally use the unit square as the bounding box. An example of such planar subdivision is illustrated in Figure~\ref{Fig:CGE}.

Doubly-connected edge list (DCEL)~\cite{Berg2008} is a data structure commonly used to store a planar subdivision. The advantage of the DCEL is that it contains topological information of individual elements. Each edge is represented by two (oriented) half-edges, where each half-edge is defined by one of the vertices (\emph{origin}) of the edge, a pointer to a half-edge within the same face such that its origin is the other vertex of the first edge (\emph{next}), and a pointer to a half-edge with an opposite orientation on the other side of the edge (\emph{twin}). Thanks to this representation, we can walk around the boundary of a given face, or visit the neighboring faces. Each half-edge has a pointer to a face that belongs to, and each face has a pointer to one of its half-edges. A fragment of a DCEL is depicted in Figure~\ref{Fig:CGE}.

Armed with the DCEL, we can now solve the line arrangement problem by incrementally inserting lines one by one. The DCEL is initialized with the unit square: four vertices, eight half-edges corresponding to four edges, one inner face, and one (unbounded) outer face. We start with intersecting the line with half-edges of the outer face. We split the half-edge and its twin by inserting a new vertex. Then, we continue to the face on the opposite side. Inside this face, we find the second intersection of the face, split the half-edge and its twin. We also split the face into two parts by inserting two additional half-edges, and then we recursively continue to the next face. This resembles 2D ray tracing finding all intersections.

The line arrangement problem can be generalized into arbitrary dimensions, where we replace lines by general hyperplanes. However, it suffers from the curse of dimensionality, which corresponds to the fact that any subset of neurons can be activated, leading to the exponential space complexity of the subdivision elements. Luckily, there are upper bounds for the planar subdivision~\cite{Berg2008}. For $N$ lines (corresponding to the $N$ neurons in a hidden layer), the number of faces is not greater than ${N \choose 2} + N + 1$, the number of edges is not greater than $N^2$, and the number of vertices is not greater than ${N \choose 2}$.

We need one additional piece of information in order to transit from one hidden layer to another. For each face, we store a mask of lines indicating whether the face lies on the positive side of a line. Note that some lines may lie completely outside the integration domain, which results in no new faces, making the above upper bounds even more conservative. However, we still need to update the line mask if the integration domain lies on the positive side.

\subsection{Integration Algorithm}
In the previous section, we showed that the integration domain subdivision induced by a single hidden layer can be formulated as the line arrangement problem. In this section, we discuss how to generalize it for multiple hidden layers, and how to compute $G$.

As we discussed in Section~\ref{Sec:DomainSubdivision}, each face needs to be processed recursively. Na\"ively, we could allocate a DCEL for each face in the current hidden layer, initialize it with the face and process it recursively. However, this would result in high memory consumption, getting worse with more hidden layers. The key observation is that we want to accumulate the $G$, and thus we do not need to store all subdomains, that have already processed, in the memory. In other words, once the face is processed, we can reuse the DCEL for the next face, requiring only one DCEL per hidden layer.

This is summarized in Algorithm~\ref{Alg:Traversal}, resembling a traversal of a wide tree such that the branching factor is equal to the number of faces. For $M$ hidden layers, we allocate $M$ DCEL structures (line 4). Since we need to visit all subdomains, the traversal can be performed in a stackless manner. We keep the current depth and the index of the current face on each level (stored in variable $trail$). We initialize the root DCEL with the unit square, add the lines corresponding to the neurons of the first hidden layer (lines 5-10), and then we enter the main loop.

If we have not reached the last layer (lines 13-34), we first check whether all faces of the current hidden layer have been processed (lines 14-18). If so, we reset the trail and decrease the depth, going to the previous hidden layer. Otherwise, we initialize a DCEL of the next hidden layer with the current face and add the corresponding lines (lines 22-33). Each such line is an affine combination of lines (plus the bias) in the current layer, excluding those lines when the face lies on the negative side of these lines (lines 25-31). This corresponds to when the neuron is not activated, and it is exactly the information the line mask stores. We increase the trail and depth (lines 20-21), going to the next layer.

If we reach the last layer (lines 34-48), we compute the output function for each face as an affine combination of lines (lines 38-44). This is very similar to when we combine the lines between hidden layers. Assuming a single neuron in the output layer, the output weights are represented as a single vector of length $M$ and the bias is a scalar value. This could be easily generalized into multiple output neurons. Once we have a face with the corresponding function, we can integrate it (line 45), which we discuss in the following section.

\begin{algorithm}[!h]
\caption{Our integration algorithm traverses the recursion tree in a stackless manner. In each step, we combine lines from the previous layer for the next layer, using weights, biases, and the line mask. At the last layer, we compute the output function as a combination of lines in the last hidden layer. Finally, we integrate the function on a particular face, accumulating the result in $G$.}
\label{Alg:Traversal}
\begin{algorithmic}[1]
\State $G \gets 0$
\State $depth \gets 0$
\State $trail[0, \dots, M-1] \gets \{0, \dots, 0\}$ 
\State $dcels[0, \dots, M-1] \gets \{\Call{dcel}{N}, \dots, \Call{dcel}{N}\}$
\State $\Call{initWithUnitSquare}{dcels[0]}$
\For{$i = 0$ to $N - 1$}
    \State $(weights, biases) \gets \Call{weights\&Biases}{0}$
    \State $line = \{weights[i, 0], weights[i, 1], biases[i] \}$
    \State $\Call{addLine}{dcels[0], i, line}$
\EndFor
\While{$depth \geq 0$}
\State $dcel \gets dcels[depth]$
\If{$depth < M - 1$}
    \If{$\textit{trail}[depth] = |dcel.faces|$}
        \State $trail[depth] \gets 0$
        \State $depth \gets depth - 1$
        \State \textbf{continue}
    \EndIf
    \State $\textit{face} \gets \textit{dcel.faces}[\textit{trail}[depth]]$
    \State $\textit{trail}[depth] \gets \textit{trail}[depth] + 1$
    \State $depth \gets depth + 1$
    \State $\Call{initWithFace}{dcels[depth], face }$
    \State $(weights, biases) \gets \Call{weights\&Biases}{depth}$
    \For{$i = 0 \to N - 1$}
        \State $line \gets \{0,0,0\}$
        \For{$j = 0 \to N - 1$}
            \If{$face.lineMask \And (1 \ll j)$}
                \State $line \gets line + weights[i, j] \cdot dcel.lines[j]$
            \EndIf
        \EndFor
        \State $line.c \gets line.c + biases[i]$
        \State $\Call{addLine}{dcels[depth], i, line}$
    \EndFor
\Else
    \State $(weights, bias) \gets \Call{weights\&Biases}{M}$
    \For{$i = 0$ to $|dcel.faces| - 1$}
        \State $face \gets dcel.faces[i]$
        \State $line \gets \{0,0,0\}$
        \For{$j = 0$ to $N - 1$}
            \If{$face.lineMask~\And~(1 \ll j)$}
                \State $line \gets line + weights[j] \cdot dcel.lines[j]$
            \EndIf
        \EndFor
        \State $line.c \gets line.c + bias$            
        \State $G \gets G + \Call{integrate}{face, line}$
    \EndFor
    \State $depth \gets depth - 1$
\EndIf
\EndWhile
\end{algorithmic}
\end{algorithm}

\subsection{Subdomain Integration}
\begin{figure}
\centering
\includegraphics[width=0.45\textwidth]{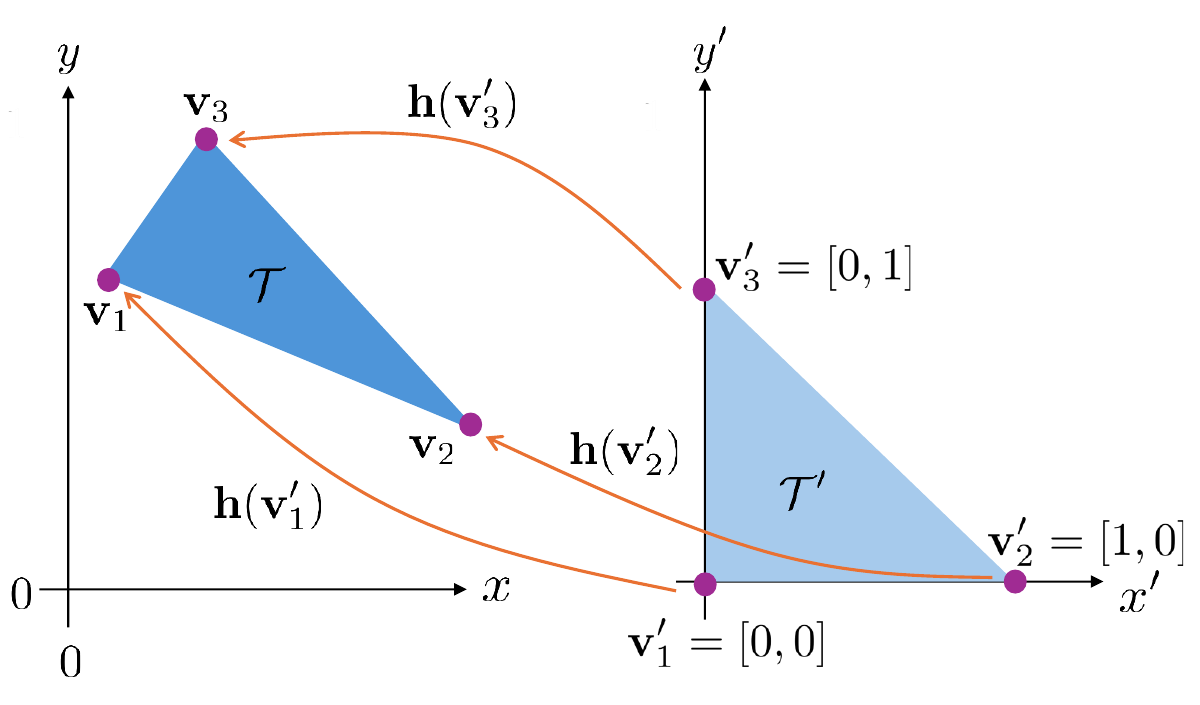}
\caption{An example illustrating function $\mathbf{h}$, mapping points in the unit triangle back to the original triangle.}
\label{Fig:Triangle}
\end{figure}

In the previous section, we described an algorithm that finds a subdomain and the corresponding function. The last missing piece is to show how to integrate such function on the subdomain. Remember that the function is affine (thanks to the closure under composition) and the subdomain is convex (thanks to the intersection of half-planes). It is straightforward to find the antiderivative of the affine function. However, it is not possible to integrate such a function as an iterative integral (i.e., as two \rev{nested} one-dimensional integrals) on a general convex subdomain, since $x$ and $y$ on the boundary are functions of each other. Thus, we need to split the subdomain into regions with a boundary where at least one of $x$ and $y$ is independent of the other one (e.g., axis-aligned rectangles or triangles). We opt for triangulation, where each triangle is transformed into a unit axis-aligned triangle. The triangulation is trivial since the subdomain is convex. For instance, we can split the subdomain into the triangle fan. After the triangulation, we can rewrite Equation~\ref{Eq:Subdomains} as follows:
\begin{equation}
G = \sum_{i=1}^n \int_{\mathcal{D}_i} g_i(x,y) \diff y \diff x = \sum_{i=1}^n \sum_{j=1}^{m_i} \int_{\mathcal{T}_{i,j}} a_ix  + b_iy + c_i \diff y \diff x,
\label{Eq:Triangulation}
\end{equation}
where $\mathcal{T}_{i,j}$ is $j$-th triangle in $\mathcal{D}_i$, $m_i$ is the number of triangles in $\mathcal{D}_i$, and $(a_i, b_i, c_i)$ are coefficients of the affine function associated with $\mathcal{D}_i$. For the sake of clarity, we will omit subscripts in the following formulas. We first define a mapping from the unit triangle to the original one:
\begin{equation}
\mathbf{h}: (x',y') \mapsto \mathbf{v}_1 + x'(\mathbf{v}_2 - \mathbf{v}_1) + y'(\mathbf{v}_3 - \mathbf{v}_1),
\label{Eq:h}
\end{equation}
where $\mathbf{v}_k$ are vertices of triangle $\mathcal{T}$, and $(x',y')$ are (local) coordinates in the space of the unit triangle. We rewrite the vector function in Equation~\ref{Eq:h} into two scalar functions:
\begin{equation}
h_1(x', y') = v_{1,1} + x'(v_{2,1} - v_{1,1}) + y'(v_{3,1} - v_{1,1}),
\label{Eq:h1}
\end{equation}
\begin{equation}
h_2(x', y') = v_{1,2} + x'(v_{2,2} - v_{1,2}) + y'(v_{3,2} - v_{1,2}),
\label{Eq:h2}
\end{equation}
where $v_{k,t}$ is $t$-th component of vertex $\mathbf{v}_k$. Mapping $\mathbf{h}$ is illustrated in Figure~\ref{Fig:Triangle}. We plug Equations~\ref{Eq:h1} and \ref{Eq:h2} into the integrand in Equation \ref{Eq:Triangulation}:
\begin{multline}
    \int_{\mathcal{T}} ax  + by + c \diff y \diff x =\\ |J_{\mathbf{h}}| \int_{0}^{1} \int_{0}^{1-x'} a\cdot h_1(x',y') + b\cdot h_2(x',y') + c \diff y' \diff x',
\end{multline}
where $|J_\mathbf{h}|$ is Jacobian determinant of function $\mathbf{h}$, which is twice an area of the triangle. Functions $h_1$ and $h_2$ are affine functions, and thus we can plug them into the affine function associated with $\mathcal{D}_i$ to obtain coefficients for new affine function (with respect to $(x',y')$):
\begin{multline}
    |J_{\mathbf{h}}| \int_{0}^{1} \int_{0}^{1-x'} a\cdot h_1(x',y') + b\cdot h_2(x',y') + c \diff y' \diff x' = \\
    2A_\mathcal{T} \int_{0}^{1} \int_{0}^{1-x'} a'x' + b'y' + c'\diff y' \diff x'
\end{multline}
where $A_\mathcal{T}$ is an area of triangle $\mathcal{T}$, and $(a',b',c')$ are coefficients of the composed affine function in the space of the unit triangle. The whole integral reduces to the following expression:
\begin{multline}
    2A_\mathcal{T} \int_{0}^{1} \int_{0}^{1-x'} a'x' + b'y' + c'\diff y' \diff x' = \frac{A_\mathcal{T}}{3}(a'+b'+3c').
\end{multline}

\section{Light Transport Simulation}
\label{Sec:LightTransport}
In this section, we formulate several light transport problems as 2D integrations with the control variates, employing the proposed integration method. Using the method described above, we are able to integrate an MLP over a single integration domain. In light transport, however, we typically solve many integrations simultaneously (e.g., per pixel). Our goal is to have a single MLP that could represent multiple integrands, and thus we condition the function represented by the MLP by additional parameters: $g(x,y\mid \phi)$,
where $\phi$ aggregates an arbitrary number of additional inputs to the MLP. These inputs are fixed with respect to a given integration domain, and thus after passing through the input layer (i.e., multiplying by the input weights), they are added to the bias. In other words, the MLP represents a higher dimensional function, where we integrate a 2D slice of that function for a fixed $\phi$. Table~\ref{Tab:Inputs} shows an overview of inputs we use in the following applications. For the integration domain, it is important not to use any non-linear encoding that would result in non-linear borders in the planar subdivision. 


\begin{table}[]
\tiny
\centering
\begin{tabular}{ccccc}
\hline
Parameter & Symbol & Encoding & Input dim. & Output dim. \\
\hline
Area light sample & $(x,y)$ & Identity & $2$ & $2$ \\
Incoming direction & $\mathbf{\omega}_i$ & Local hemisph. coords. & $2$ & $2$ \\
\hline
Position & $\mathbf{x}$ & Hashgrid & $3$ ($2$) & $2 \times 4$ \\
Outgoing direction & $\mathbf{\omega}_o$ & Sph. coords. / one-blob & $2$ & $2 \times 4$ \\
Surface normal & $\mathbf{n}(\mathbf{x})$ & Sph. coords. / one-blob & $2$ & $2 \times 4$ \\
Diffuse albedo & $\sigma(\mathbf{x})$ & Identity & 3 & 3 \\
Surface roughness & $\rho(\mathbf{x})$ & Identity & 1 & 1 \\
\hline
\end{tabular}
\caption{A table showing inputs of the MLP and their encodings with input and output dimensions that we use in the light transport simulation. The inputs in the top correspond to the integration domain, while the inputs in the bottom are parameters fixed with respect to the integration.}
\label{Tab:Inputs}
\end{table}

\subsection{Ambient Occlusion}
The first application is ambient occlusion, which expresses relative occlusion by the geometry itself regardless of light sources and materials. Formally, it can be described as an integral over the upper hemisphere:
\begin{equation}
AO(\mathbf{x} \mid r) = \int_{\mathbf{\omega}_i \in \Omega} \frac{V(\mathbf{x}, \mathbf{\omega}_i \mid r)(\mathbf{n}(\mathbf{x}) \cdot \mathbf{\omega}_i)}{\pi} \diff\mathbf{\omega}_i,
\end{equation}
where $\Omega$ is the upper hemisphere, $r$ is a radius (a parameter), and $V(\mathbf{x}, \mathbf{\omega}_i \mid r)$ is a visibility function that returns $1$ if there is no occluder from point $\mathbf{x}$ along direction $\mathbf{\omega}_i$ at a distance less than or equal to $r$, and $0$ otherwise. Notice that ambient occlusion is only dependent on the position $\mathbf{x}$, and the surface normal $\mathbf{n}$ is a function of $\mathbf{x}$. We train an MLP to approximate the integrand which can be used as a control variate:
\begin{equation}
g(\mathbf{\omega}_i \mid \mathbf{x}, \mathbf{n}(\mathbf{x})) \approx f(\mathbf{x}, \mathbf{\omega}_i \mid r) = \frac{V(\mathbf{x}, \mathbf{\omega}_i \mid r)(\mathbf{n}(\mathbf{x}) \cdot \mathbf{\omega}_i)}{\pi}.
\end{equation}
We use the surface normal $\mathbf{n}$ as an explicit additional guiding input, encoded by the one-blob encoding in spherical coordinates. Since ambient occlusion depends only on geometry, we do not include the material parameters (e.g., albedo or roughness).

We need to embed the incoming direction $\mathbf{\omega}_i$ into the unit square to use our analytic integration. The direction is typically converted to normalized spherical coordinates in world space. Since an MLP can provide arbitrary values on the integration domain, we need to ensure that the probability density function we use for sampling is non-zero across the whole integration domain to match the analytic integration in Equation~\ref{Eq:ControlVariates}. The spherical system\rev{, however,} covers the whole sphere while the Monte Carlo integration used for the residual integral accounts for the upper hemisphere (e.g., using the cosine-weighted sampling). Therefore, we first transform the direction to local space, and then we convert it to hemispherical coordinates (i.e., the azimuthal angle and the height) \rev{such that the integration domain is the hemisphere in both cases}. The result of the analytic integration needs to be scaled by the volume of the integration domain (i.e., $2\pi$ in the case of the upper hemisphere).

\subsection{Direct Illumination}
The second application is direct lighting with rectangular area light source $A$ such as an emissive rectangle (or parallelogram) defined by an anchor vertex $\mathbf{v}$ and two linearly independent vectors $\mathbf{e}_1$ and $\mathbf{e}_2$, where each point $\mathbf{y}$ has the form $\mathbf{y} = \mathbf{v} + x\mathbf{e}_1 + y\mathbf{e}_2$ for $(x, y) \in [0,1]^2$. This way, the problem of direct lighting can be formulated as an integral over the area of the light source:
\begin{multline}
L_o^{direct}(\mathbf{x}, \mathbf{\omega}_o) =\\
\int_{\mathbf{y} \in A} f_r(\mathbf{x}, \mathbf{\omega}_{\mathbf{x} \xrightarrow{} \mathbf{y}}, \mathbf{\omega}_o) L_e(\mathbf{y}, \mathbf{\omega}_{\mathbf{y} \xrightarrow{} \mathbf{x}}) G(\mathbf{x}, \mathbf{y}) V(\mathbf{x}, \mathbf{y}) \diff \mathbf{y},
\end{multline}
where $L_o^{direct}(\mathbf{x}, \mathbf{\omega}_o)$ is outgoing radiance from point $\mathbf{x}$ in direction $\mathbf{\omega}_o$, $L_e(\mathbf{y}, \mathbf{\omega}_{\mathbf{y} \xrightarrow{} \mathbf{x}})$ is emitted radiance from point $\mathbf{y}$ in direction $\mathbf{\omega}_{\mathbf{y} \xrightarrow{} \mathbf{x}}$, $f_r$ is the bidirectional reflectance function (BRDF), $\mathbf{\omega}_{\mathbf{x} \xrightarrow{} \mathbf{y}} = \frac{\mathbf{y} - \mathbf{x}}{\Vert\mathbf{y} - \mathbf{x} \rVert}$ is a unit direction pointing from $\mathbf{x}$ to $\mathbf{y}$, $G(\mathbf{x}, \mathbf{y}) = \frac{(\mathbf{n}(\mathbf{x}) \cdot \mathbf{\omega}_{\mathbf{x} \xrightarrow{} \mathbf{y}})(\mathbf{n}(\mathbf{y}) \cdot \mathbf{\omega}_{\mathbf{y} \xrightarrow{} \mathbf{x}})}{\Vert\mathbf{y} - \mathbf{x} \rVert^2}$ is the geometry term, and $V(\mathbf{x}, \mathbf{y})$ is the visibility function indicating binary visibility between points $\mathbf{x}$ and $\mathbf{y}$. Similarly as for ambient occlusion, we want to approximate the integrand by an MLP:
\begin{equation}
\mathbf{g}(x,y \mid \mathbf{x}, \mathbf{\omega}_o, \mathbf{n}(\mathbf{x}), \sigma(\mathbf{x}), \rho(\mathbf{x})) \approx \mathbf{f}(\mathbf{x}, \mathbf{v} + x\mathbf{e}_1 + y\mathbf{e}_2, \mathbf{\omega}_o) = \mathbf{f}(\mathbf{x}, \mathbf{y}, \mathbf{\omega}_o).
\end{equation}

\begin{equation}
\mathbf{f}(\mathbf{x}, \mathbf{y}, \mathbf{\omega}_o) = f_r(\mathbf{x}, \mathbf{\omega}_{\mathbf{x} \xrightarrow{} \mathbf{y}}, \mathbf{\omega}_o) L_e(\mathbf{y}, \mathbf{\omega}_{\mathbf{y} \xrightarrow{} \mathbf{x}}) G(\mathbf{x}, \mathbf{y}) V(\mathbf{x}, \mathbf{y}),
\end{equation}
In contrast to monochromatic ambient occlusion, radiance is wavelength dependent, and thus the whole integrand is a vector function representing RGB values. Radiance is also view-dependent, and thus we need to include the outgoing direction $\mathbf{\omega}_o$ as an input encoded by the one-blob encoding in spherical coordinates. For the primary hits, the outgoing direction is uniquely determined by the position, and thus can be omitted. We also use diffuse albedo $\sigma(\mathbf{x})$ and surface roughness $\rho(\mathbf{x})$ as additional guiding inputs since the integrand includes the BRDF. The result of the analytic integration needs to be scaled by the area of the light source.

\subsection{Global Illumination}
Last, we discuss global illumination, described by the rendering equation, which is typically solved by path tracing:
\begin{multline}
L_o(\mathbf{x}, \mathbf{\omega}_o) = L_e(\mathbf{x}, \mathbf{\omega}_o) + \\
\int_{\mathbf{\omega}_i \in \Omega} f_r(\mathbf{x}, \mathbf{\omega}_i, \mathbf{\omega}_o) L_i(\mathbf{x}, \mathbf{\omega}_i) (\mathbf{n}(\mathbf{x}) \cdot \mathbf{\omega}_i) \diff \mathbf{\omega}_i,
\end{multline}
where $L_o(\mathbf{x}, \mathbf{\omega}_o)$ is outgoing radiance from point $\mathbf{x}$ in direction $\mathbf{\omega}_o$ and $L_i(\mathbf{x}, \mathbf{\omega}_i)$ is incoming (incidence) radiance at point $\mathbf{x}$ from direction $\mathbf{\omega}_i$. We approximate the integrand by an MLP:
\begin{multline}
\mathbf{g}(\mathbf{\omega}_i \mid \mathbf{x}, \mathbf{\omega}_o, \mathbf{n}(\mathbf{x}), \sigma(\mathbf{x}), \rho(\mathbf{x})) \approx \\\mathbf{f}(\mathbf{x}, \mathbf{\omega}_i, \mathbf{\omega}_o) = f_r(\mathbf{x}, \mathbf{\omega}_i, \mathbf{\omega}_o) L_i(\mathbf{x}, \mathbf{\omega}_i) (\mathbf{n}(\mathbf{x}) \cdot \mathbf{\omega}_i),
\end{multline}
At first glance, everything appears to work correctly; nonetheless, there is a few caveats. The MLP can learn the radiance field quite credibly even from the noisy estimates \rev{(thanks to the Adam optimizer~\cite{Kingma2014})}. Note that in this case, we need to learn the whole 7D dimensional triple product, not only incidence radiance. However, for control variates, we also need access to $\mathbf{f}(\mathbf{x}, \mathbf{\omega}_i, \mathbf{\omega}_o)$, which is a product including $L_i(\mathbf{x}, \mathbf{\omega}_i)$\rev{, which itself is a nested integral)}, to integrate the residual integral. Typically, we have only a one-sample estimate of $L_i(\mathbf{x}, \mathbf{\omega}_i)$ available, which could be very far from the actual value. 

We also need to make sure that the domains of analytic integration and the residual integration are the same. In path tracing, we sample the incoming direction, for example, according to BRDF, which might not cover the whole hemisphere (e.g., highly specular surfaces). Consequently, some parts of the hemisphere have zero probability of being sampled where the original integrand has zero contribution. This might not be the case for the MLP as the MLP for such samples might be non-zero, and thus such samples are included in the analytic integration. Thus, we need to sample according to a probability density that is non-zero on the whole hemisphere \rev{to match the analytic integration}. Note that for the sake of brevity, we do not consider transparent materials (i.e., refraction).

\section{Experiments and Results}
We discuss the experiments we conducted to evaluate the control variates \rev{using} our geometric integration. In all experiments, we assume that $\alpha = 1$ (i.e., perfect correlation). We use mean squared error (MSE) as an error metric as well as a loss function in all cases.

\subsection{Analytic Functions}
\label{Sec:AnalyticFunctions}
\begin{figure*}
\centering

\begin{overpic}[height=0.225\linewidth]{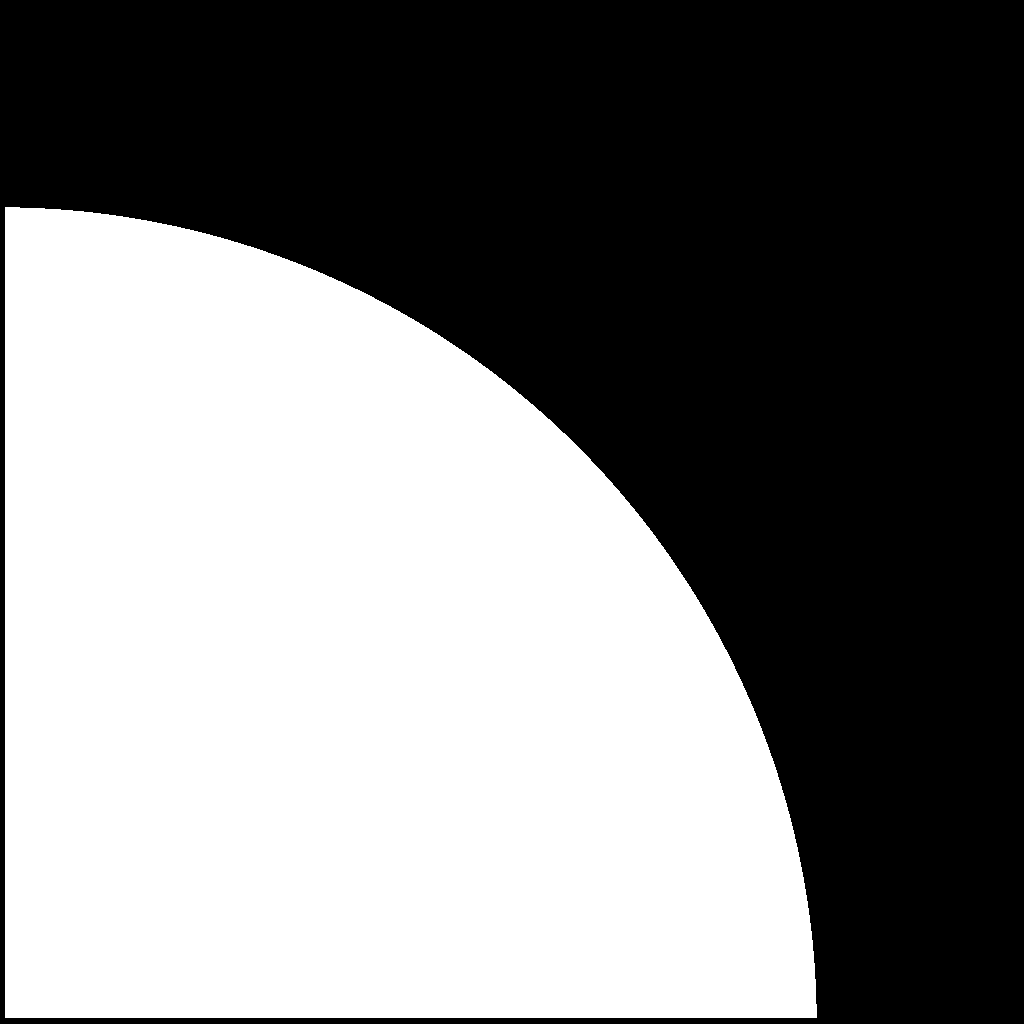}
\put(-8, 50){\makebox(0,0){\rotatebox{90}{$2$ if $x^2 + y^2 < \frac{2}{\pi}$}}}
\end{overpic}
\includegraphics[height=0.225\linewidth]{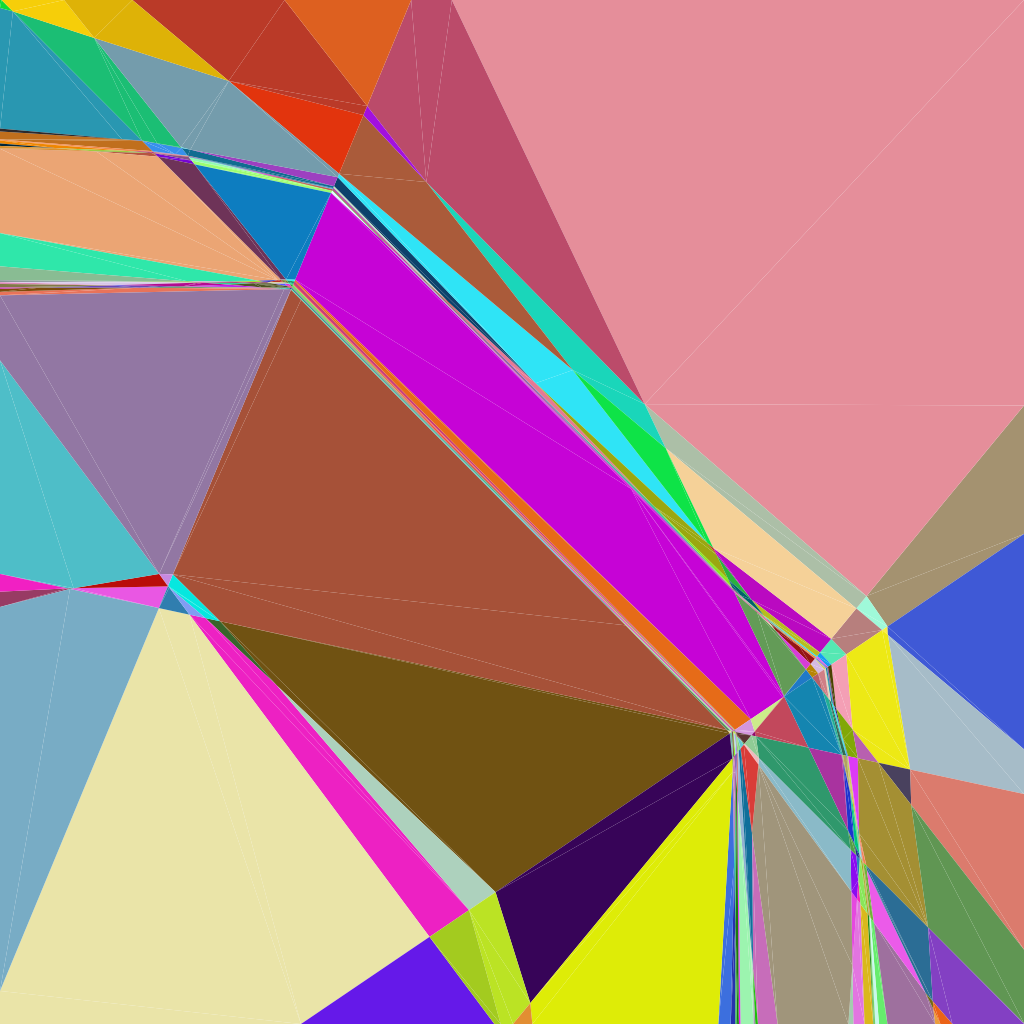}
\includegraphics[height=0.225\linewidth]{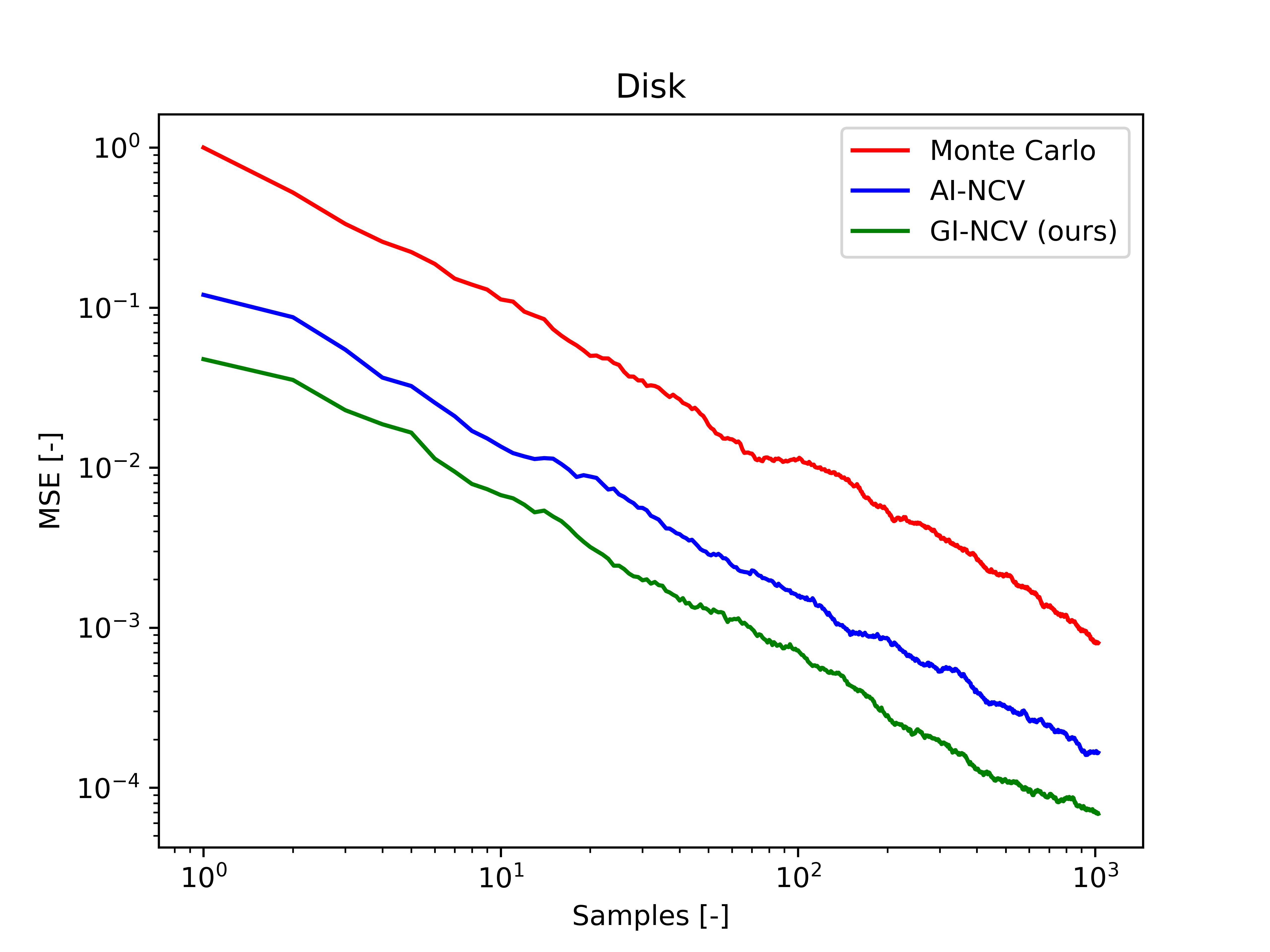}

\begin{overpic}[height=0.225\linewidth]{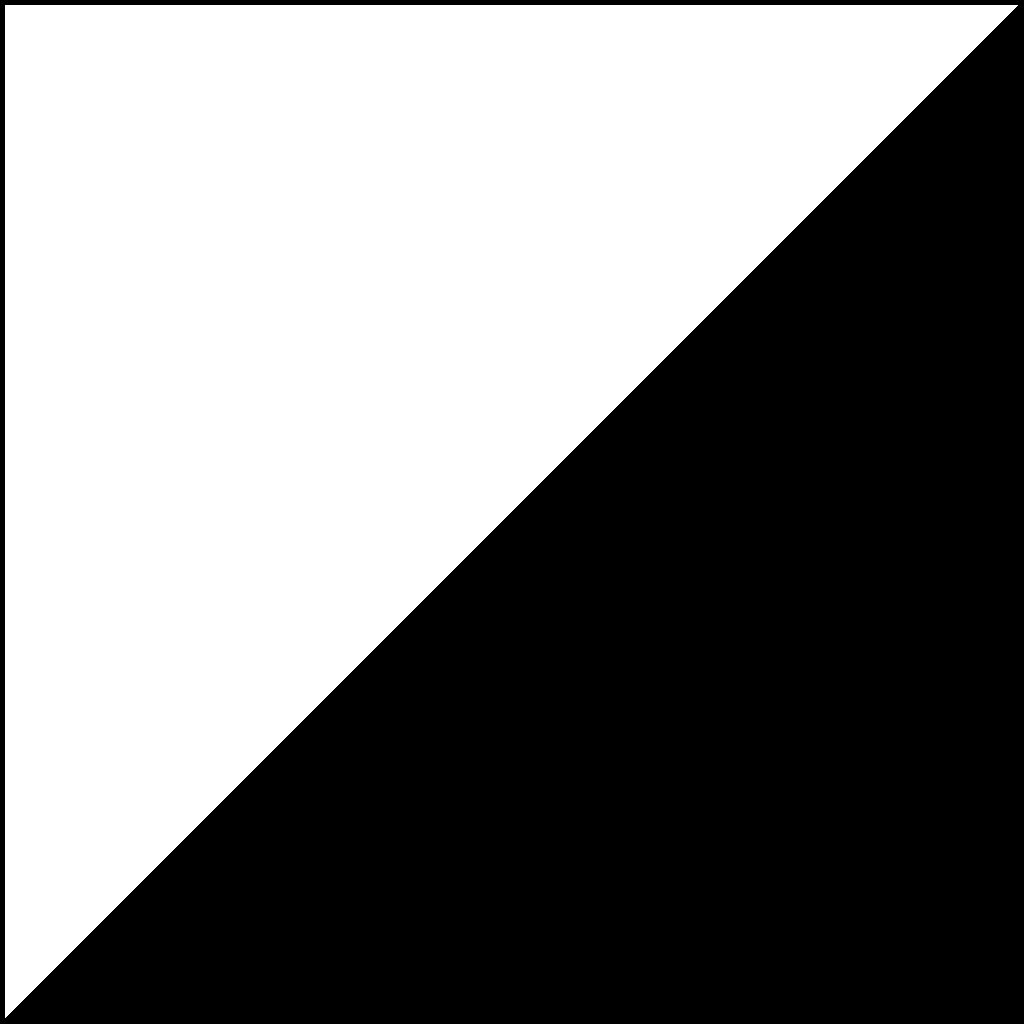}
\put(-8, 50){\makebox(0,0){\rotatebox{90}{$2$ if $x^2 + y^2 < \frac{2}{\pi}$}}}
\end{overpic}
\includegraphics[height=0.225\linewidth]{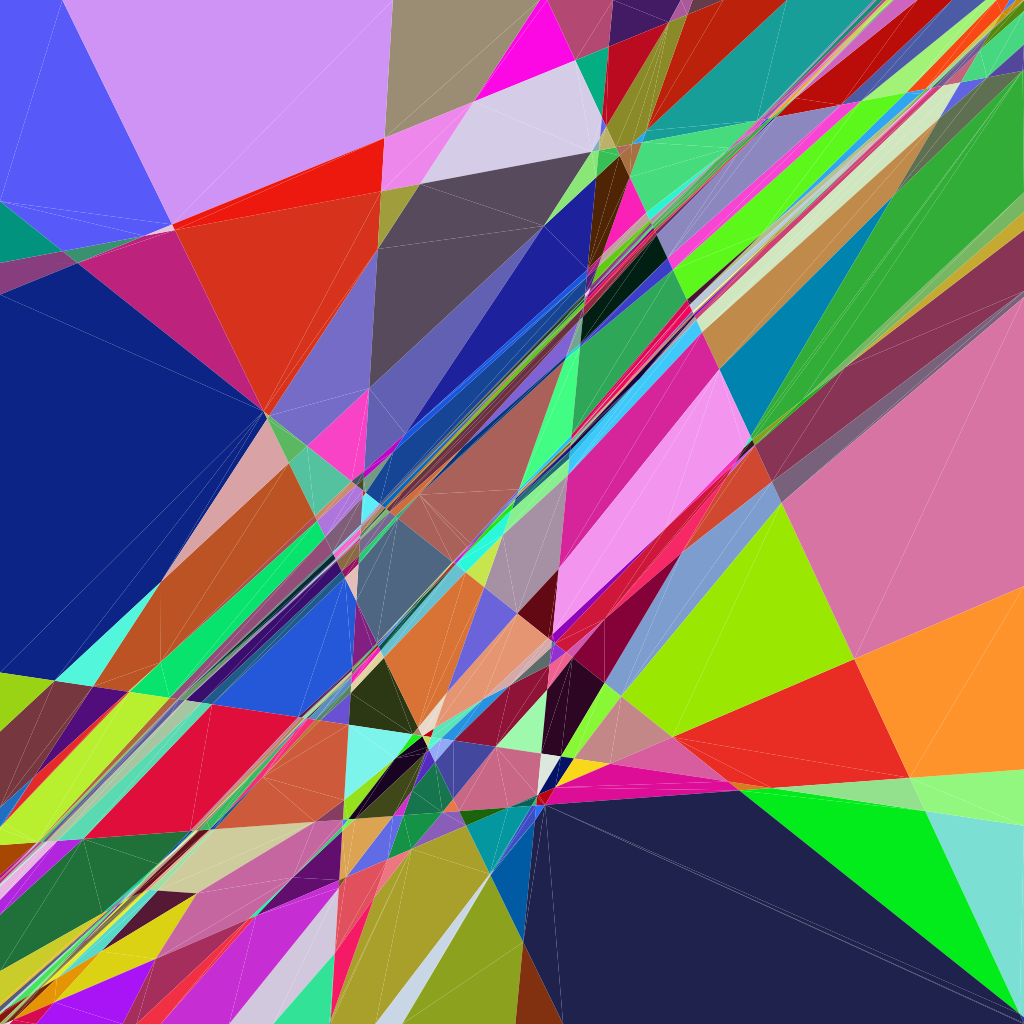}
\includegraphics[height=0.225\linewidth]{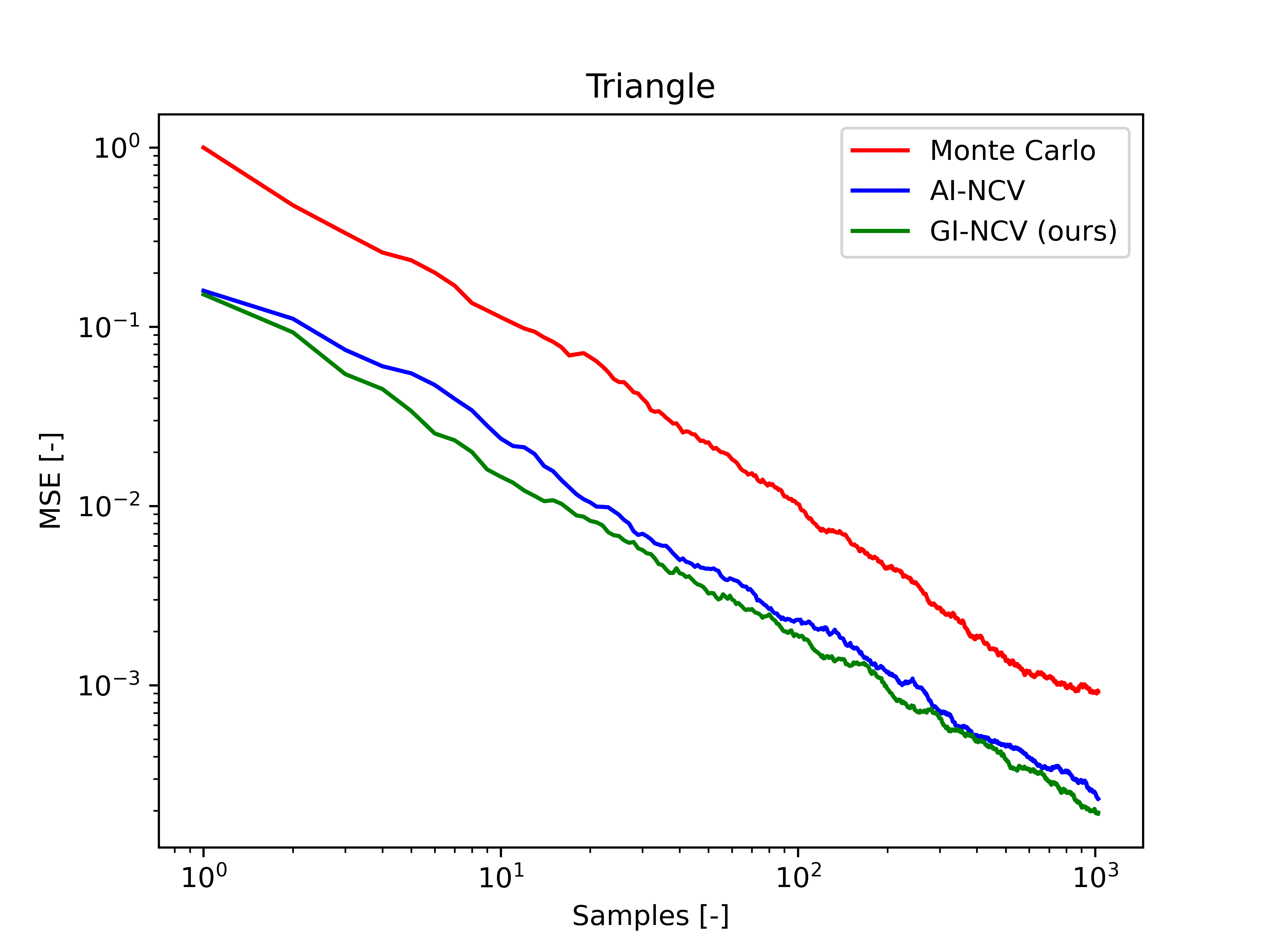}

\begin{overpic}[height=0.225\linewidth]{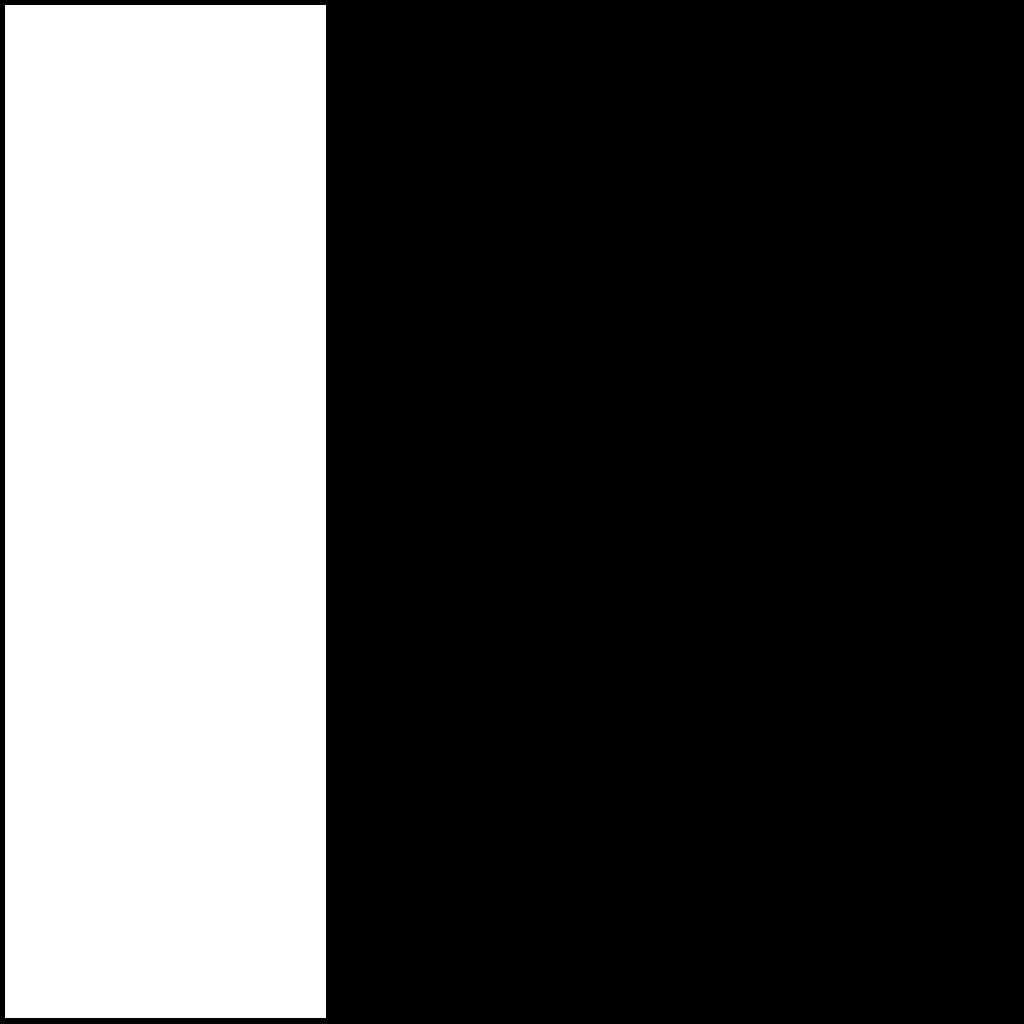}
\put(-8, 50){\makebox(0,0){\rotatebox{90}{$\pi$ if $x < \frac{1}{\pi}$}}}
\end{overpic}
\includegraphics[height=0.225\linewidth]{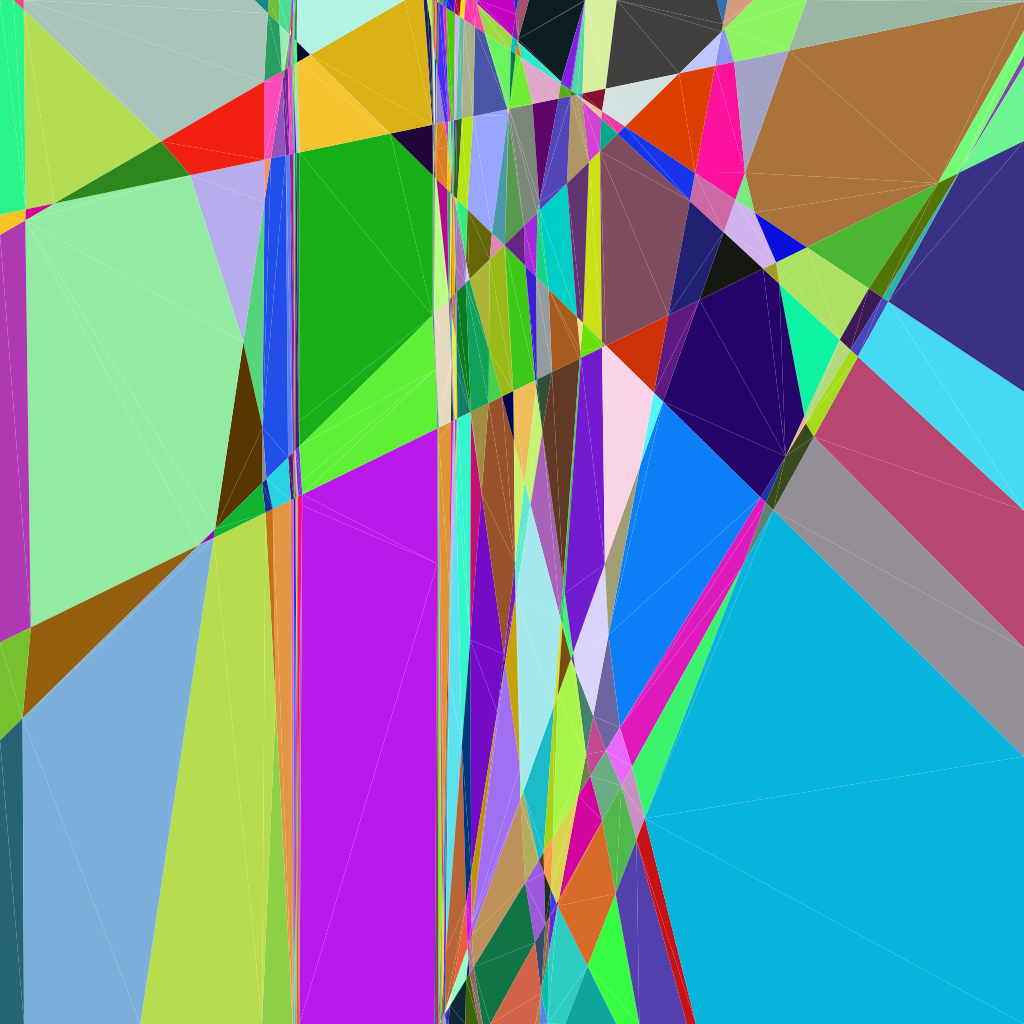}
\includegraphics[height=0.225\linewidth]{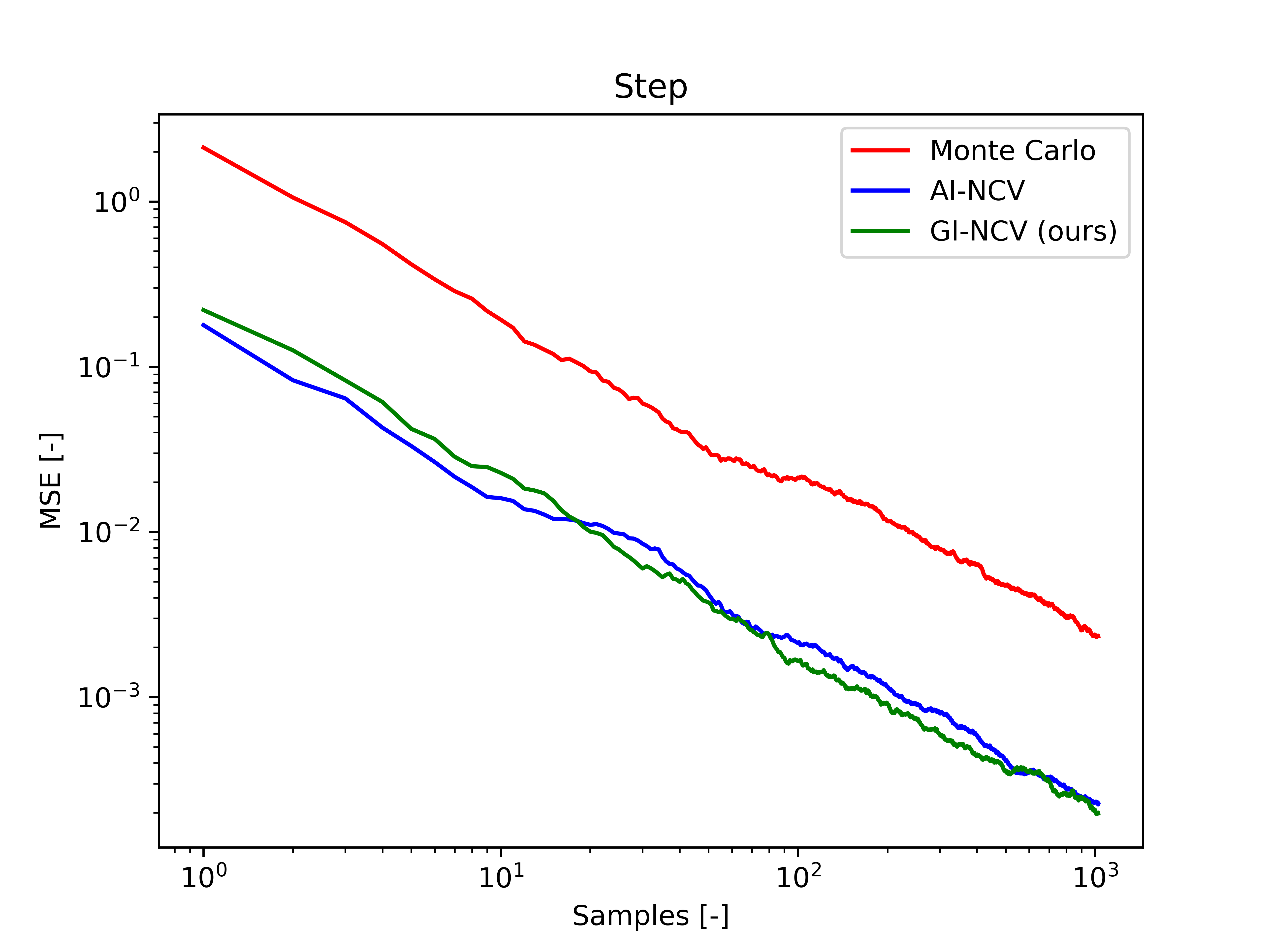}

\begin{overpic}[height=0.225\linewidth]{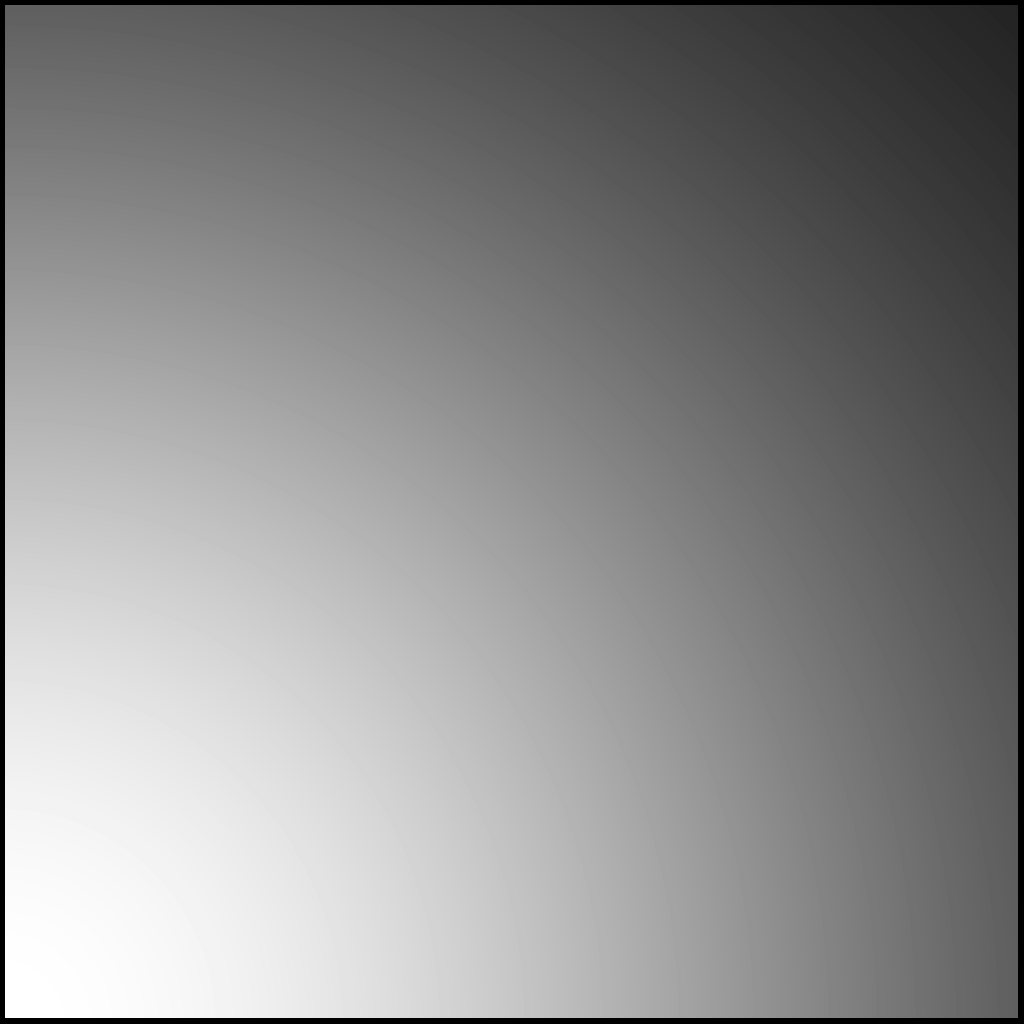}
\put(-8, 50){\makebox(0,0){\rotatebox{90}{$\frac{4}{\pi\mathrm{erf}^2(1)} \exp(-x^2 - y^2)$}}}
\end{overpic}
\includegraphics[height=0.225\linewidth]{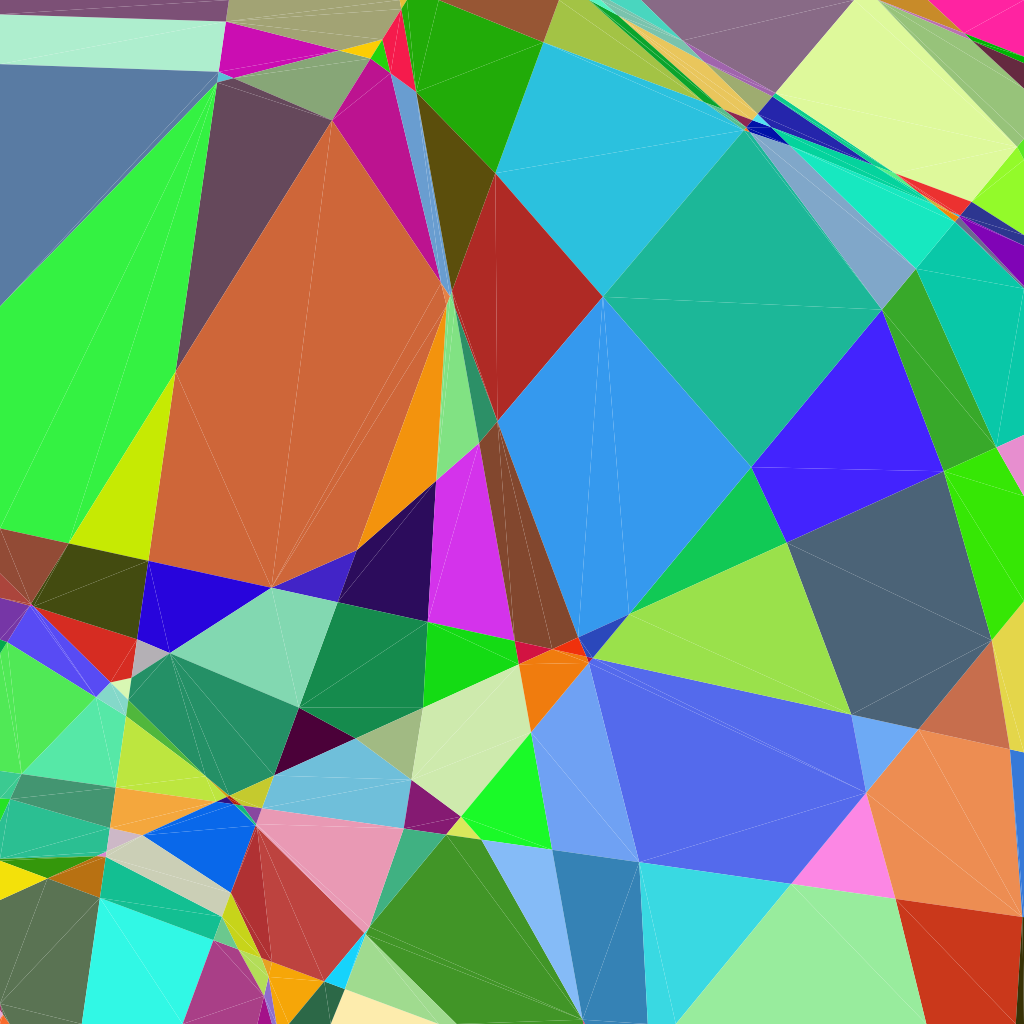}
\includegraphics[height=0.225\linewidth]{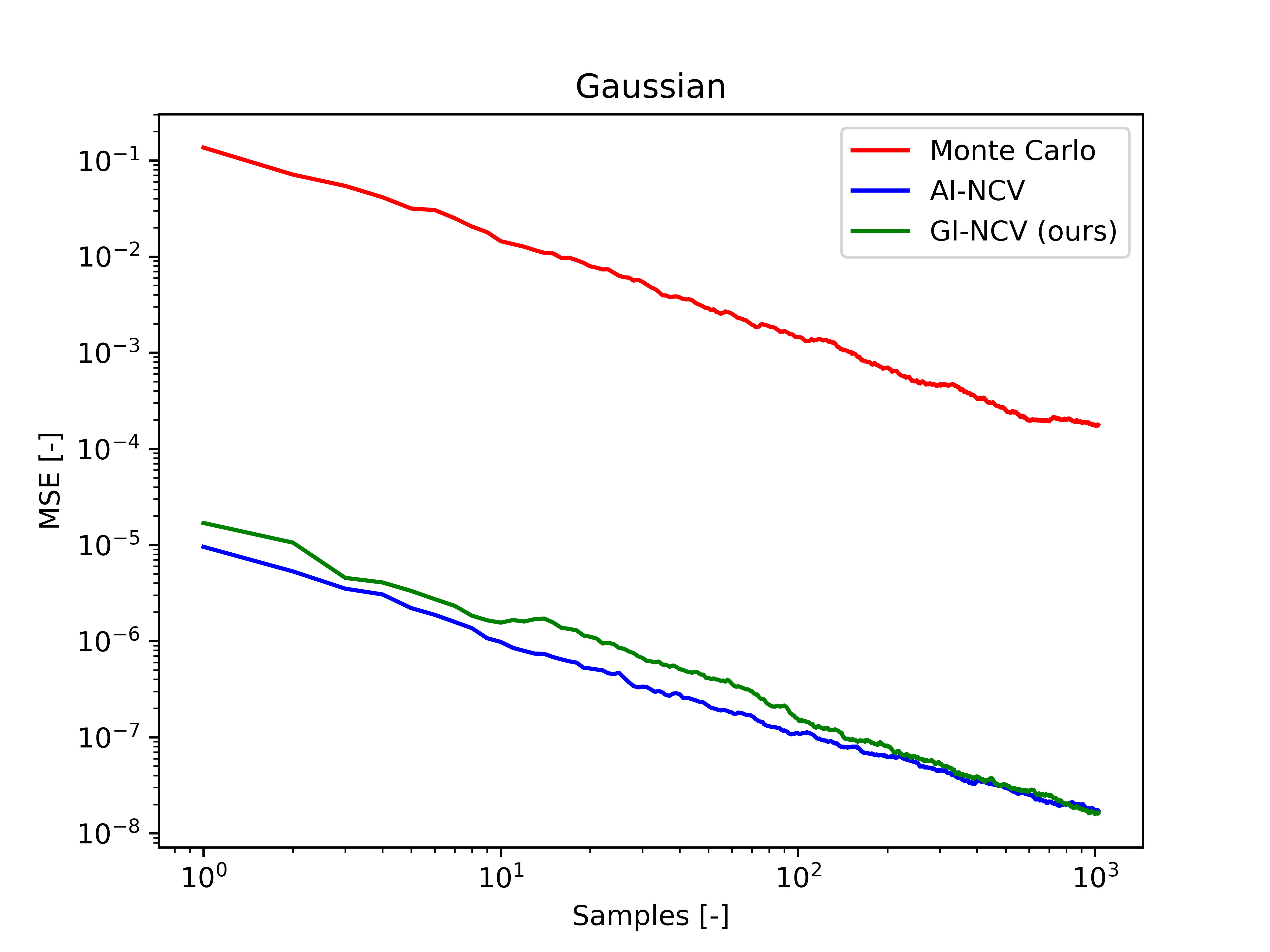}

\begin{overpic}[height=0.225\linewidth]{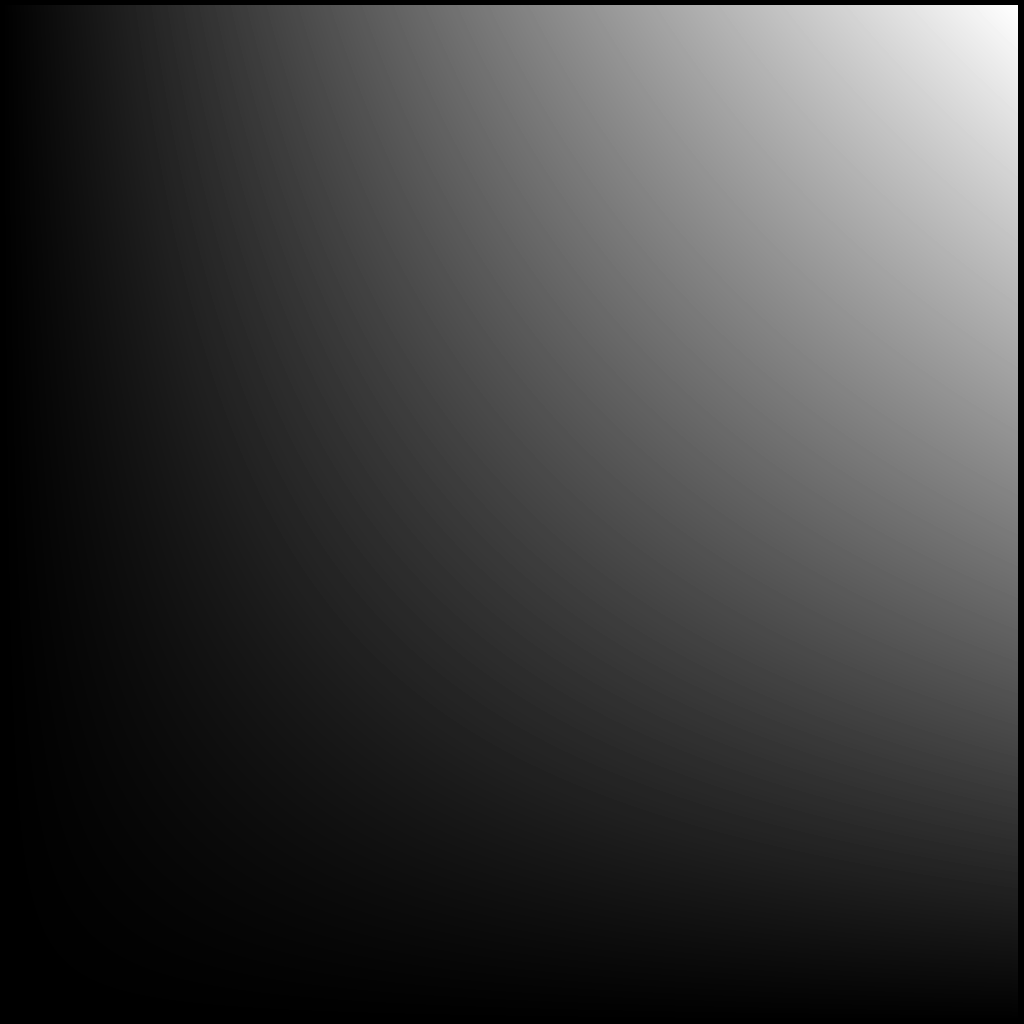}
\put(-8, 50){\makebox(0,0){\rotatebox{90}{$4xy$}}}
\end{overpic}
\includegraphics[height=0.225\linewidth]{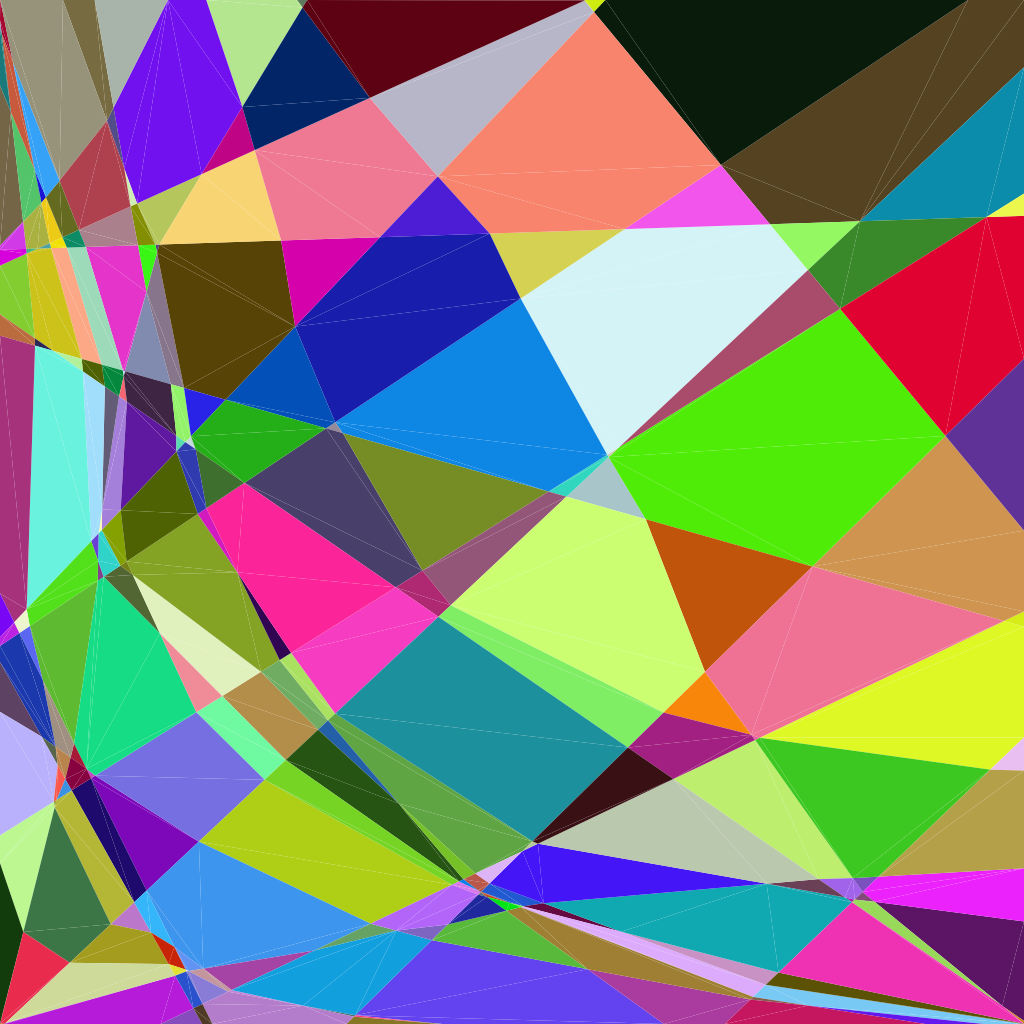}
\includegraphics[height=0.225\linewidth]{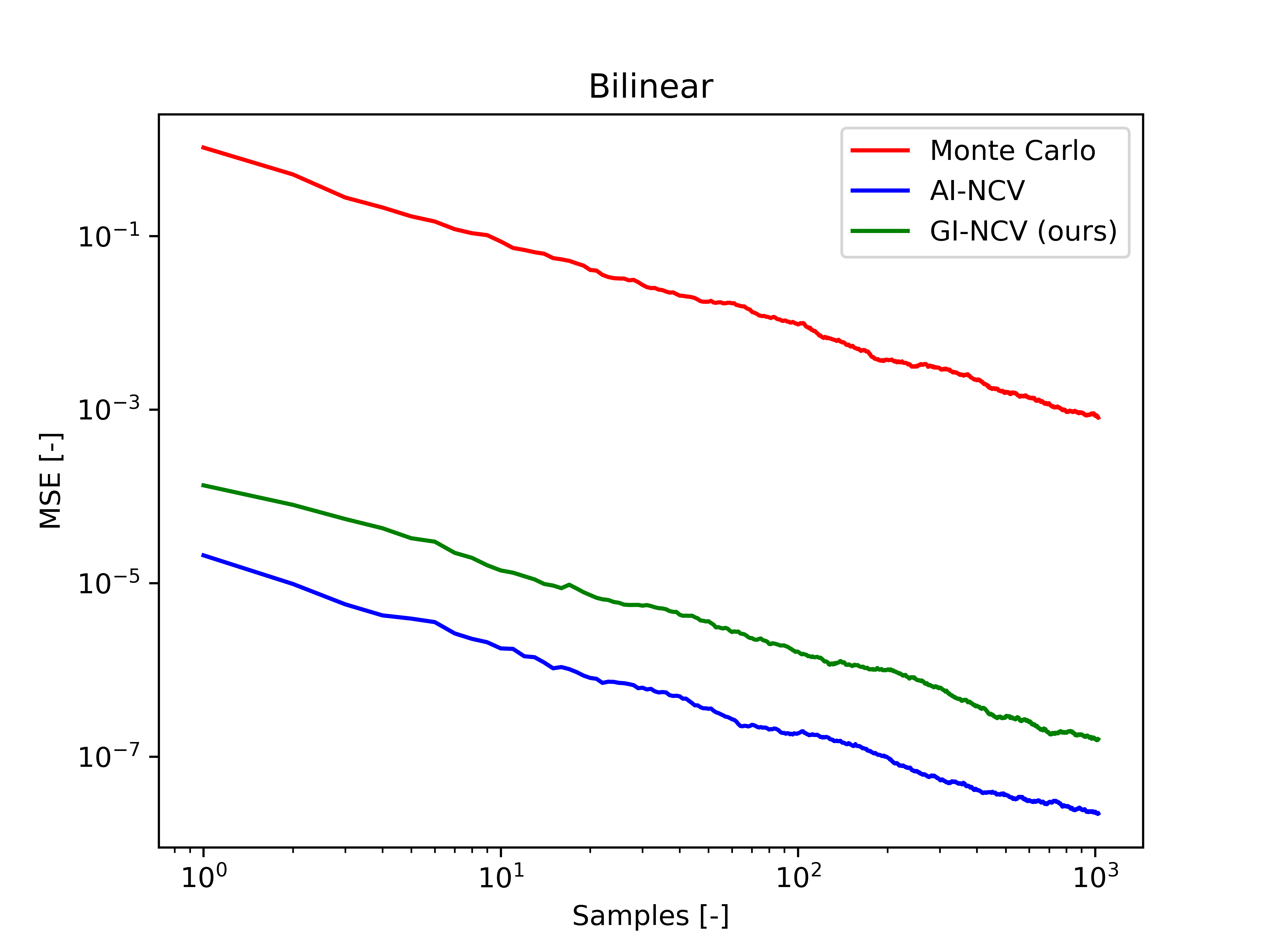}

\caption{Comparison of the vanilla Monte Carlo method, Monte Carlo with neural control variates using the automatic integration~\cite{Lindell2021,Li2024} (AI-NCV), and Monte Carlo with neural control variates using our geometric integration (GI-NCV), integrating simple analytic 2D functions~\cite{Christensen2018,Burley2020}: function visualization (left), geometric subdivision of our method highlighting piecewise affine regions (middle), and the convergence graphs (right). We used 128 trials to estimate the variance and up to 1024 samples for the convergence graphs. Note that all functions are normalized such that they integrate to one.}
\label{Fig:AnalyticFunctions}
\end{figure*}

\begin{figure}
\begin{center}
\begin{minipage}{0.2\textwidth}
    \begin{tikzpicture}
        \node[anchor=south west, inner sep=0] (image) at (0,0) {\includegraphics[width=\linewidth]{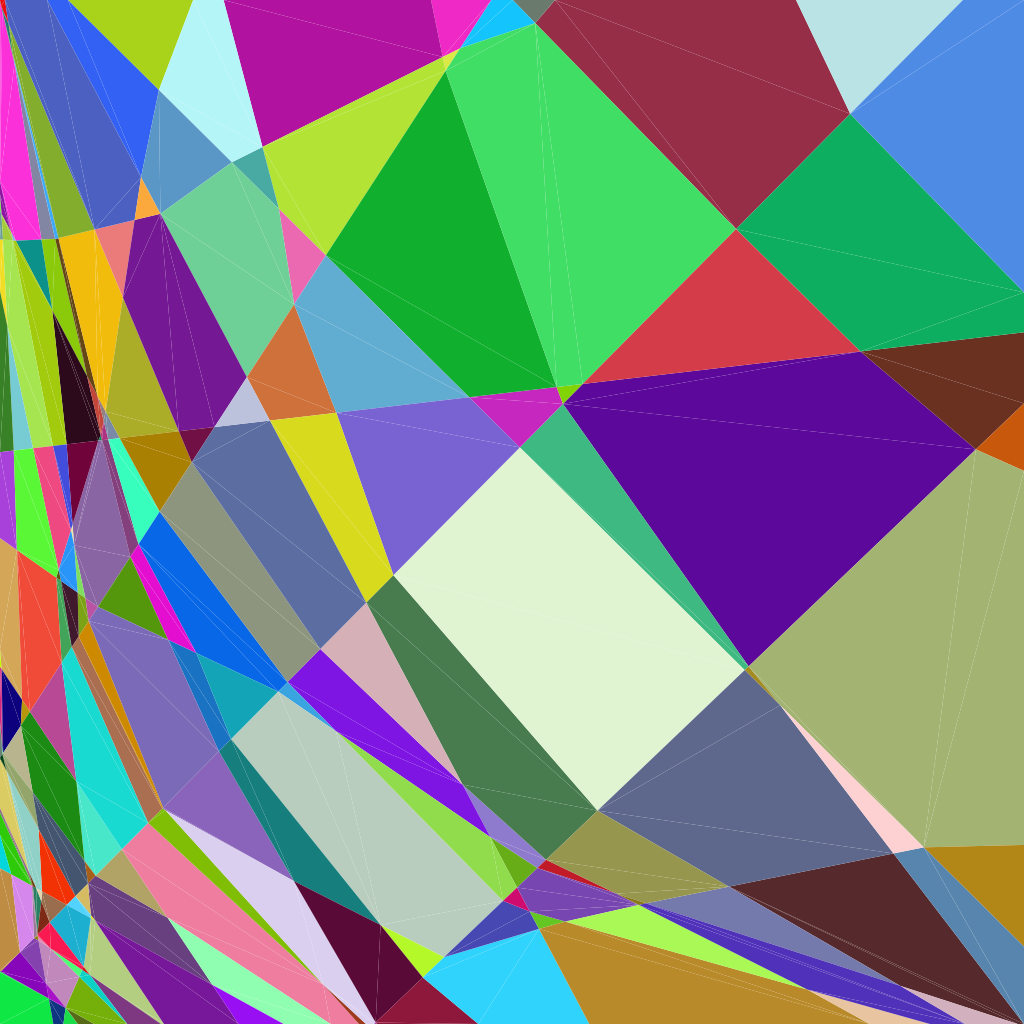}};
        \node[anchor=south east, xshift=-2mm, yshift=2mm, fill=white, opacity=0.7, text opacity=1, font=\scriptsize] 
            at (image.south east) {2 hidden layers / 32 neurons};
        \node[anchor=north west, xshift=2mm, yshift=-2mm, fill=white, opacity=0.7, text opacity=1, font=\scriptsize] 
            at (image.north west) {85 faces};
    \end{tikzpicture}
\end{minipage}%
\begin{minipage}{0.2\textwidth}
    \begin{tikzpicture}
        \node[anchor=south west, inner sep=0] (image) at (0,0) {\includegraphics[width=\linewidth]{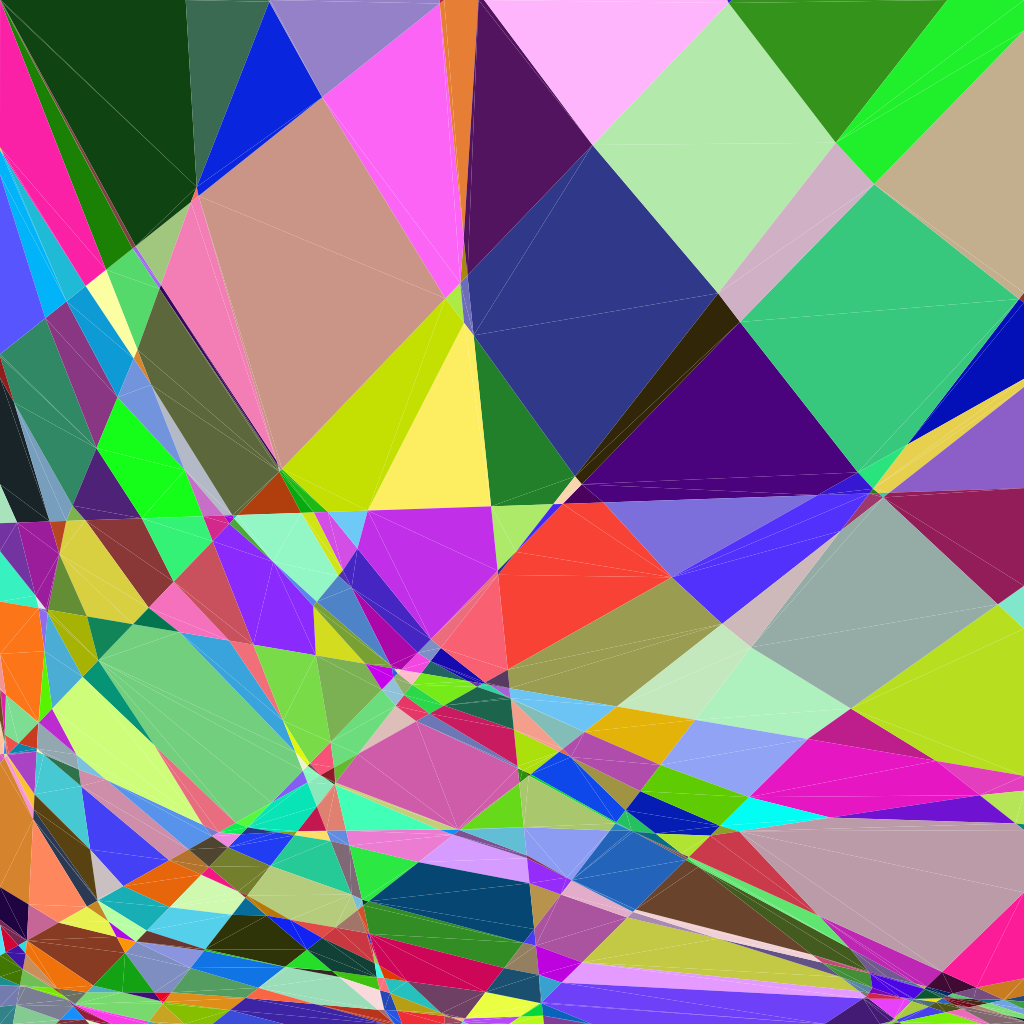}};
        \node[anchor=south east, xshift=-2mm, yshift=2mm, fill=white, opacity=0.7, text opacity=1, font=\scriptsize] 
            at (image.south east) {3 hidden layers / 32 neurons};
        \node[anchor=north west, xshift=2mm, yshift=-2mm, fill=white, opacity=0.7, text opacity=1, font=\scriptsize] 
            at (image.north west) {542 faces};
    \end{tikzpicture}
\end{minipage}

\begin{minipage}{0.2\textwidth}
    \begin{tikzpicture}
        \node[anchor=south west, inner sep=0] (image) at (0,0) {\includegraphics[width=\linewidth]{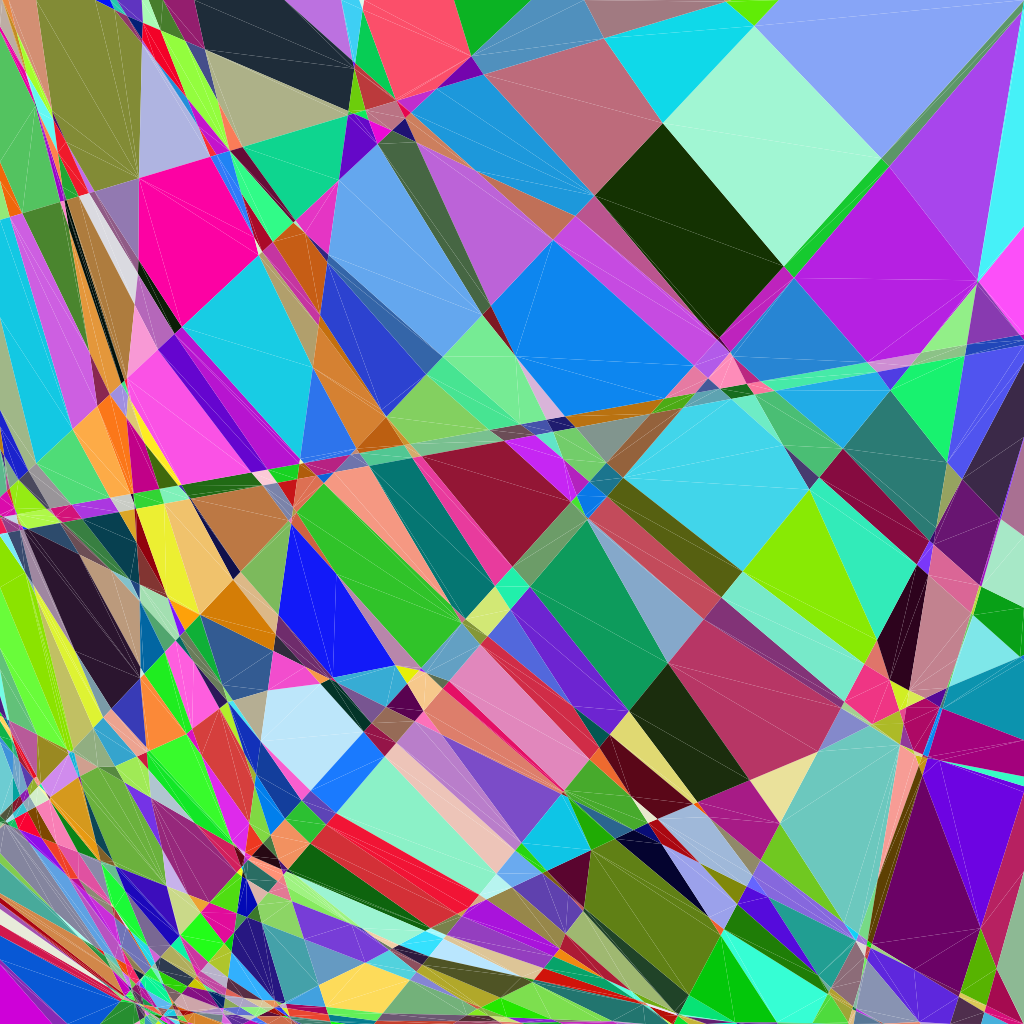}};
        \node[anchor=south east, xshift=-2mm, yshift=2mm, fill=white, opacity=0.7, text opacity=1, font=\scriptsize] 
            at (image.south east) {2 hidden layers / 64 neurons};
        \node[anchor=north west, xshift=2mm, yshift=-2mm, fill=white, opacity=0.7, text opacity=1, font=\scriptsize] 
            at (image.north west) {528 faces};
    \end{tikzpicture}
\end{minipage}%
\begin{minipage}{0.2\textwidth}
    \begin{tikzpicture}
        \node[anchor=south west, inner sep=0] (image) at (0,0) {\includegraphics[width=\linewidth]{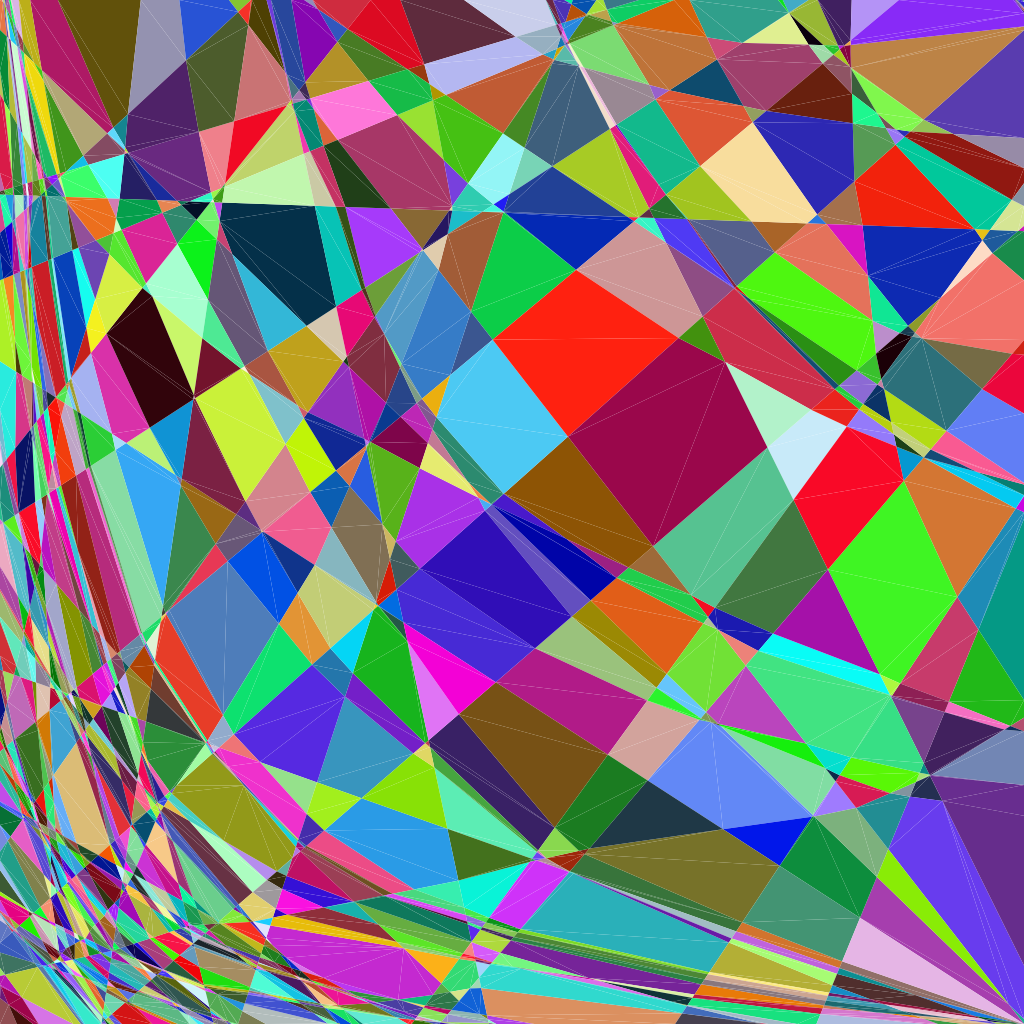}};
        \node[anchor=south east, xshift=-2mm, yshift=2mm, fill=white, opacity=0.7, text opacity=1, font=\scriptsize] 
            at (image.south east) {3 hidden layers / 64 neurons};
        \node[anchor=north west, xshift=2mm, yshift=-2mm, fill=white, opacity=0.7, text opacity=1, font=\scriptsize] 
            at (image.north west) {1570 faces};
    \end{tikzpicture}
\end{minipage}%
\end{center}
\caption{\rev{The number of faces of subdivisions of the bilinear function for different MLP configurations: the number of hidden layers and the number of neurons in each hidden layer.}}
\label{Fig:FaceCount}
\end{figure}

As a proof-of-concept, we implemented our method in PyTorch, integrating a set of analytic 2D functions~\cite{Christensen2018, Burley2020}. We compare control variates with our geometric integration (GI-NCV) to \rev{control variates with} the automatic integration~\cite{Lindell2021,Li2024} (AI-NCV) and vanilla Monte Carlo. We use an MLP with two hidden layers with 32 neurons in each layer for both control variates methods. For GI-NCV, we use the ReLU activation function, while for AI-NCV, we use the sigmoid activation function, which has non-zero second order gradients compared to ReLU. We pre-train each MLP using 5000 epochs. The results are summarized in Figure~\ref{Fig:AnalyticFunctions}. The convergence graphs show the variance estimates up to for 1024 samples, averaged over 128 trials for each sample count. GI-NCV and AI-NCV are competitive, the only difference is due to the different activation functions: GI-NCV with ReLU approximates better discontinuous step-like functions (e.g., the disk function), while AI-NCV better approximates smooth functions (e.g., the bilinear function). \rev{In Figure~\ref{Fig:FaceCount}, we show the number of faces of subdivisions corresponding to the bilinear function for different MLP configurations: the number of hidden layers and the number of neurons per hidden layer. Note that the exact number of faces depends on a particular MLP weights and biases.}

\newcommand{\erf}[1]{\mathrm{erf}\left(#1\right)}

\subsection{Light Transport Simulation}
\label{Sec:ResultsLT}
We integrated our method into our in-house renderer and MLP framework to evaluate it on the light transport problems discussed in Section~\ref{Sec:LightTransport}. We use an MLP with two hidden layers, each with 32 neurons, and the ReLU activation function. We employ multiresolution hashgrid encoding~\cite{Muller2022} with a base resolution of 64, four levels, and two features per level to encode the position. All images are rendered at a resolution of $1024\times1024$ (unless stated otherwise).

Our rendering pipeline consists of three phases. (1) We pre-train the MLP before it is actually used as a control variate. (2) Once the MLP is pre-trained, we integrate the MLP using our analytic integration. (3) We use Monte Carlo integration to integrate the residual integral. We use one training sample per pixel in each epoch. Training samples can be used either only for training or for both training and rendering, introducing correlation but not the bias (see Appendix~\ref{Sec:Proof}). 

In Figure~\ref{Fig:LightTransport}, we present the results when we use the training samples only for training to determine whether using control variates brings any improvement compared to vanilla Monte Carlo regardless of the training overhead. The corresponding curves in the convergence graph are parallel to those of Monte Carlo shifted by some offset in the logarithmic scale. We observe that the control variates can reduce the error for direct lighting and ambient occlusion but not for the indirect lighting (we omit the direct lighting component from the global illumination discussed in Section~\ref{Sec:LightTransport} to emphasize the indirect component) due to the noisy estimates of incidence radiance. \rev{We analyze this problem in  Appendix~\ref{Sec:NoisyEstrimates}}.

In Figure~\ref{Fig:EqualSampleCount}, we show an equal-sample-count comparison that takes into account both training samples (used for both training and rendering) and the rendering samples. We offset the corresponding curve by the number of training samples in the convergence graph (see Figure~\ref{Fig:EqualSampleCount}). Our method using control variates can achieve lower error compared to vanilla Monte Carlo. As a sanity check, we show that Monte Carlo integration of the MLP converges to our analytic integration, which is a necessary condition to avoid bias.


\begin{figure*}
\begin{center}
\begin{minipage}{0.22\textwidth}
    \begin{tikzpicture}
        \node[anchor=south west, inner sep=0] (image) at (0,0) {\includegraphics[width=\linewidth]{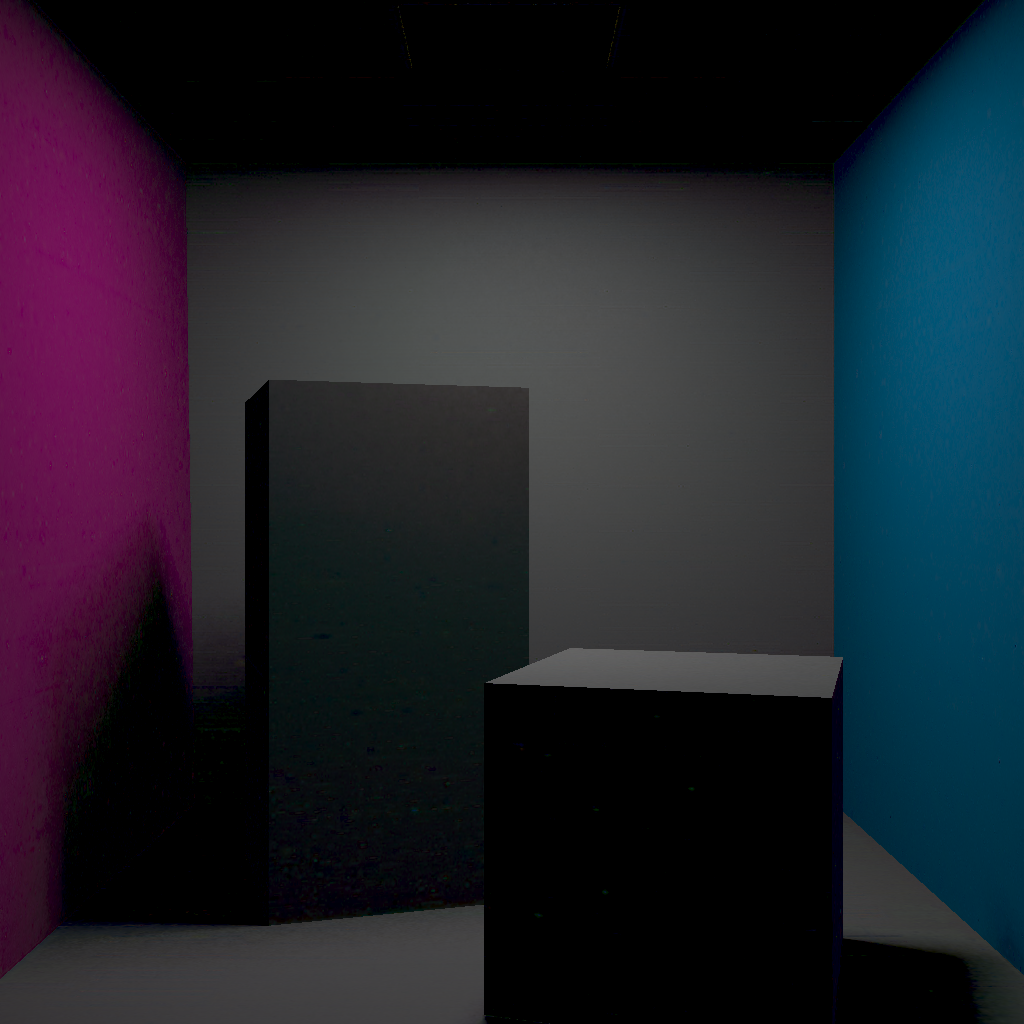}};
        \node[anchor=north west, xshift=2mm, yshift=-2mm, fill=white, opacity=0.7, text opacity=1, font=\scriptsize] 
            at (image.north west) {$G$ (ours)};
    \end{tikzpicture}
\end{minipage}%
\begin{minipage}{0.22\textwidth}
    \begin{tikzpicture}
        \node[anchor=south west, inner sep=0] (image) at (0,0) {\includegraphics[width=\linewidth]{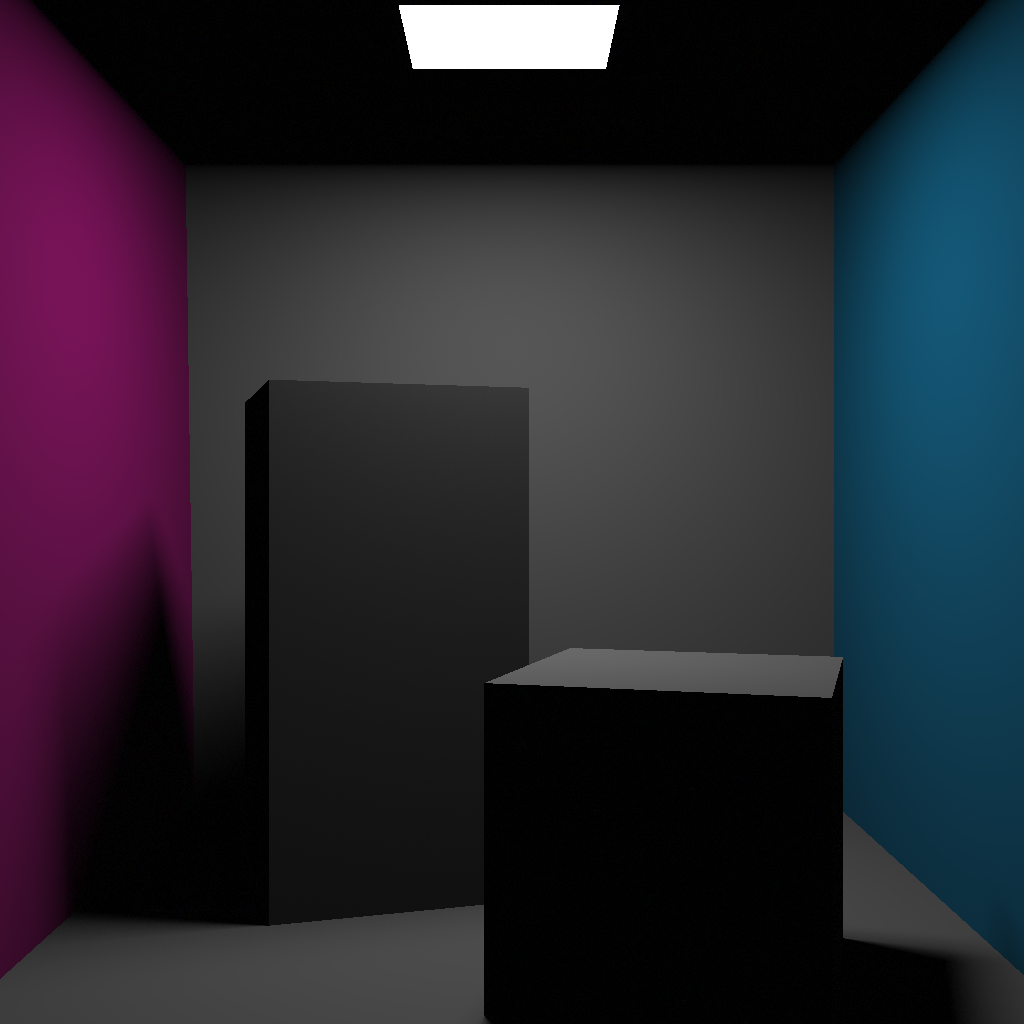}};
        \node[anchor=south east, xshift=-2mm, yshift=2mm, fill=white, opacity=0.7, text opacity=1, font=\scriptsize] 
            at (image.south east) {MSE $6.8 \cdot 10^{-8}$};
        \node[anchor=north west, xshift=2mm, yshift=-2mm, fill=white, opacity=0.7, text opacity=1, font=\scriptsize] 
            at (image.north west) {GI-NCV (ours)};
    \end{tikzpicture}
\end{minipage}%
\begin{minipage}{0.22\textwidth}
    \begin{tikzpicture}
        \node[anchor=south west, inner sep=0] (image) at (0,0) {\includegraphics[width=\linewidth]{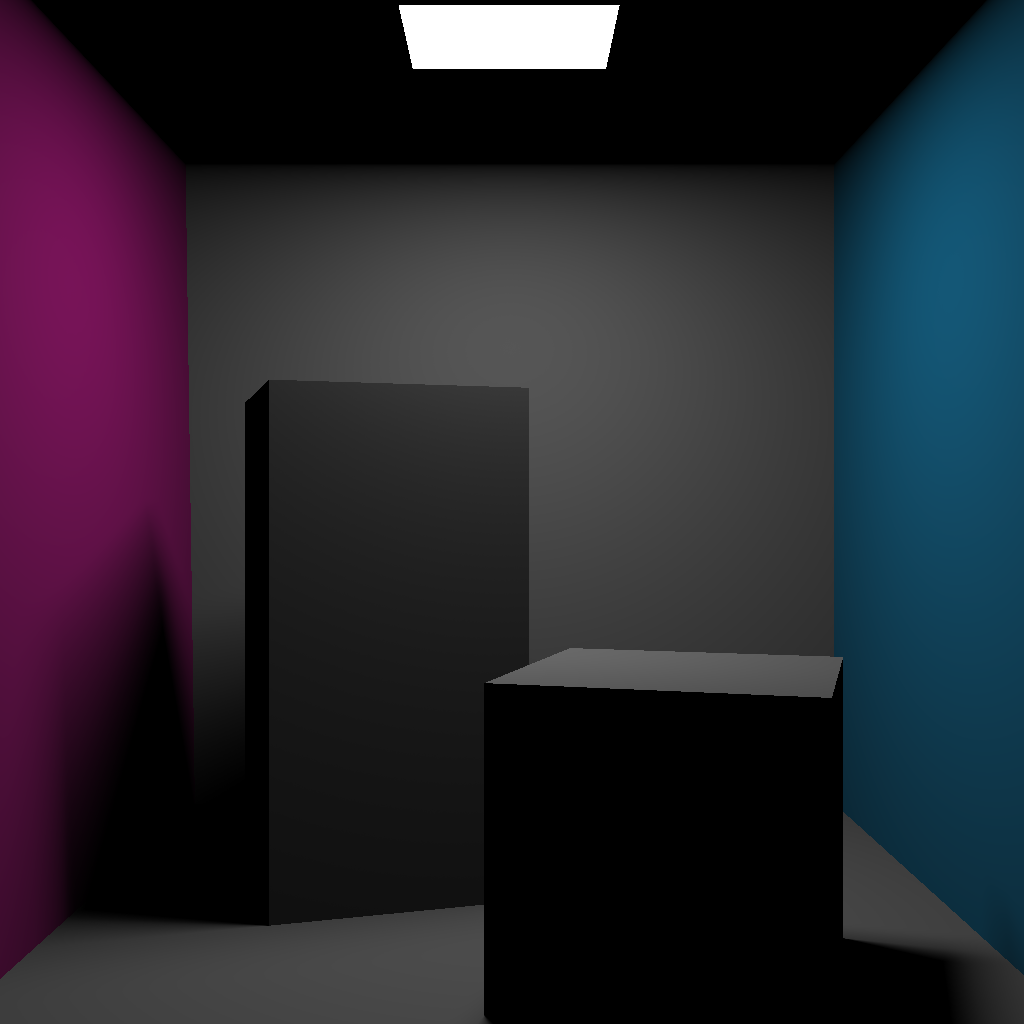}};
        \node[anchor=north west, xshift=2mm, yshift=-2mm, fill=white, opacity=0.7, text opacity=1, font=\scriptsize] 
            at (image.north west) {Reference};
    \end{tikzpicture}
\end{minipage}%
\begin{minipage}{0.3\textwidth}
    \begin{tikzpicture}
        \node[anchor=south west, inner sep=0] (image) at (0,0) {\includegraphics[width=\linewidth]{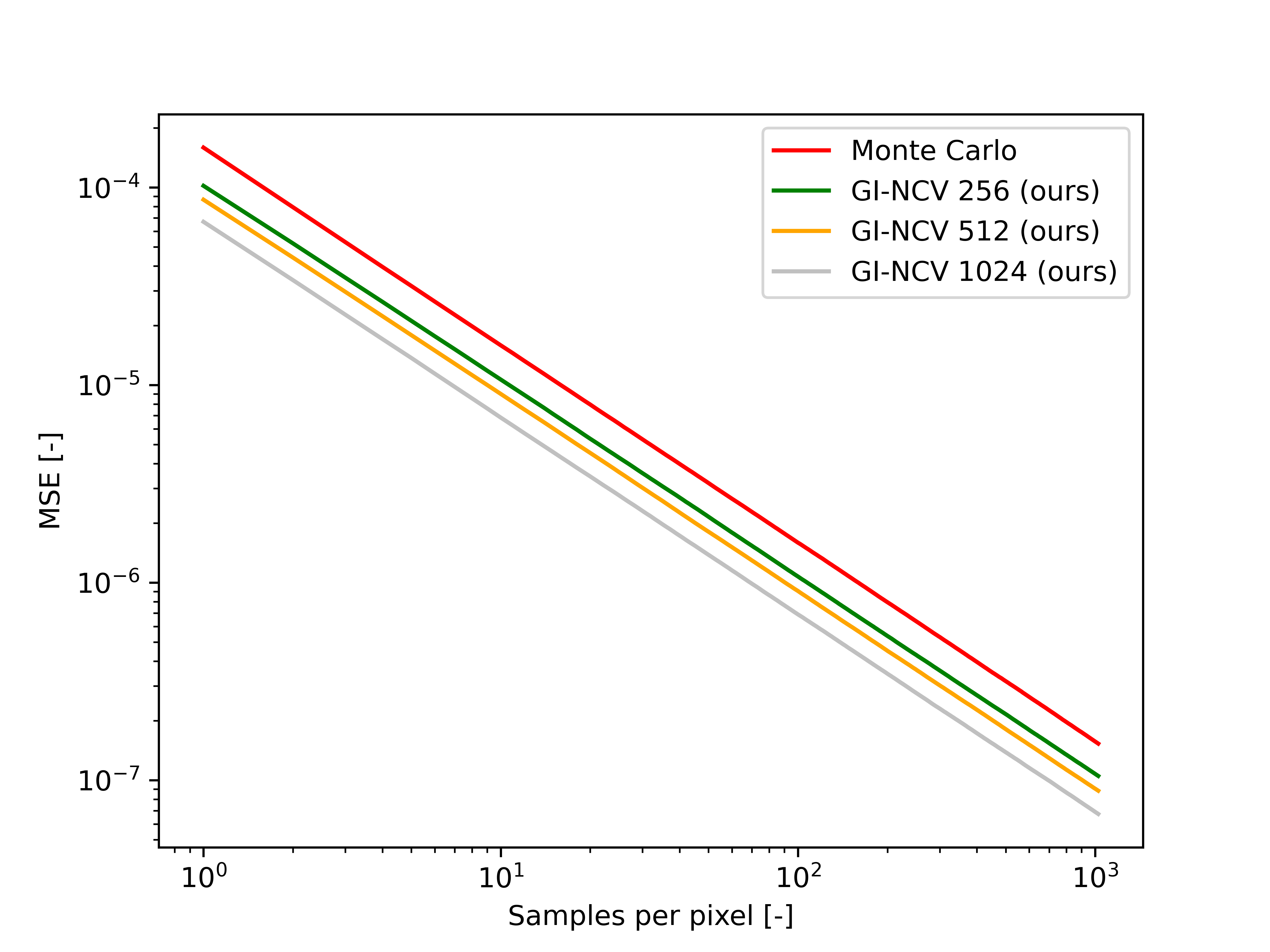}};
    \end{tikzpicture}
\end{minipage}

\begin{minipage}{0.22\textwidth}
    \begin{tikzpicture}
        \node[anchor=south west, inner sep=0] (image) at (0,0) {\includegraphics[width=\linewidth]{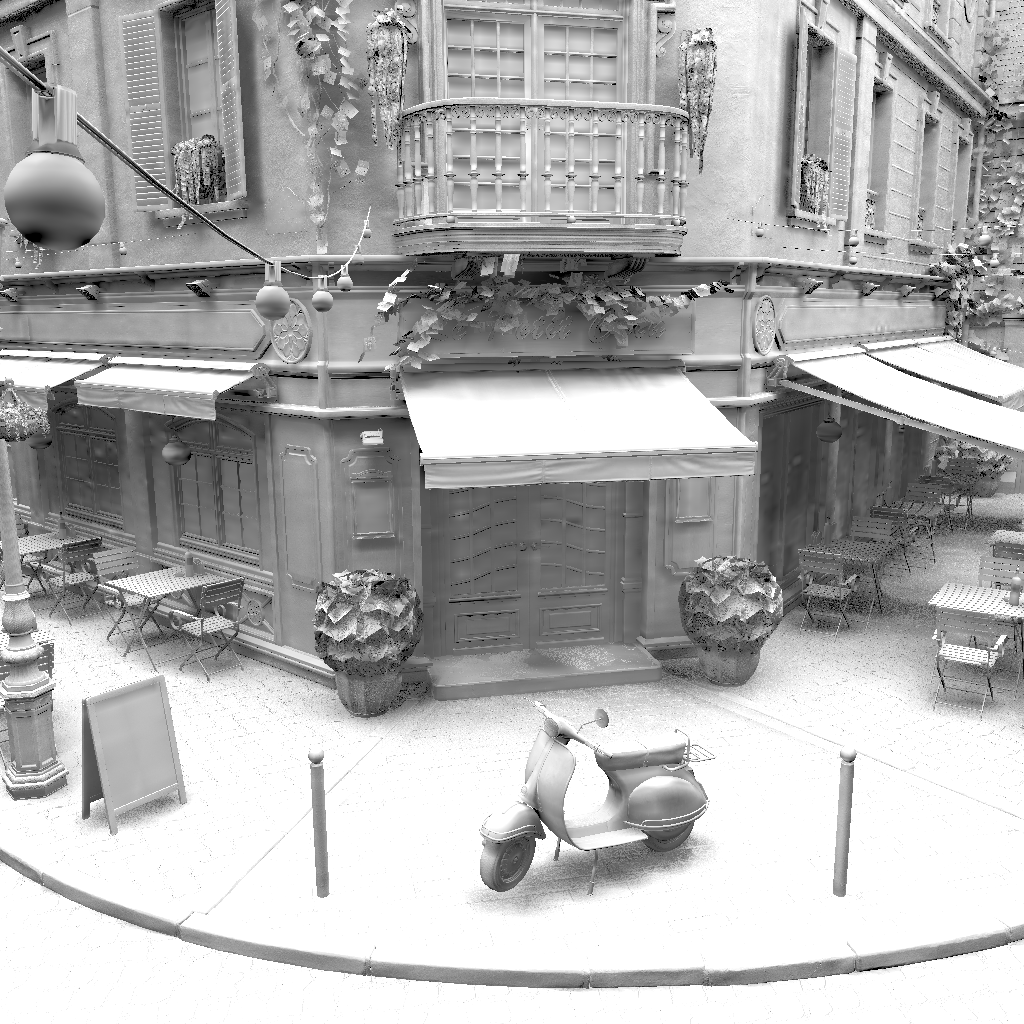}};
        \node[anchor=north west, xshift=2mm, yshift=-2mm, fill=white, opacity=0.7, text opacity=1, font=\scriptsize] 
            at (image.north west) {$G$ (ours)};
    \end{tikzpicture}
\end{minipage}%
\begin{minipage}{0.22\textwidth}
    \begin{tikzpicture}
        \node[anchor=south west, inner sep=0] (image) at (0,0) {\includegraphics[width=\linewidth]{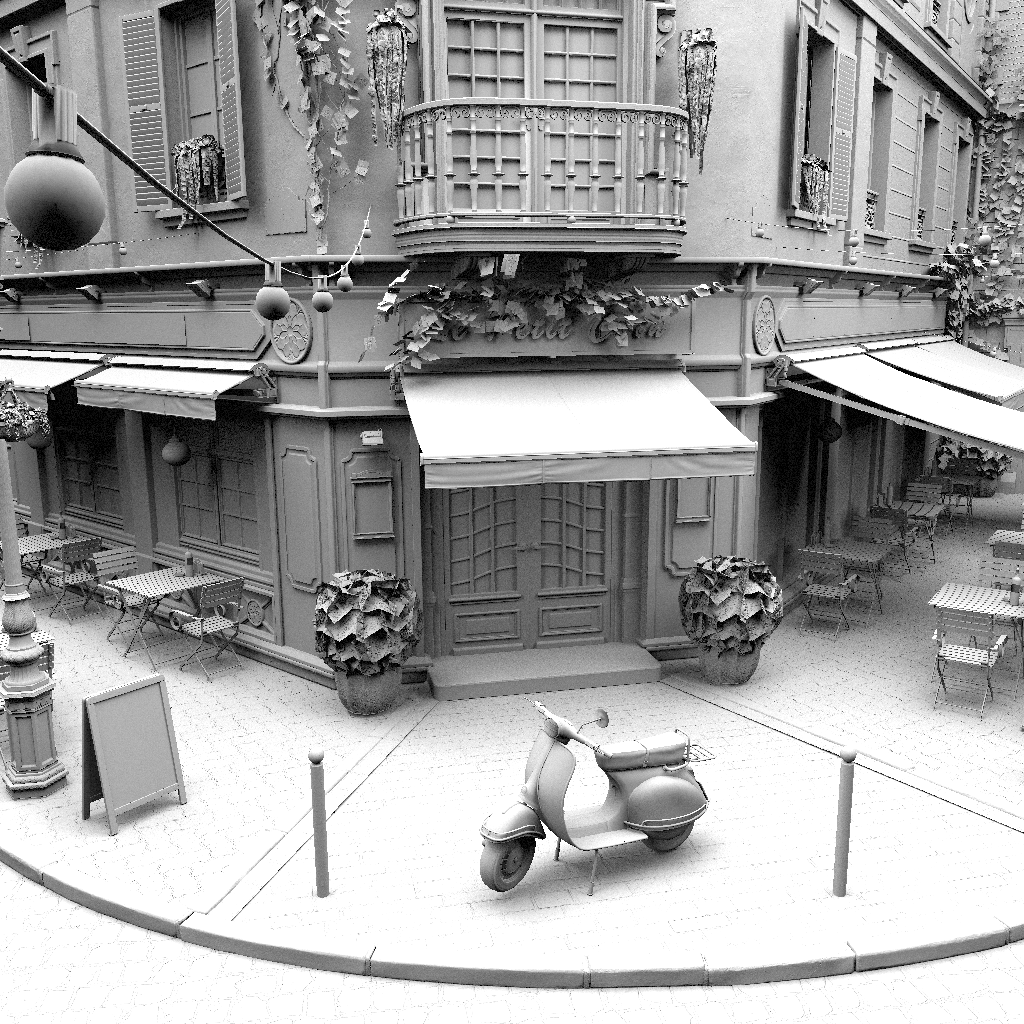}};
        \node[anchor=south east, xshift=-2mm, yshift=2mm, fill=white, opacity=0.7, text opacity=1, font=\scriptsize] 
            at (image.south east) {MSE $7.7 \cdot 10^{-4}$};
        \node[anchor=north west, xshift=2mm, yshift=-2mm, fill=white, opacity=0.7, text opacity=1, font=\scriptsize] 
            at (image.north west) {GI-NCV (ours)};
    \end{tikzpicture}
\end{minipage}%
\begin{minipage}{0.22\textwidth}
    \begin{tikzpicture}
        \node[anchor=south west, inner sep=0] (image) at (0,0) {\includegraphics[width=\linewidth]{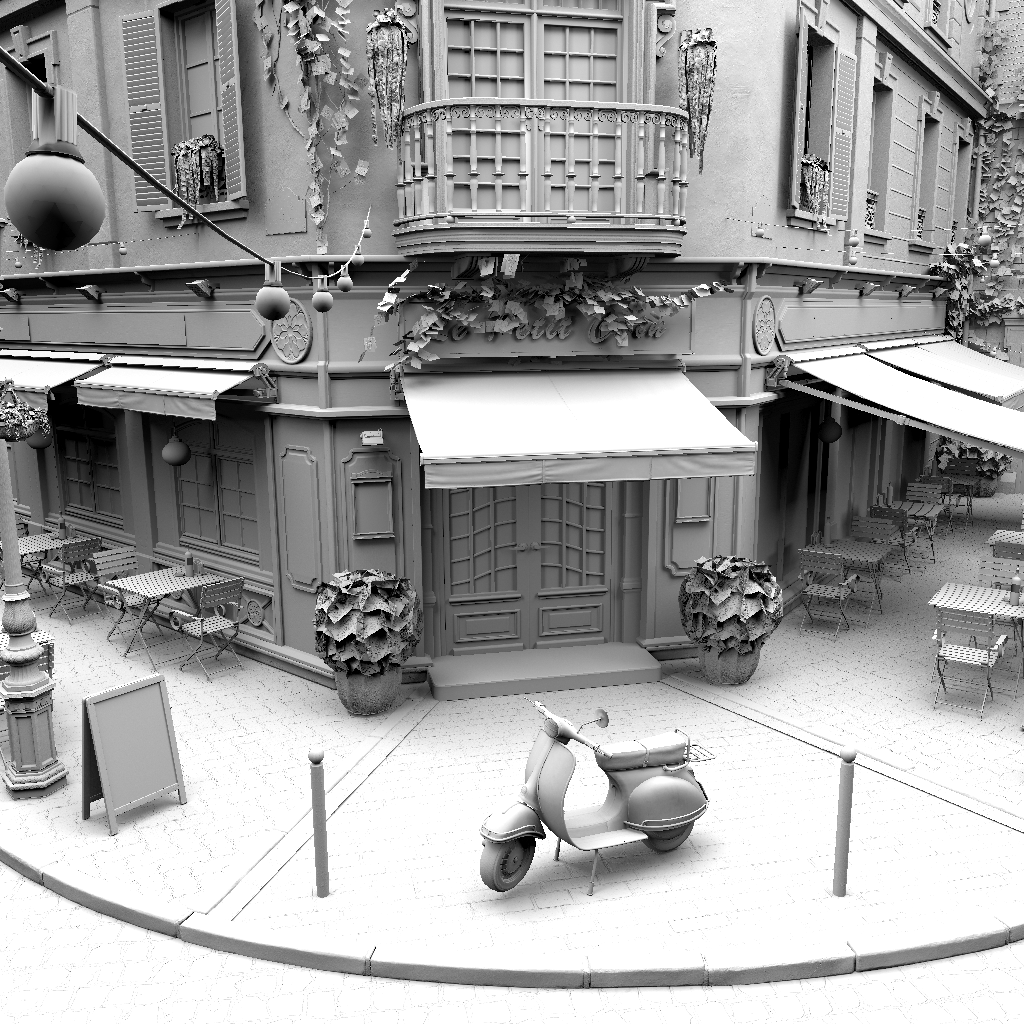}};
        \node[anchor=north west, xshift=2mm, yshift=-2mm, fill=white, opacity=0.7, text opacity=1, font=\scriptsize] 
            at (image.north west) {Reference};
    \end{tikzpicture}
\end{minipage}%
\begin{minipage}{0.3\textwidth}
    \begin{tikzpicture}
        \node[anchor=south west, inner sep=0] (image) at (0,0) {\includegraphics[width=\linewidth]{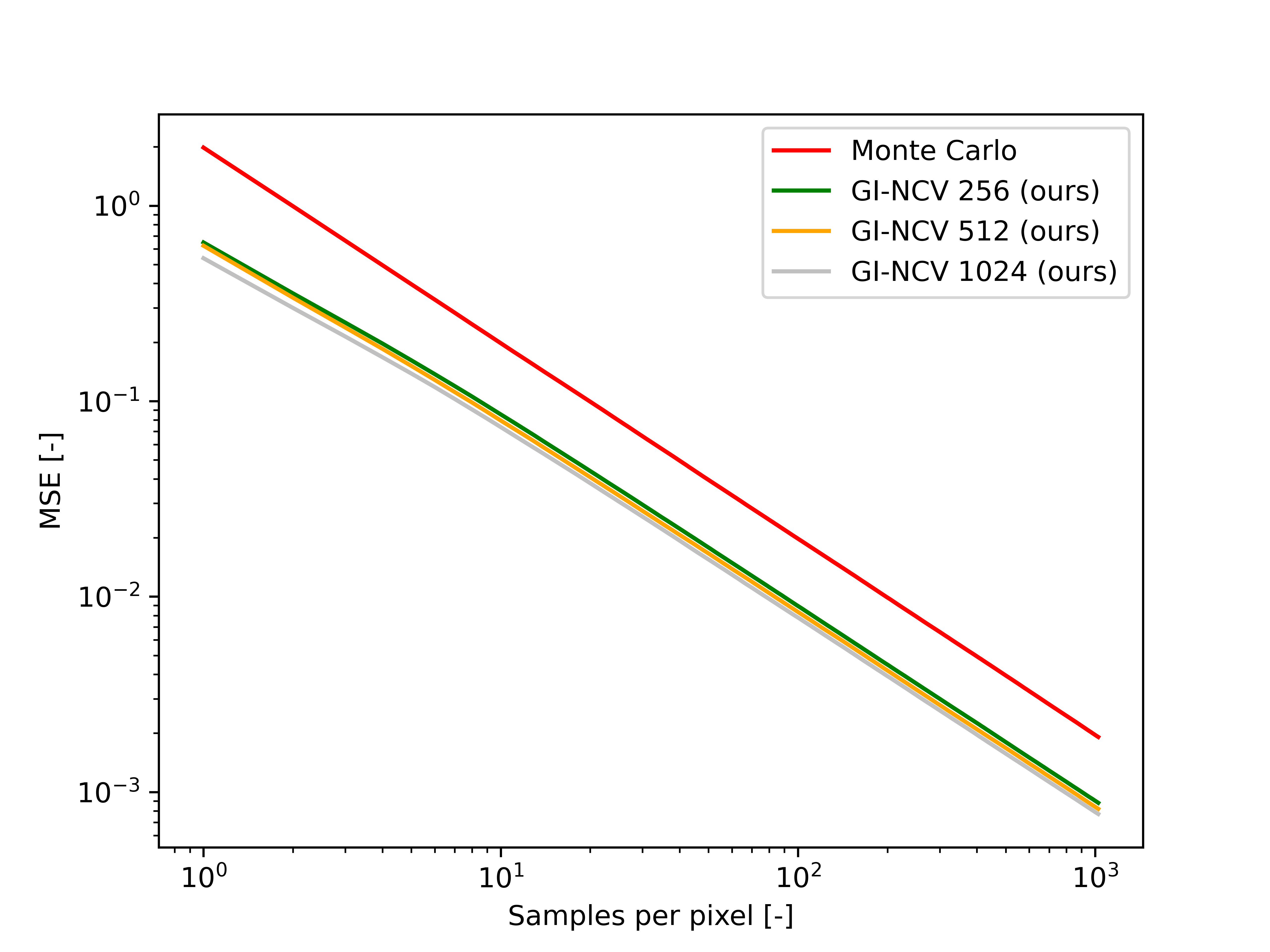}};
    \end{tikzpicture}
\end{minipage}

\begin{minipage}{0.22\textwidth}
    \begin{tikzpicture}
        \node[anchor=south west, inner sep=0] (image) at (0,0) {\includegraphics[width=\linewidth]{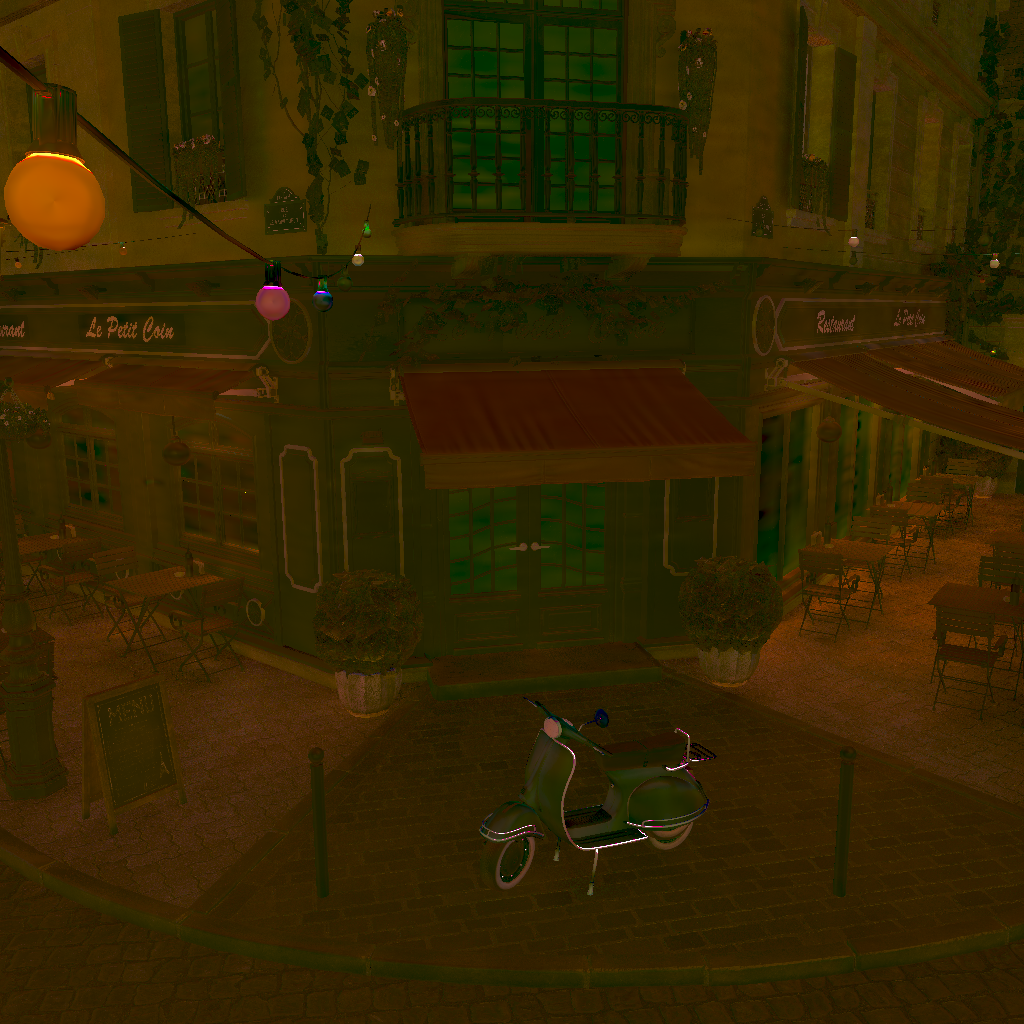}};
        \node[anchor=north west, xshift=2mm, yshift=-2mm, fill=white, opacity=0.7, text opacity=1, font=\scriptsize] 
            at (image.north west) {$G$ (ours)};
    \end{tikzpicture}
\end{minipage}%
\begin{minipage}{0.22\textwidth}
    \begin{tikzpicture}
        \node[anchor=south west, inner sep=0] (image) at (0,0) {\includegraphics[width=\linewidth]{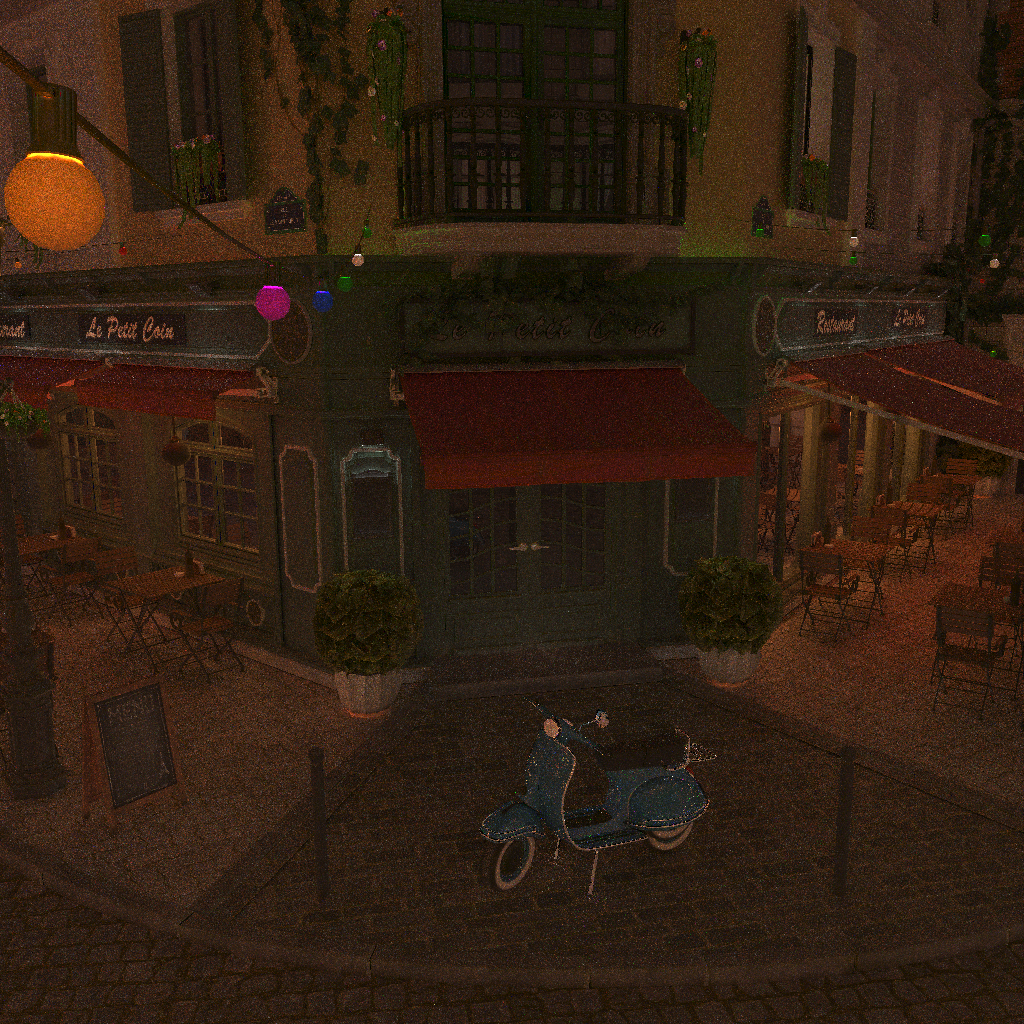}};
        \node[anchor=south east, xshift=-2mm, yshift=2mm, fill=white, opacity=0.7, text opacity=1, font=\scriptsize] 
            at (image.south east) {MSE $3.7 \cdot 10^{-4}$};
        \node[anchor=north west, xshift=2mm, yshift=-2mm, fill=white, opacity=0.7, text opacity=1, font=\scriptsize] 
            at (image.north west) {GI-NCV (ours)};
    \end{tikzpicture}
\end{minipage}%
\begin{minipage}{0.22\textwidth}
    \begin{tikzpicture}
        \node[anchor=south west, inner sep=0] (image) at (0,0) {\includegraphics[width=\linewidth]{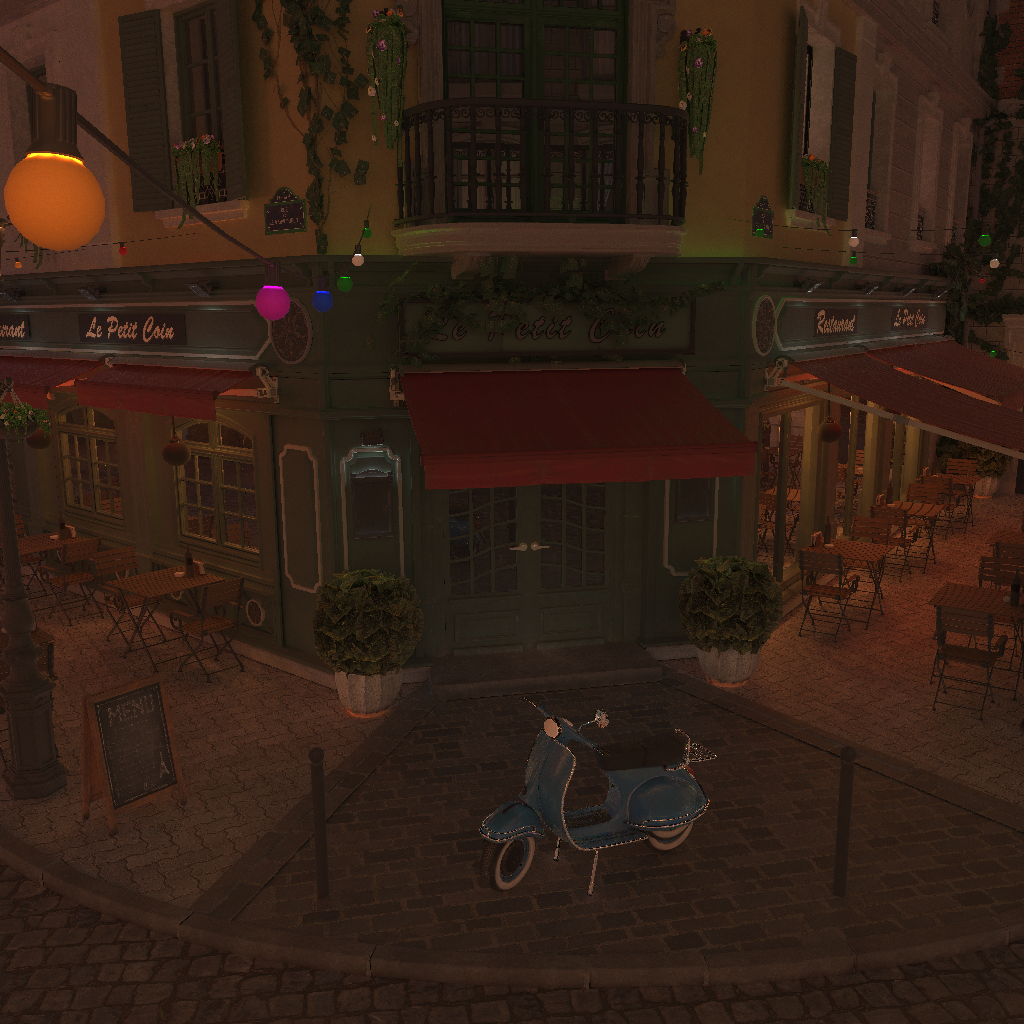}};
        \node[anchor=north west, xshift=2mm, yshift=-2mm, fill=white, opacity=0.7, text opacity=1, font=\scriptsize] 
            at (image.north west) {Reference};
    \end{tikzpicture}
\end{minipage}%
\begin{minipage}{0.3\textwidth}
    \begin{tikzpicture}
        \node[anchor=south west, inner sep=0] (image) at (0,0) {\includegraphics[width=\linewidth]{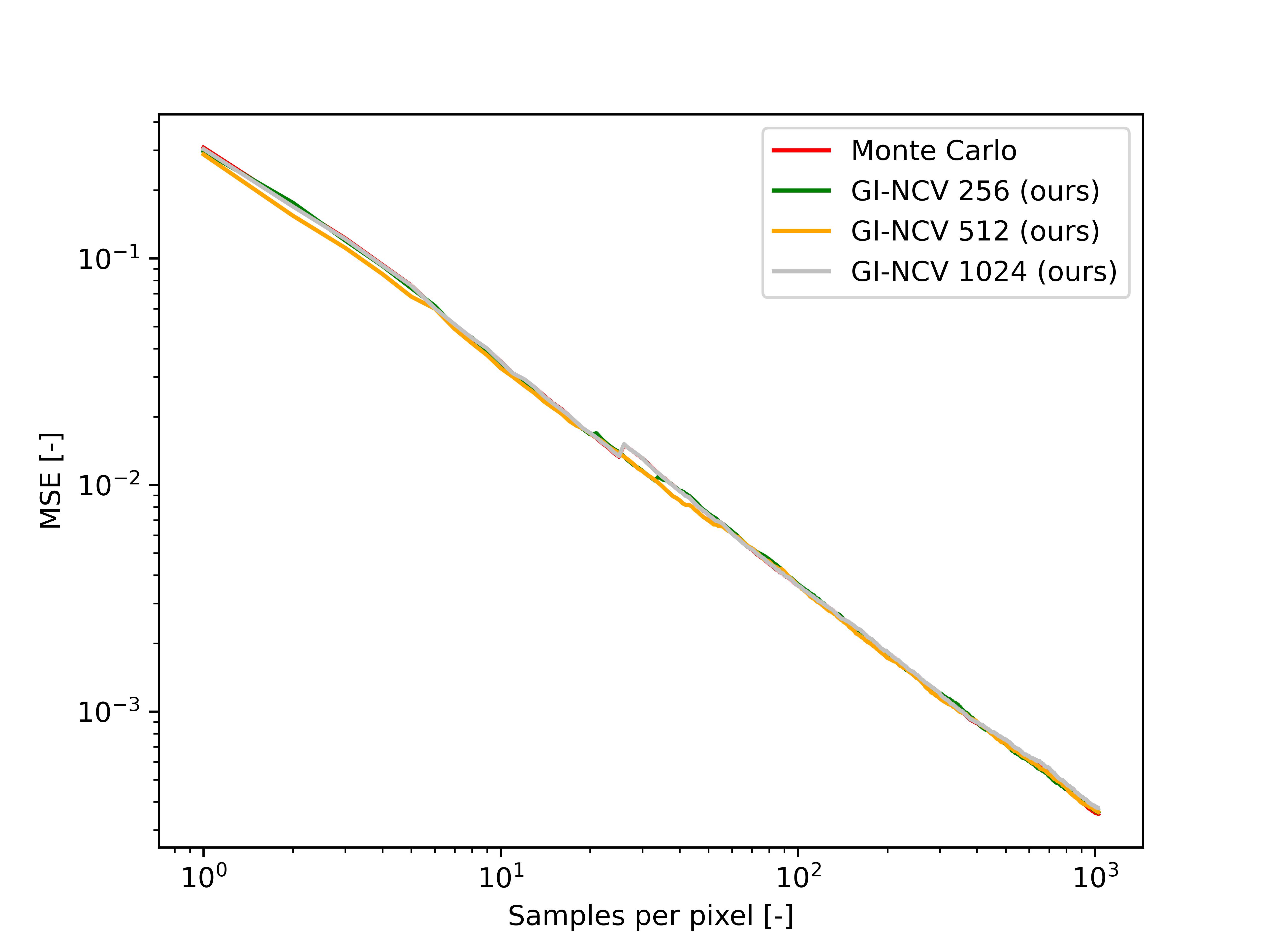}};
    \end{tikzpicture}
\end{minipage}
\end{center}
\caption{Comparison of control variates with our geometric integration (GI-NCV) and vanilla Monte Carlo for different numbers of training samples per pixel: direct lighting in the Cornell box (top), ambient occlusion in the Bistro scene (center), and indirect lighting in the Bistro scene (bottom). Notice that the control variates do not help in reducing the error for indirect lighting due to the noisy estimates of incidence radiance (see Section~\ref{Sec:LightTransport}). The images correspond to the result using 1024 training samples per pixel (epochs).}
\label{Fig:LightTransport}
\end{figure*}

\begin{figure*}
\begin{center}
\begin{minipage}{0.24\textwidth}
    \begin{tikzpicture}
        \node[anchor=south west, inner sep=0] (image) at (0,0) {\includegraphics[width=\linewidth]{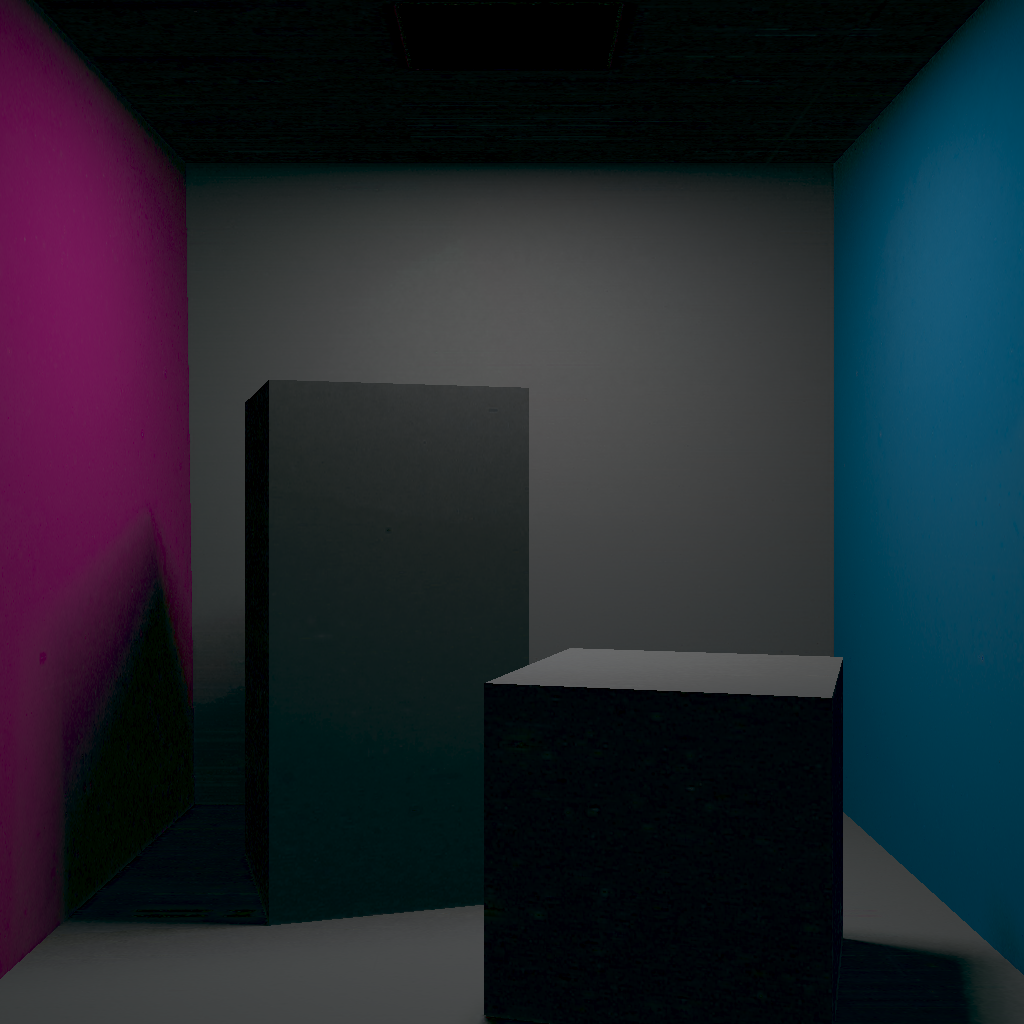}};
        \node[anchor=north west, xshift=2mm, yshift=-2mm, fill=white, opacity=0.7, text opacity=1, font=\scriptsize] 
            at (image.north west) {$G$ (ours)};
    \end{tikzpicture}
\end{minipage}%
\begin{minipage}{0.24\textwidth}
    \begin{tikzpicture}
        \node[anchor=south west, inner sep=0] (image) at (0,0) {\includegraphics[width=\linewidth]{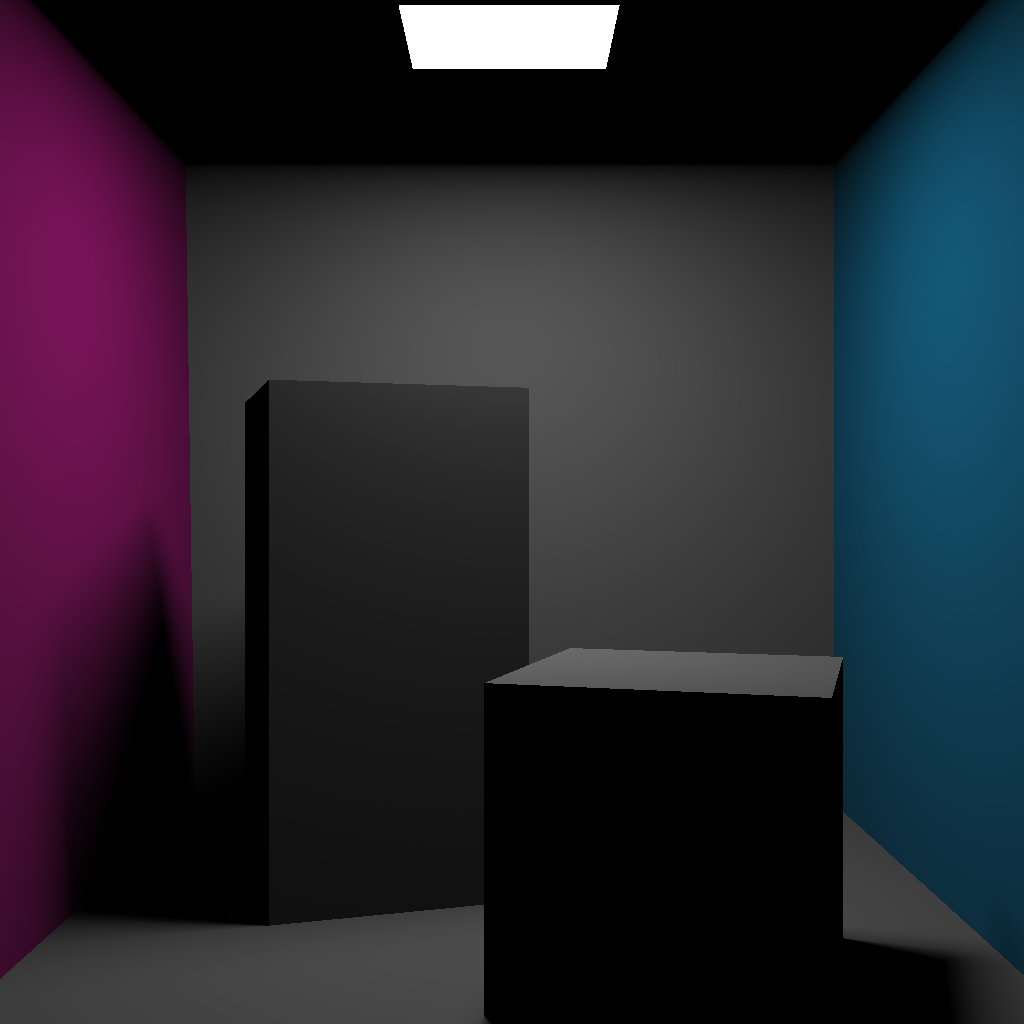}};
        \node[anchor=south east, xshift=-2mm, yshift=2mm, fill=white, opacity=0.7, text opacity=1, font=\scriptsize] 
            at (image.south east) {2048 + 6144 SPP};
        \node[anchor=north west, xshift=2mm, yshift=-2mm, fill=white, opacity=0.7, text opacity=1, font=\scriptsize] 
            at (image.north west) {GI-NCV (ours)};
    \end{tikzpicture}
\end{minipage}%
\begin{minipage}{0.24\textwidth}
    \begin{tikzpicture}
        \node[anchor=south west, inner sep=0] (image) at (0,0) {\includegraphics[width=\linewidth]{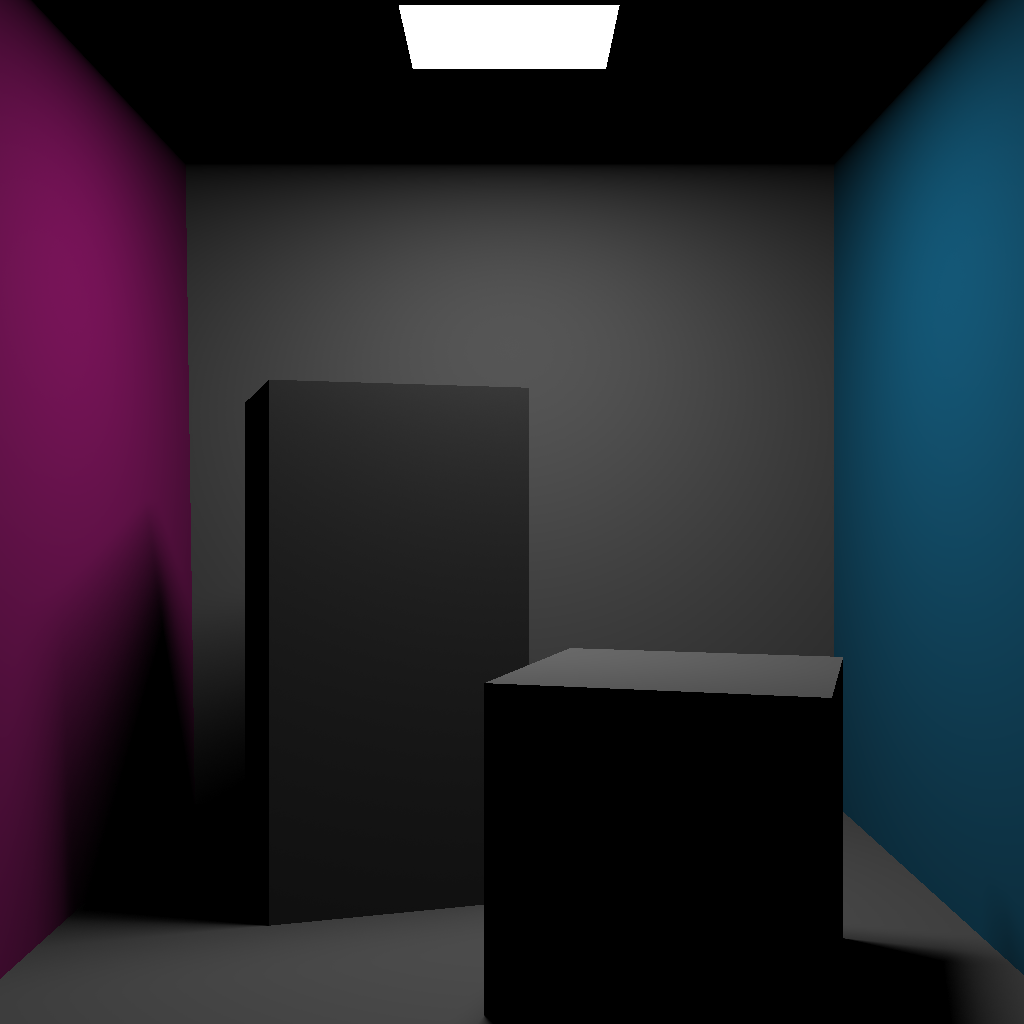}};
        \node[anchor=south east, xshift=-2mm, yshift=2mm, fill=white, opacity=0.7, text opacity=1, font=\scriptsize] 
            at (image.south east) {8192 SPP};
        \node[anchor=north west, xshift=2mm, yshift=-2mm, fill=white, opacity=0.7, text opacity=1, font=\scriptsize] 
            at (image.north west) {Monte Carlo};
    \end{tikzpicture}
\end{minipage}%
\begin{minipage}{0.24\textwidth}
    \begin{tikzpicture}
        \node[anchor=south west, inner sep=0] (image) at (0,0) {\includegraphics[width=\linewidth]{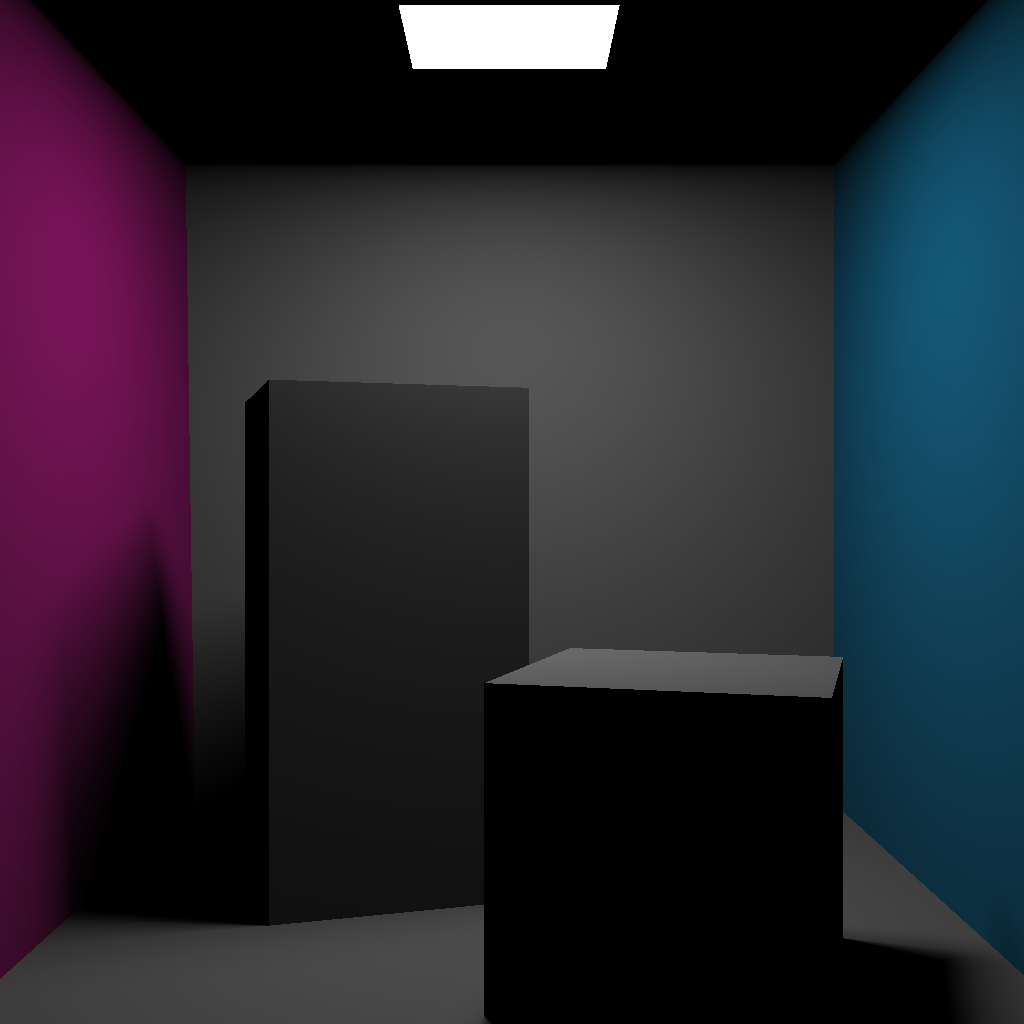}};
        \node[anchor=north west, xshift=2mm, yshift=-2mm, fill=white, opacity=0.7, text opacity=1, font=\scriptsize] 
            at (image.north west) {Reference};
    \end{tikzpicture}
\end{minipage}%

\begin{minipage}{0.24\textwidth}
    \begin{tikzpicture}
        \node[anchor=south west, inner sep=0] (image) at (0,0) {\includegraphics[width=\linewidth]{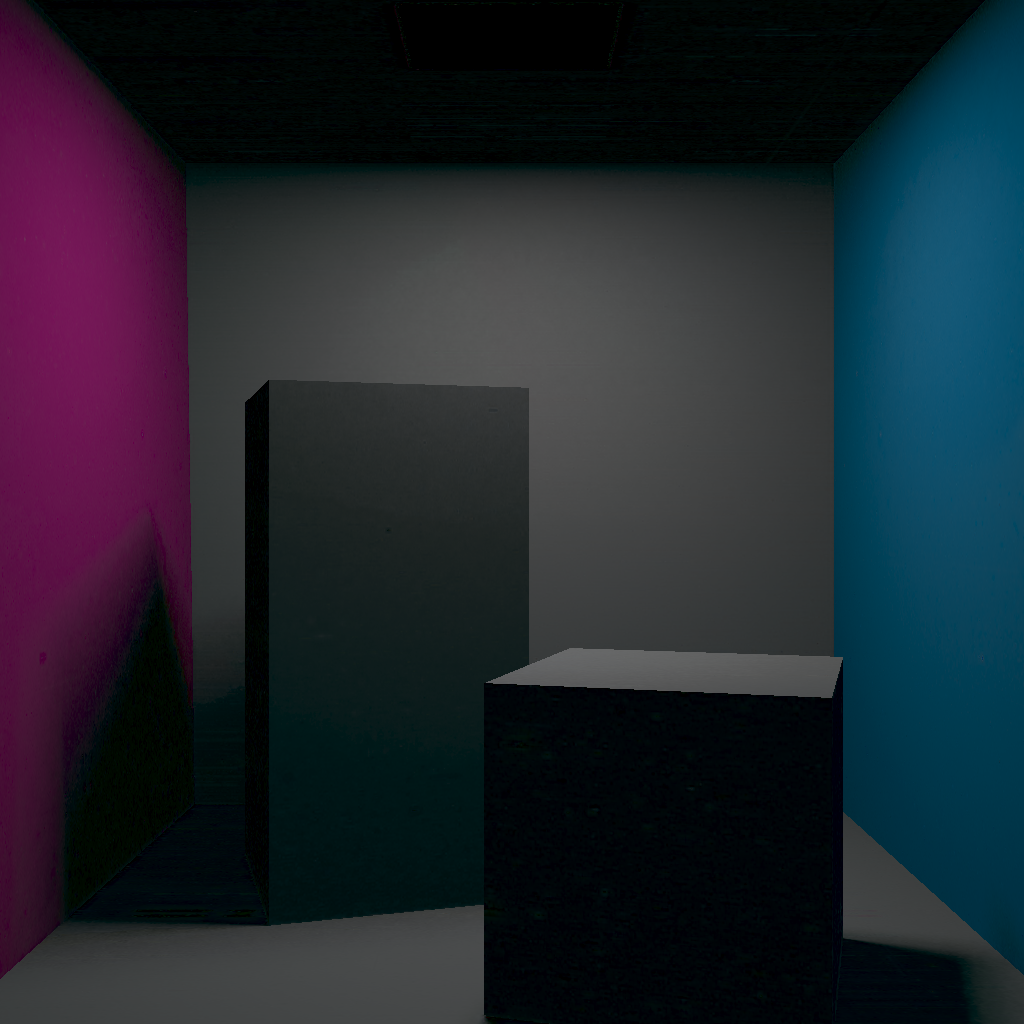}};
        \node[anchor=north west, xshift=2mm, yshift=-2mm, fill=white, opacity=0.7, text opacity=1, font=\scriptsize] 
            at (image.north west) {$\hat{G} = \sum_{i=1}^n \frac{g(x_i)}{p(x_i)}$ (ours)};
    \end{tikzpicture}
\end{minipage}%
\begin{minipage}{0.24\textwidth}
    \begin{tikzpicture}
        \node[anchor=south west, inner sep=0] (image) at (0,0) {\includegraphics[width=\linewidth]{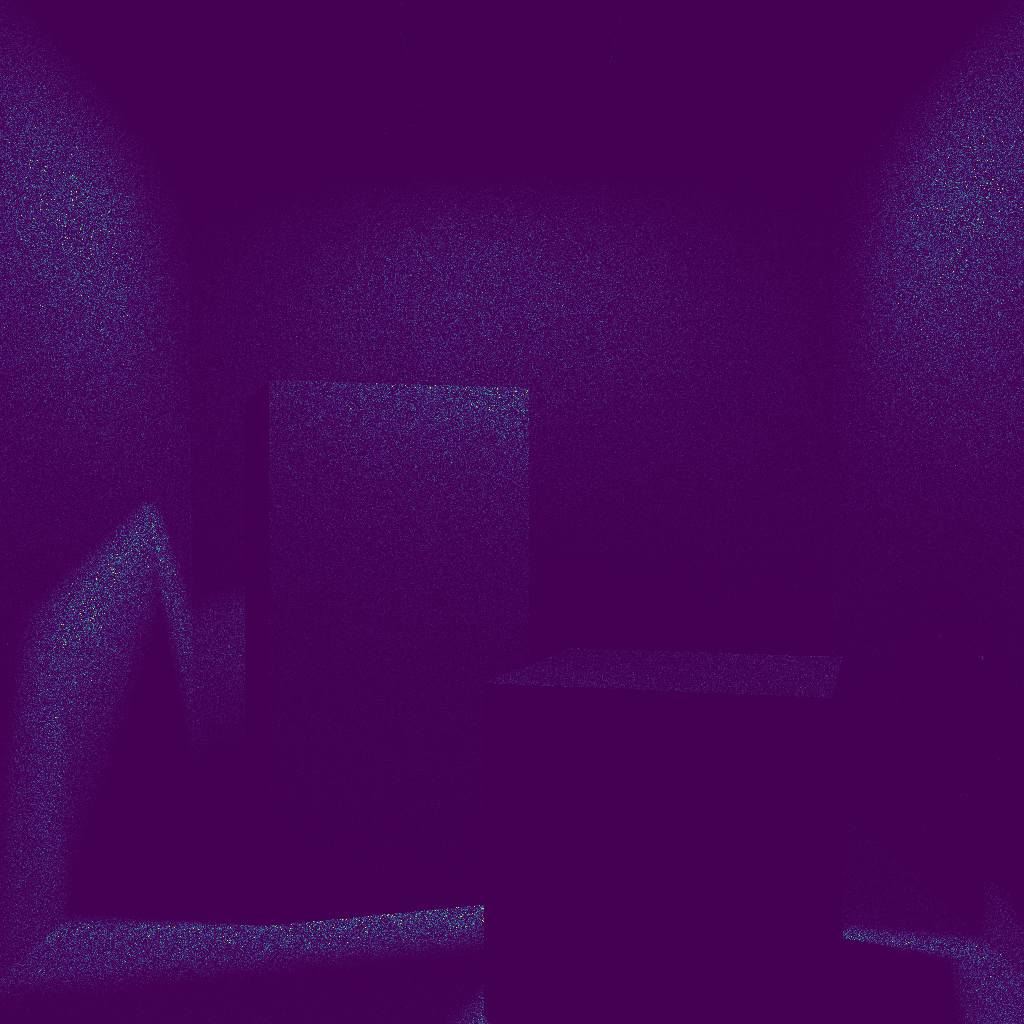}};
        \node[anchor=south east, xshift=-2mm, yshift=2mm, fill=white, opacity=0.7, text opacity=1, font=\scriptsize] 
            at (image.south east) {MSE $8.7 \cdot 10^{-9}$};
        \node[anchor=north west, xshift=2mm, yshift=-2mm, fill=white, opacity=0.7, text opacity=1, font=\scriptsize] 
            at (image.north west) {GI-NCV (ours)};
    \end{tikzpicture}
\end{minipage}%
\begin{minipage}{0.24\textwidth}
    \begin{tikzpicture}
        \node[anchor=south west, inner sep=0] (image) at (0,0) {\includegraphics[width=\linewidth]{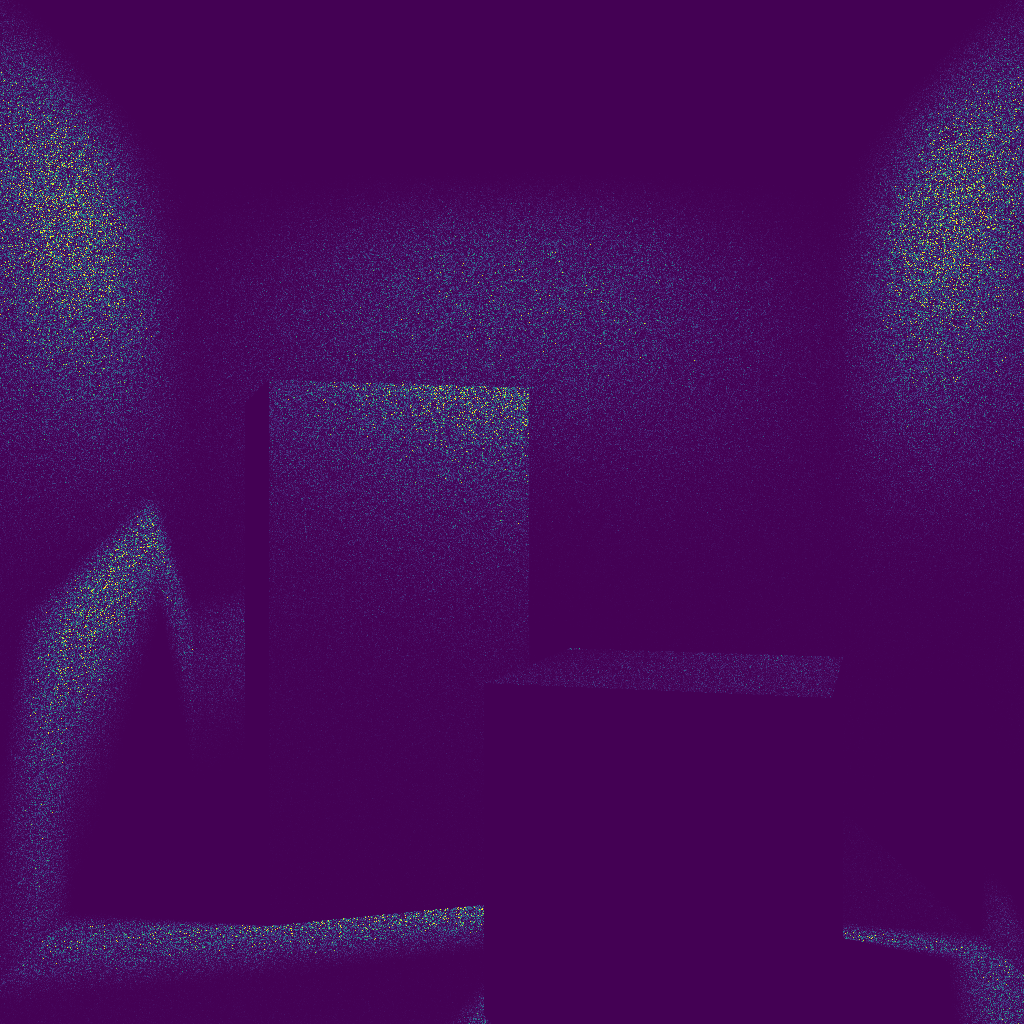}};
        \node[anchor=south east, xshift=-2mm, yshift=2mm, fill=white, opacity=0.7, text opacity=1, font=\scriptsize] 
            at (image.south east) {MSE $16.9 \cdot 10^{-9}$};
        \node[anchor=north west, xshift=2mm, yshift=-2mm, fill=white, opacity=0.7, text opacity=1, font=\scriptsize] 
            at (image.north west) {Monte Carlo};
    \end{tikzpicture}
\end{minipage}%
\begin{minipage}{0.24\textwidth}
    \begin{tikzpicture}
        \node[anchor=south west, inner sep=0] (image) at (0,0) {\includegraphics[width=\linewidth]{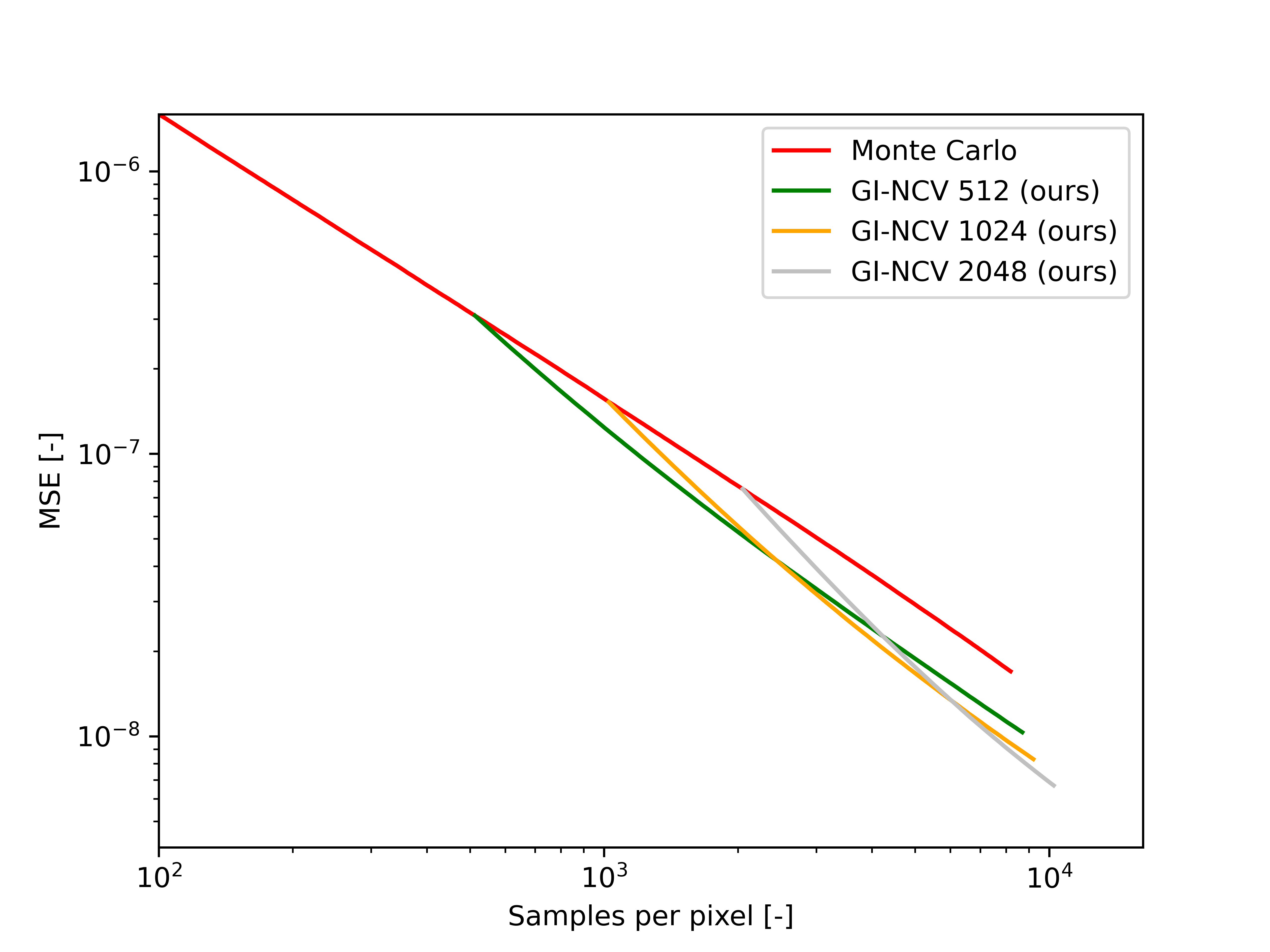}};
    \end{tikzpicture}

    \begin{tikzpicture}
    \node[anchor=south west, inner sep=0] (image) at (0,0) {\includegraphics[width=\linewidth]{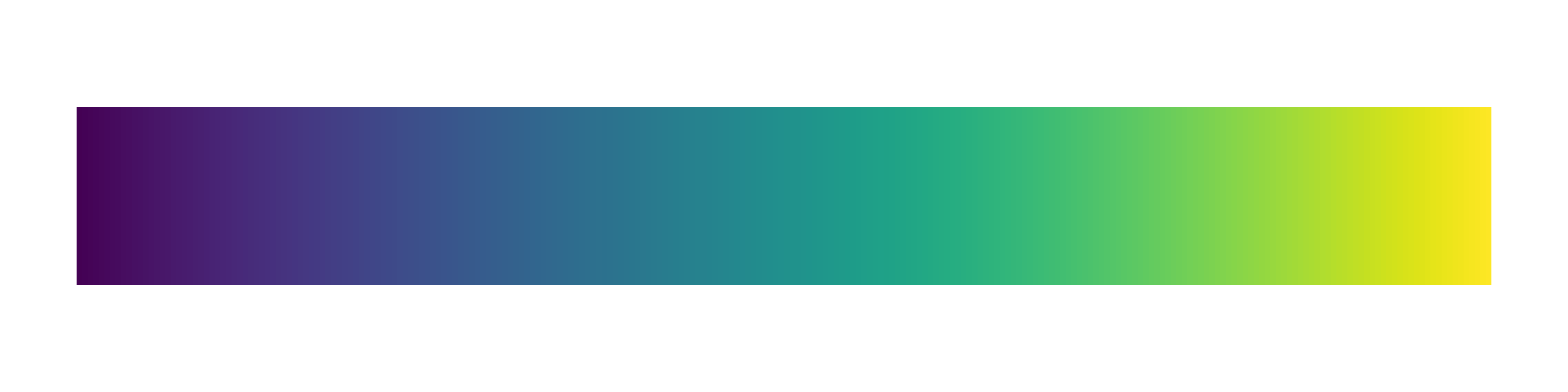}};
   
    \node[anchor=center, xshift=-0.4\linewidth, fill=white, opacity=0.2, text opacity=1, font=\scriptsize] 
       at (image.center) {0};
    
    
    \node[anchor=center, xshift=0.35\linewidth, fill=white, opacity=0.2, text opacity=1, font=\scriptsize] 
       at (image.center) {$5\cdot10^{-7}$};
\end{tikzpicture}
\end{minipage}%
\end{center}

\begin{center}
\begin{minipage}{0.24\textwidth}
    \begin{tikzpicture}
        \node[anchor=south west, inner sep=0] (image) at (0,0) {\includegraphics[width=\linewidth]{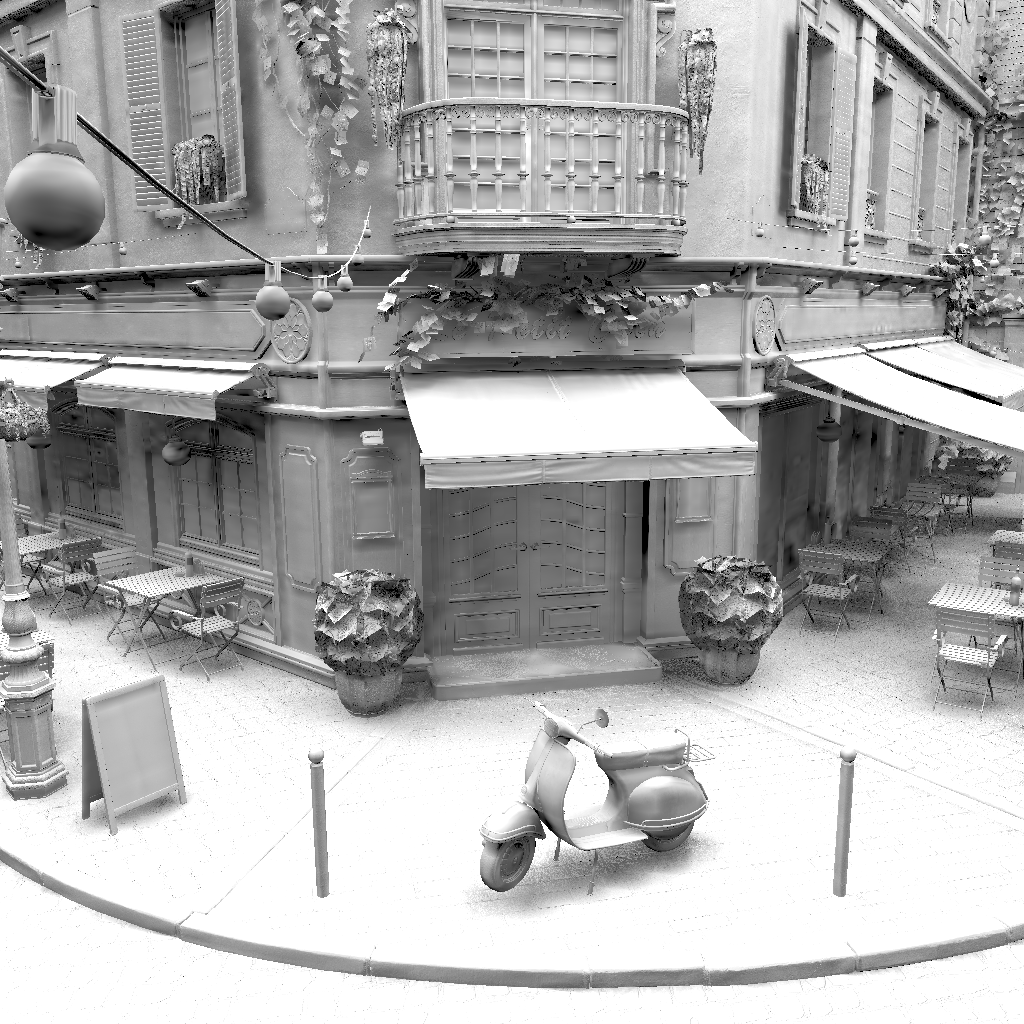}};
        \node[anchor=north west, xshift=2mm, yshift=-2mm, fill=white, opacity=0.7, text opacity=1, font=\scriptsize] 
            at (image.north west) {$G$ (ours)};
    \end{tikzpicture}
\end{minipage}%
\begin{minipage}{0.24\textwidth}
    \begin{tikzpicture}
        \node[anchor=south west, inner sep=0] (image) at (0,0) {\includegraphics[width=\linewidth]{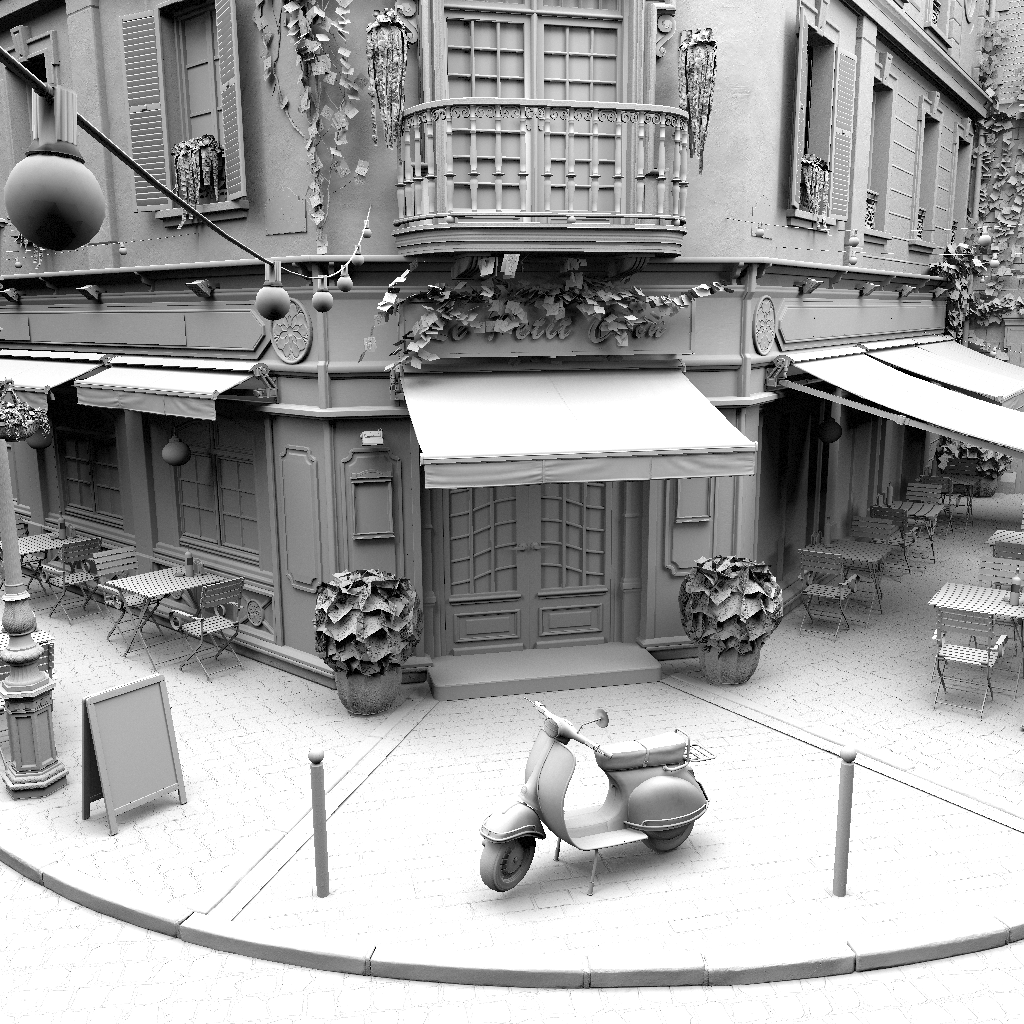}};
        \node[anchor=south east, xshift=-2mm, yshift=2mm, fill=white, opacity=0.7, text opacity=1, font=\scriptsize] 
            at (image.south east) {2048 + 6144 SPP};
        \node[anchor=north west, xshift=2mm, yshift=-2mm, fill=white, opacity=0.7, text opacity=1, font=\scriptsize] 
            at (image.north west) {GI-NCV (ours)};
    \end{tikzpicture}
\end{minipage}%
\begin{minipage}{0.24\textwidth}
    \begin{tikzpicture}
        \node[anchor=south west, inner sep=0] (image) at (0,0) {\includegraphics[width=\linewidth]{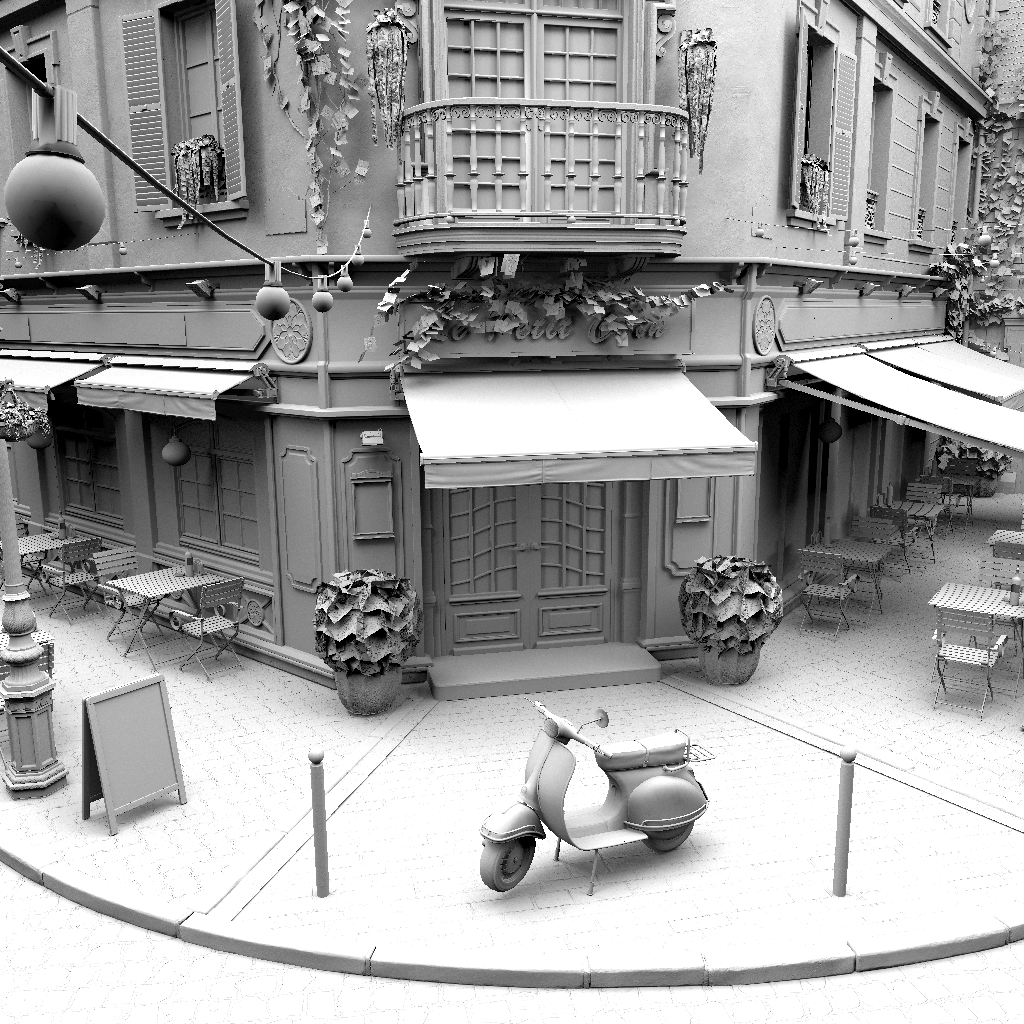}};
        \node[anchor=south east, xshift=-2mm, yshift=2mm, fill=white, opacity=0.7, text opacity=1, font=\scriptsize] 
            at (image.south east) {8192 SPP};
        \node[anchor=north west, xshift=2mm, yshift=-2mm, fill=white, opacity=0.7, text opacity=1, font=\scriptsize] 
            at (image.north west) {Monte Carlo};
    \end{tikzpicture}
\end{minipage}%
\begin{minipage}{0.24\textwidth}
    \begin{tikzpicture}
        \node[anchor=south west, inner sep=0] (image) at (0,0) {\includegraphics[width=\linewidth]{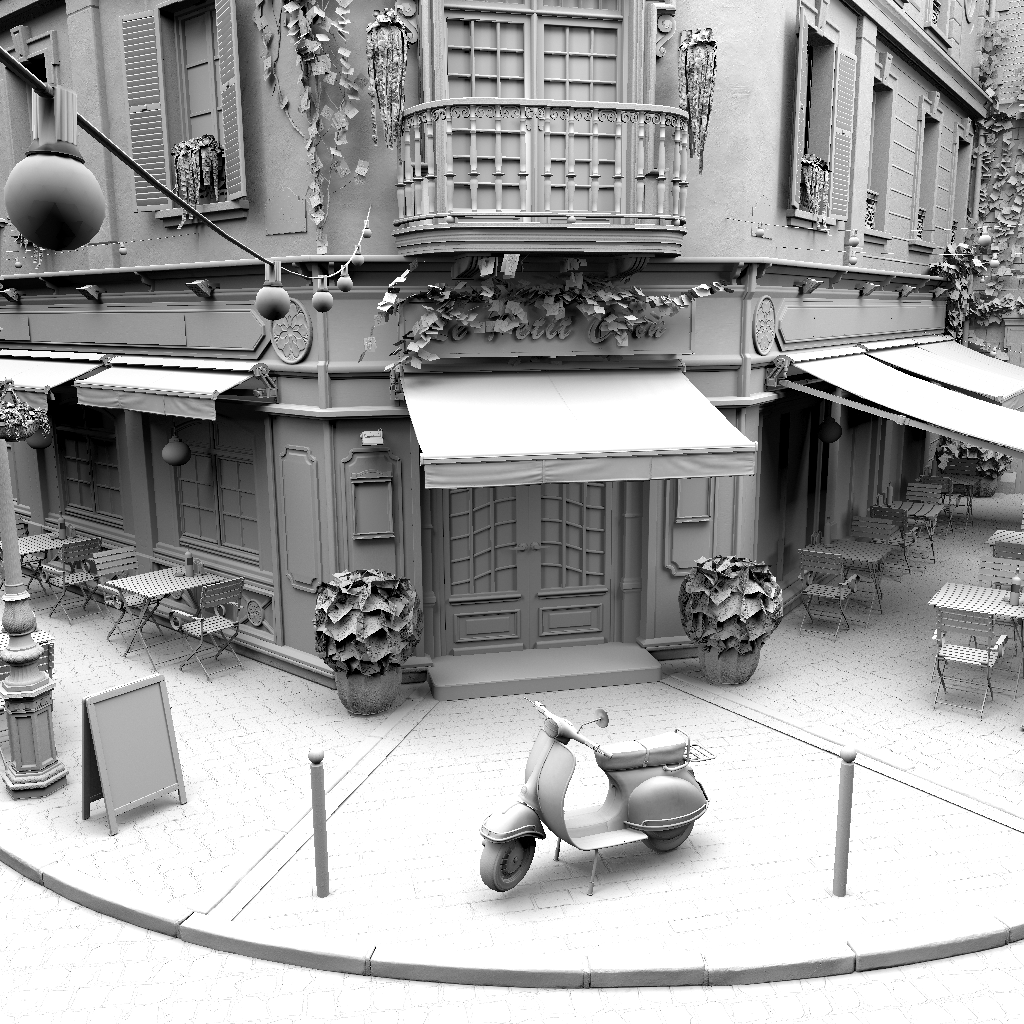}};
        \node[anchor=north west, xshift=2mm, yshift=-2mm, fill=white, opacity=0.7, text opacity=1, font=\scriptsize] 
            at (image.north west) {Reference};
    \end{tikzpicture}
\end{minipage}%

\begin{minipage}{0.24\textwidth}
    \begin{tikzpicture}
        \node[anchor=south west, inner sep=0] (image) at (0,0) {\includegraphics[width=\linewidth]{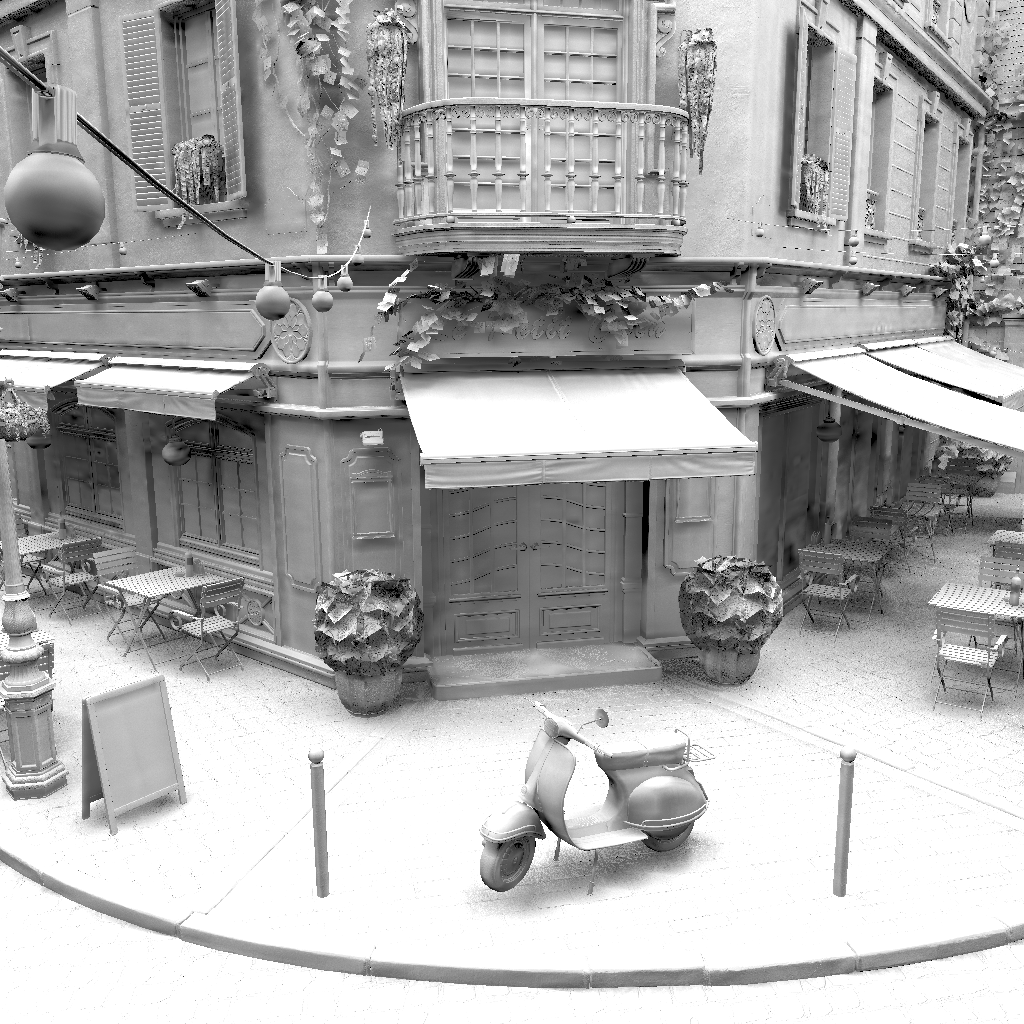}};
        \node[anchor=north west, xshift=2mm, yshift=-2mm, fill=white, opacity=0.7, text opacity=1, font=\scriptsize] 
            at (image.north west) {$\hat{G} = \sum_{i=1}^n \frac{g(x_i)}{p(x_i)}$ (ours)};
    \end{tikzpicture}
\end{minipage}%
\begin{minipage}{0.24\textwidth}
    \begin{tikzpicture}
        \node[anchor=south west, inner sep=0] (image) at (0,0) {\includegraphics[width=\linewidth]{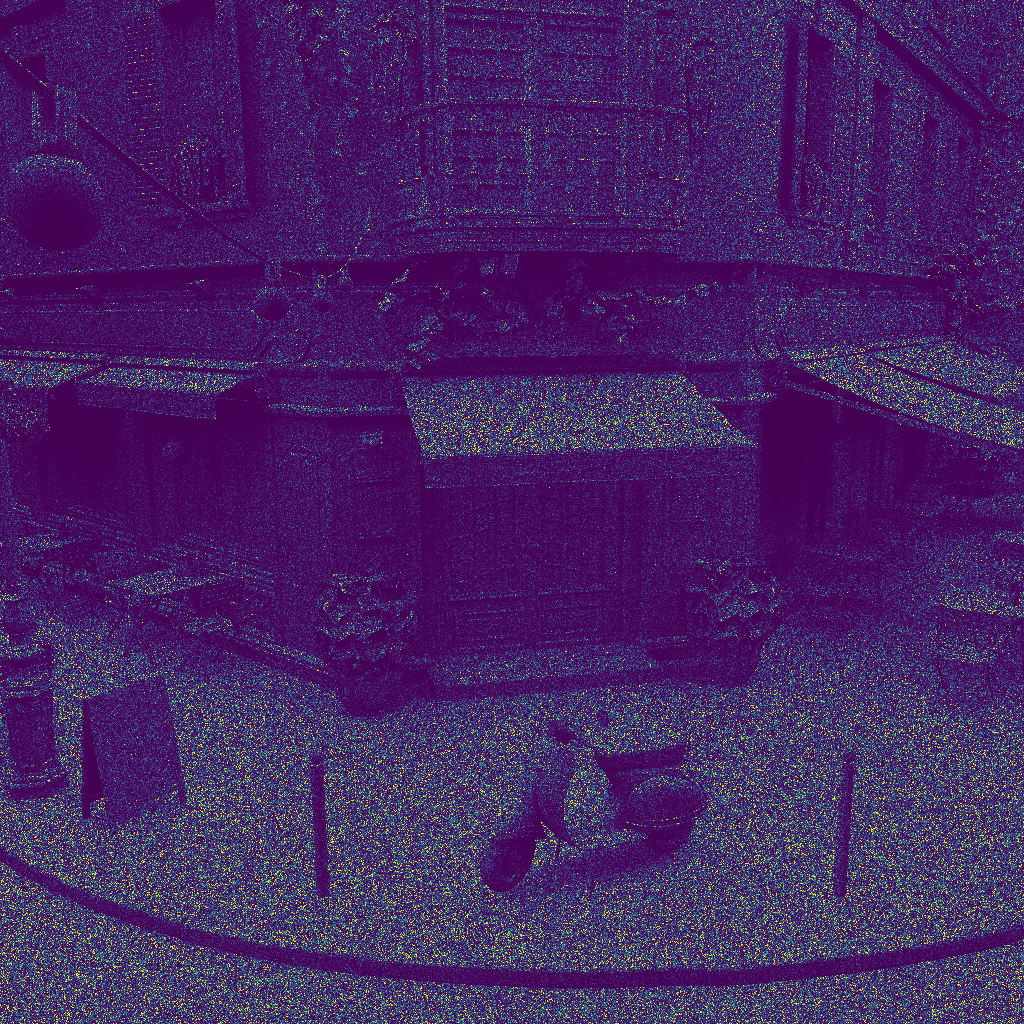}};
        \node[anchor=south east, xshift=-2mm, yshift=2mm, fill=white, opacity=0.7, text opacity=1, font=\scriptsize] 
            at (image.south east) {MSE $1.2 \cdot 10^{-4}$};
        \node[anchor=north west, xshift=2mm, yshift=-2mm, fill=white, opacity=0.7, text opacity=1, font=\scriptsize] 
            at (image.north west) {GI-NCV (ours)};
    \end{tikzpicture}
\end{minipage}%
\begin{minipage}{0.24\textwidth}
    \begin{tikzpicture}
        \node[anchor=south west, inner sep=0] (image) at (0,0) {\includegraphics[width=\linewidth]{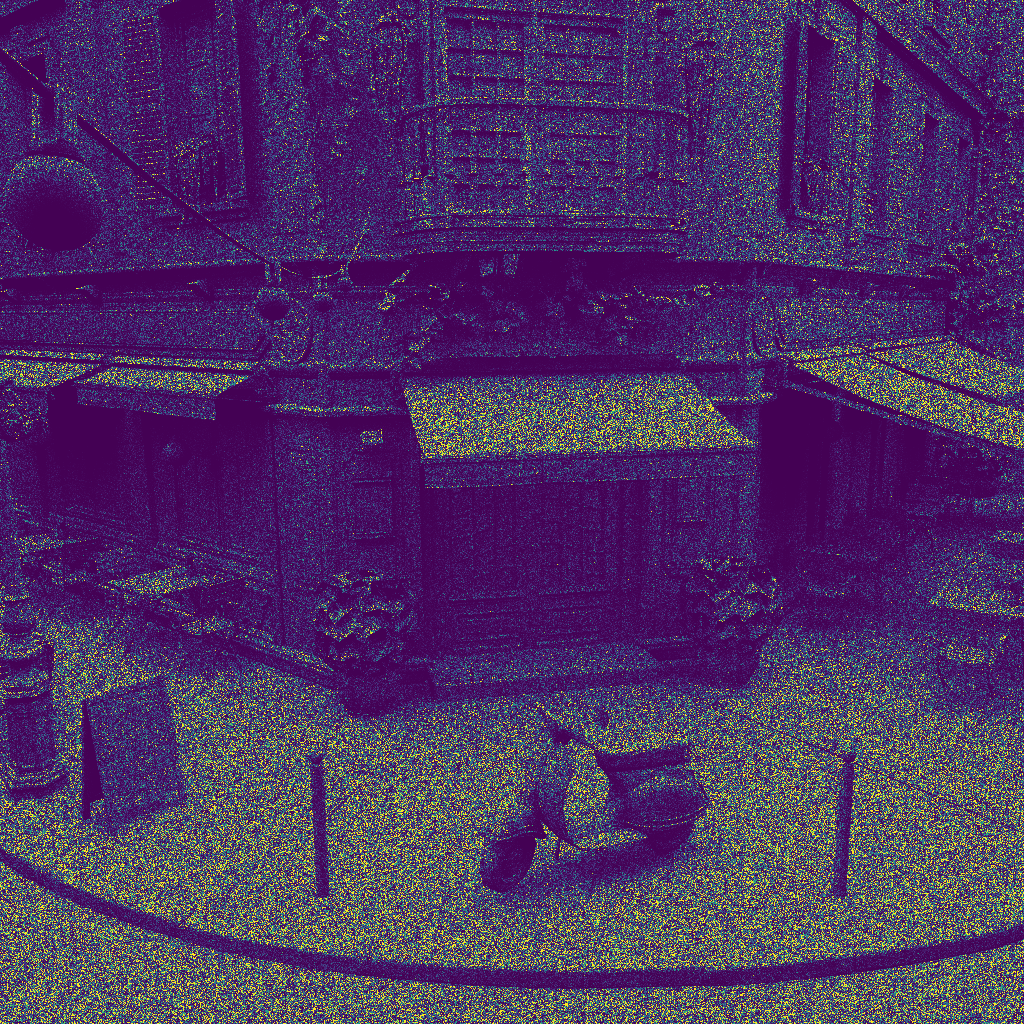}};
        \node[anchor=south east, xshift=-2mm, yshift=2mm, fill=white, opacity=0.7, text opacity=1, font=\scriptsize] 
            at (image.south east) {MSE $2.1 \cdot 10^{-4}$};
        \node[anchor=north west, xshift=2mm, yshift=-2mm, fill=white, opacity=0.7, text opacity=1, font=\scriptsize] 
            at (image.north west) {Monte Carlo};
    \end{tikzpicture}
\end{minipage}%
\begin{minipage}{0.24\textwidth}
    \begin{tikzpicture}
        \node[anchor=south west, inner sep=0] (image) at (0,0) {\includegraphics[width=\linewidth]{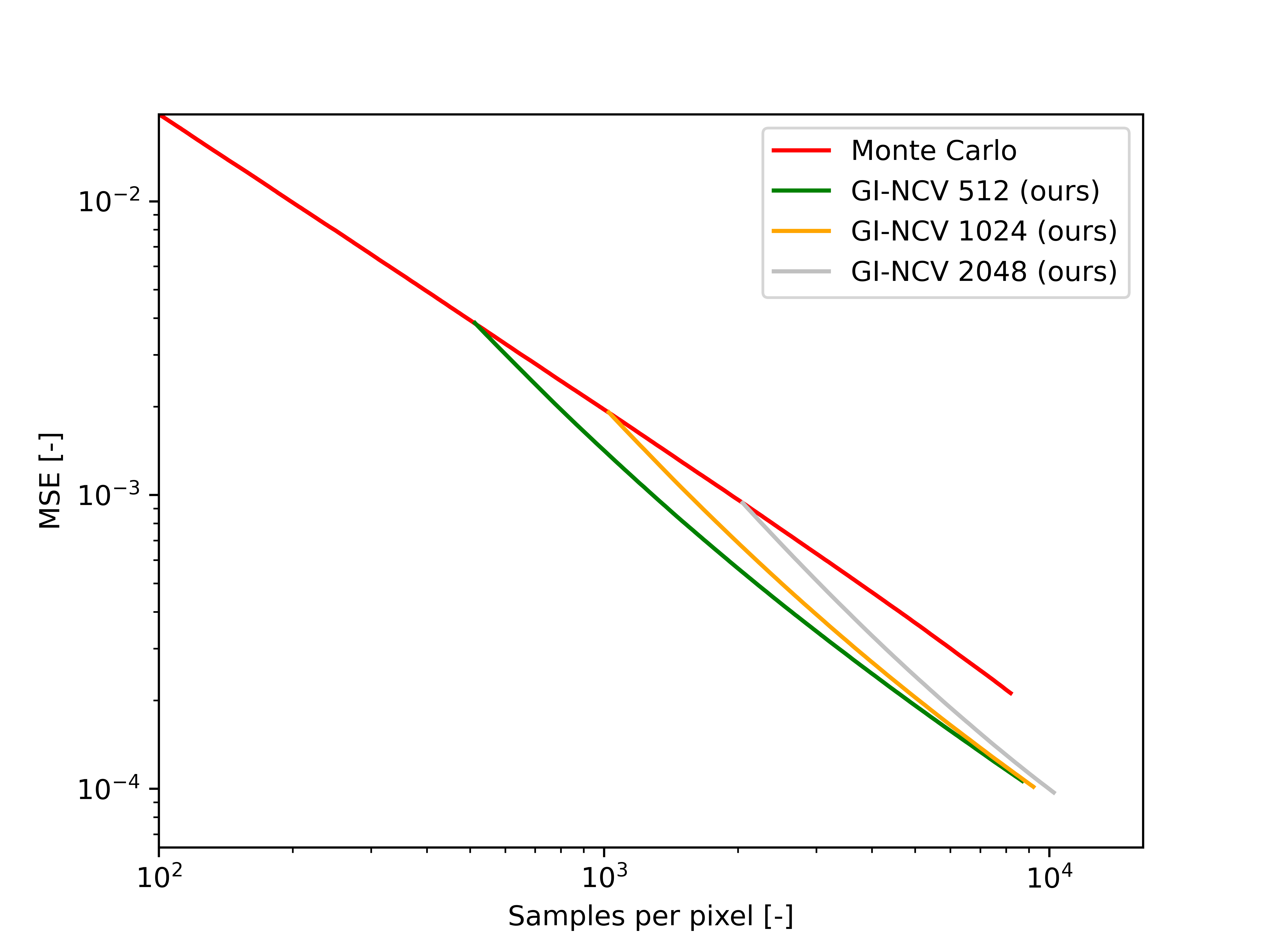}};
    \end{tikzpicture}

    \begin{tikzpicture}
    \node[anchor=south west, inner sep=0] (image) at (0,0) {\includegraphics[width=\linewidth]{images/err_bar.png}};
   
    \node[anchor=center, xshift=-0.4\linewidth, fill=white, opacity=0.2, text opacity=1, font=\scriptsize] 
       at (image.center) {0};
    
    
    \node[anchor=center, xshift=0.34\linewidth, fill=white, opacity=0.2, text opacity=1, font=\scriptsize] 
       at (image.center) {$10^{-3}$};
\end{tikzpicture}
\end{minipage}%
\end{center}
\caption{Equal-sample-count comparison of neural control variates using our geometric integration (GI-NCV) and vanilla Monte Carlo for direct lighting in the Cornell box (top) and ambient occlusion in the Bistro scene (bottom). In the left column, we show the results of the analytic integration (left - first and third rows) and Monte Carlo integration (left - second and fourth rows) of the MLP, demonstrating that Monte Carlo converges to the analytic solution. The images correspond to the result using 2048 training samples per pixel (epochs). For the sake of fair comparison, the lines of GI-NCV in the convergence graphs are offset by the corresponding training samples that are used for both for training and rendering.}
\label{Fig:EqualSampleCount}
\end{figure*}

\paragraph*{Implementation Details} Both the renderer and the MLP framework are implemented in HIP (ROCm 6.3), using HIPRT 2.5~\cite{Meister2024} and FP32 for the weights and the hashgrid latent vectors, optimized by the Adam optimizer~\cite{Kingma2014}, for the MLP. The geometric integration is implemented in a separate kernel, where each pixel is processed by a single thread. Each thread requires relatively large memory space to store DCELs in order to compute the analytic integration (see Section~\ref{Sec:LineArrangement}). To reduce the memory usage, we use persistent threads, and thus the allocated memory for one thread will be reused for multiple pixels. We use 16-bit integers for the half-edge indices since there is no more than 65k half-edges in a single DCEL. We store weights of the MLP in shared memory as the weights are common for all threads. We use an exponential moving average \cite{Muller2021} for the weights (but not for the latent vectors stored in the hashgrid) in order to make training more stable.

\begin{figure*}
\begin{center}
\begin{minipage}{0.22\textwidth}
    \begin{tikzpicture}
        \node[anchor=south west, inner sep=0] (image) at (0,0) {\includegraphics[width=\linewidth]{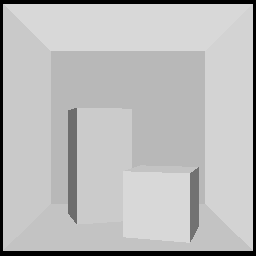}};
        \node[anchor=south east, xshift=-2mm, yshift=2mm, fill=white, opacity=0.7, text opacity=1, font=\scriptsize] 
            at (image.south east) {MSE $3.6 \cdot 10^{-4}$};
        \node[anchor=north west, xshift=2mm, yshift=-2mm, fill=white, opacity=0.7, text opacity=1, font=\scriptsize] 
            at (image.north west) {GI-NCV $G$ (ours)};
    \end{tikzpicture}
\end{minipage}%
\begin{minipage}{0.22\textwidth}
    \begin{tikzpicture}
        \node[anchor=south west, inner sep=0] (image) at (0,0) {\includegraphics[width=\linewidth]{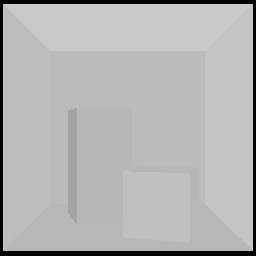}};
        \node[anchor=south east, xshift=-2mm, yshift=2mm, fill=white, opacity=0.7, text opacity=1, font=\scriptsize] 
            at (image.south east) {MSE $5.2 \cdot 10^{-4}$};
        \node[anchor=north west, xshift=2mm, yshift=-2mm, fill=white, opacity=0.7, text opacity=1, font=\scriptsize] 
            at (image.north west) {AI-NCV $G$};
    \end{tikzpicture}
\end{minipage}%
\begin{minipage}{0.22\textwidth}
    \begin{tikzpicture}
        \node[anchor=south west, inner sep=0] (image) at (0,0) {\includegraphics[width=\linewidth]{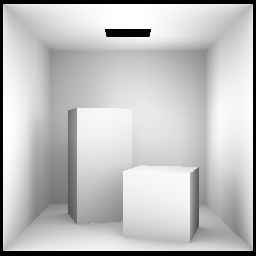}};
        \node[anchor=north west, xshift=2mm, yshift=-2mm, fill=white, opacity=0.7, text opacity=1, font=\scriptsize] 
            at (image.north west) {Reference};
    \end{tikzpicture}
\end{minipage}%
\begin{minipage}{0.3\textwidth}
    \begin{tikzpicture}
        \node[anchor=south west, inner sep=0] (image) at (0,0) {\includegraphics[width=\linewidth]{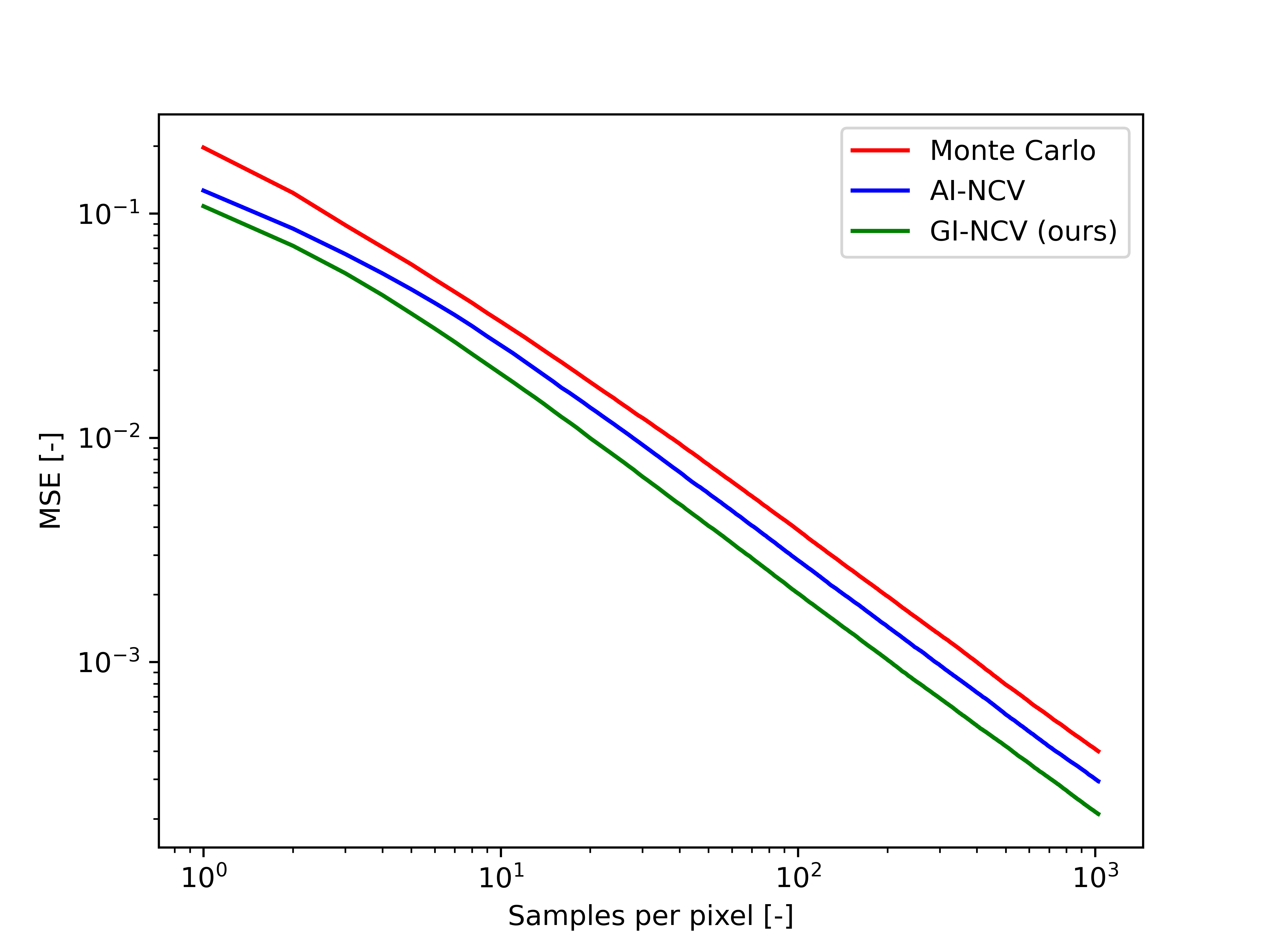}};
    \end{tikzpicture}
\end{minipage}
\end{center}
\caption{Comparison of control variates with our geometric integration (GI-NCV) and the automatic integration~\cite{Lindell2021,Li2024} (AI-NCV) for ambient occlusion in the Cornell box. Our method allows us to train the MLP directly, and thus it can approximate ambient occlusion more credibly, achieving lower error overall.}
\label{Fig:TorchPT}
\end{figure*}

\paragraph*{Comparison with Automatic Integration} We implemented a simple renderer in PyTorch to compare our method (GI-NCV) with the automatic integration~\cite{Lindell2021,Li2024} (AI-NCV), which requires autodiff. We use the same setting as in the HIP framework except the hashgrid encoding; we use frequency encoding~\cite{Mildenhall2020} instead of the hashgrid~\cite{Muller2022} for both methods. For AI-NCV, we use sigmoid as an activation function instead of ReLU for the same reasons discussed in Section~\ref{Sec:AnalyticFunctions}, and one-blob encoding~\cite{Muller2019} for the integration domain (spherical domain in the case of ambient occlusion). We use 5000 epochs to pre-train the MLP of each method, and the images are rendered at a resolution of 256$\times$256. In Figure~\ref{Fig:TorchPT}, we can observe that training the MLP directly (not via the antiderivative as the automatic integration does) using our approach can achieve better approximation, and thus lower overall error.

\subsection{Discussion and Limitations}
In this section, we discuss the limitations of the proposed method. Similar to the automatic integration~\cite{Li2024}, the geometric integration is limited to low dimensions (two dimensions in our case). In contrast to high-dimensional primary samples space control variates \cite{Muller2020,Crespo2021,Salaun2022}, we can employ our method only for simplified scenarios in light transport, not to fully-recursive path tracing, nor even to simpler techniques such as multiple importance sampling or Russian roulette.

The proposed integration method is a relatively expensive (both in time and space) part in our pipeline. Therefore, we amortize its cost through using many samples such that it is evaluated only once after the MLP is pre-trained. Note that this is not possible for real-time scenarios, where the MLP is continuously trained. The time complexity depends on the resolution and the MLP architecture. To reduce the complexity, we use a smaller MLP and more complex hashgrid, that does not influence the integration algorithm. 

In Table~\ref{Tab:Times}, we show the breakdown of absolute times needed for different phases measured on AMD Radeon PRO W7900. We achieve the same error at slightly higher times compared to vanilla Monte Carlo. With a large number of samples, the integration part becomes marginal, but the training and inference becomes dominant. We need to use significantly more training data to achieve variance reduction as we evaluate the MLP on the primary hits in contrast to the neural radiance cache~\cite{Muller2021}, which hides inaccuracies by deeper bounces. We tried to use FP16 for the network weights and latent vectors in the hashgrid, but with the larger hashgrid, we can observe the problem of vanishing gradients. We leave further optimizations of training, inference, and integration as a future work.


Another limitation specific to our approach is that we cannot use non-linear encoding for a 2D integration domain to keep the lines straight. However, we can use arbitrary encoding for other inputs (e.g., a larger hashgrid). In Figure~\ref{Fig:TorchPT}, we can see that our method achieves lower error even without encoding than the automatic integration~\cite{Lindell2021,Li2024} with one-blob encoding in the spherical domain.

\begin{table}[]
    \centering
    \small
    \begin{tabular}{c||cc}
         & Monte Carlo & GI-NCV (ours)\\
         \hline
         \hline
         SPP [-] & 32768 & 1024 + 18432\\
         MSE [-] & $4.54 \cdot 10^{-5}$ & $4.53 \cdot 10^{-5}$\\ 
         \hline
         Render time [s] & 136 & 78\\
         Train time [s] & - & 33\\
         Integration time [s] & - & 10\\
         Inference time [s] & - & 34\\
         \hline
         Total time [s] & 136 & 155\\
    \end{tabular}
    \caption{Breakdown of times for equal-quality comparison of vanilla Monte Carlo and neural control variates with our geometric integration (GI-NCV) for ambient occlusion in the Bistro scene measured on AMD Radeon PRO W7900. For a larger number of samples per pixel, the integration part becomes marginal compared to training and inference, while for a relatively a smaller number of training samples per pixel (epochs), the training phase becomes the major bottleneck.}
    \label{Tab:Times}
\end{table}

\section{Conclusion and Future Work} 
We proposed a method for integrating the multilayered perceptron with a continuous piecewise linear activation function over a 2D domain. We formulated the integration problem as geometric subdivision of the integration domain, and we showed how to solve such a problem by means of computational geometry. The proposed method can be used in combination with control variates to reduce variance in Monte Carlo integration. We demonstrated the application of our method in physically-based rendering, achieving lower error than vanilla Monte Carlo. We also showed that our method is a viable alternative to the automatic integration~\cite{Lindell2021,Li2024}, achieving similar or even lower error.

There are several interesting directions for future work. There are two inherent problems to control variates that we have not touched in the paper: how to estimate $\alpha$ and how to importance sample the residual integral. We assumed that $\alpha=1$, which corresponds to perfect correlation between $f(x)$ and $g(x)$, which is not true in practice. We employed a variant of uniform sampling (e.g., the area light source in direct lighting) of the integration domain; however, it would be beneficial to design an importance sampling technique that would prioritize samples where the difference between the integrand and the approximation by the MLP is high. Both problems could be tackled by additional neural networks similarly to the work by Muller et al.~\cite{Muller2020}. Notice that the difference could be negative, which brings another source of variance~\cite{Belhe2024}.

Another interesting idea is to use the geometric subdivision to sample the integrand itself. This would work in two steps. In the first step, we would sample a region based on the integral of the associated affine function on that region. In the second step, we sample a point within the region (the affine function is invertible and can be easily normalized). Last, we would like to evaluate the proposed method in the context of geometric processing to solve Poisson or Laplace equations, using algorithms such as walk-on-spheres~\cite{Sawhney2020} or walk-on-boundary~\cite{Sugimoto2023}.


\section*{Acknowledgements}
We thank Shin Fujieda, Chih-Chen Kao, and Atsushi Yoshimura for their continuous help and valuable feedback.

\appendix
\section{Correlated Training and Rendering Samples}
\label{Sec:Proof}
The estimator used in Section~\ref{Sec:ResultsLT} can be defined as a weighted average of two unbiased estimators $\langle I_{MC}\rangle$ and $\langle I_{CV}\rangle$:
\begin{equation}
\langle I\rangle = \frac{m}{m+n}\langle I_{MC}\rangle + \frac{n}{m+n}\langle I_{CV}\rangle,
\end{equation}
\begin{equation}
\langle I_{MC} \rangle = \frac{1}{m}\sum_{i=1}^m \frac{f(x_i)}{p(x_i)}
\end{equation}
\begin{equation}
\langle I_{CV}\rangle = G + \frac{1}{n}\sum_{i=1}^n\frac{f(x_{m+i}) - g(x_{m+i})}{p(x_{m+i})}
\end{equation}
where $m$ is the number of training samples (used also for rendering) and $n$ is the number of rendering samples. We can easily show that the estimator is unbiased:
\begin{align}
\mathbb{E}[\langle I\rangle] &= \mathbb{E}\left[\frac{m}{m+n}\langle I_{MC}\rangle\right] + \mathbb{E}\left[\frac{n}{m+n}\langle I_{CV}\rangle\right], \\
&= \frac{mF}{m+n} + \frac{n}{m+n}\mathbb{E}\left[G + \frac{1}{n}\sum_{i=1}^n\frac{f(x_{m+i}) - g(x_{m+i})}{p(x_{m+i})}\right], \\
&= \frac{mF}{m+n} + \frac{n}{m+n}\left(G + \mathbb{E}\left[\frac{1}{n}\sum_{i=1}^n\frac{f(x_{m+i}) - g(x_{m+i})}{p(x_{m+i})}\right]\right), \\
&= \frac{mF}{m+n} + \frac{n}{m+n}(G + F - G) = \frac{(m+n)F}{m+n} = F.
\end{align}
Note that we use a 1D function to make notation simpler, but the same reasoning can be applied to higher-dimensional function such those used in Section~\ref{Sec:LightTransport}.

\begin{figure}
    \centering
    \includegraphics[width=0.8\linewidth]{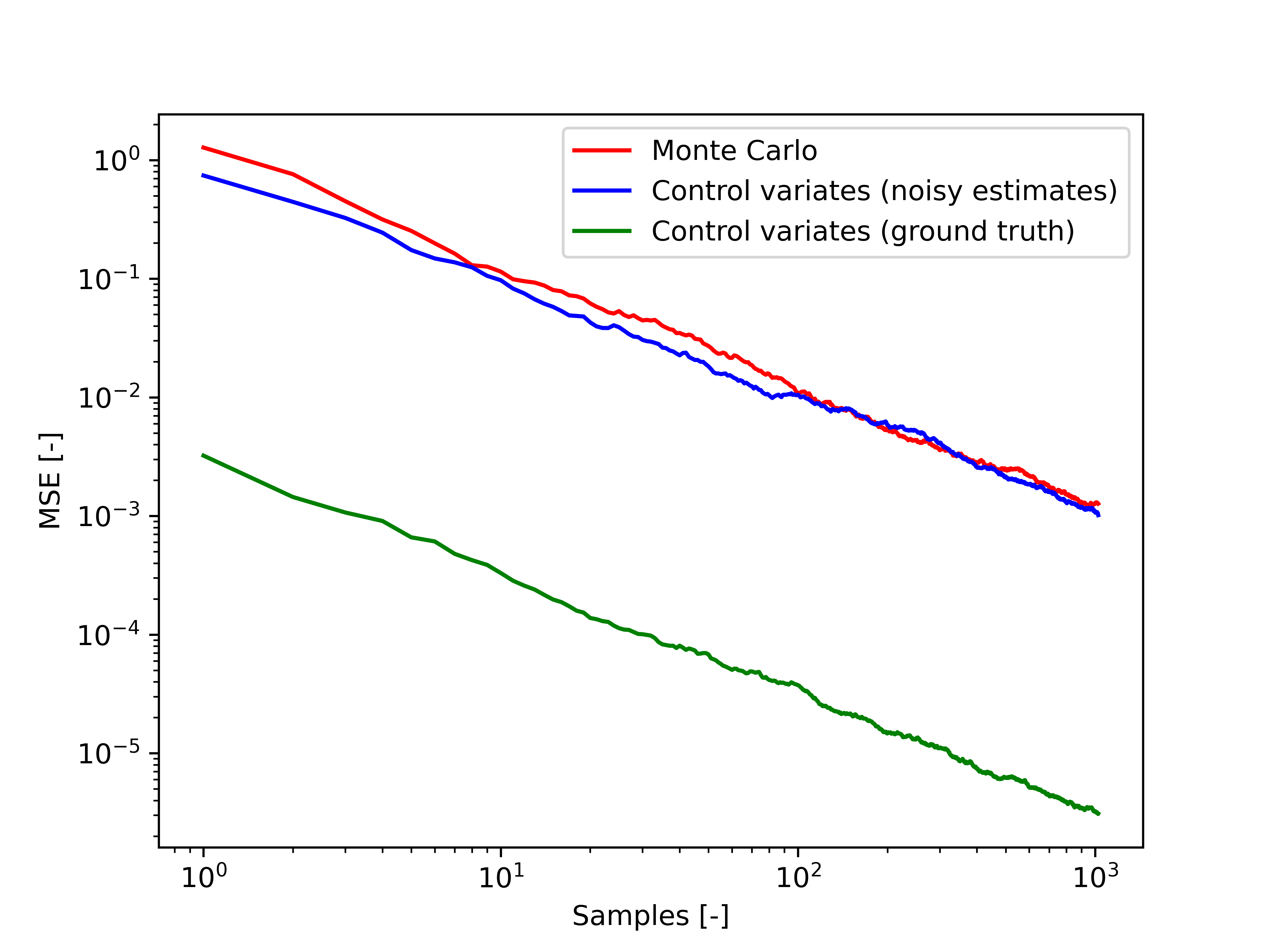}
    \caption{\rev{Comparison of control variates using ground truth $f(x) = 2x$ and noisy estimates $\bar{f}(x) = 8xy_iz_i$, using $g(x) = 0.9 \cdot f(x)$ as a control variate in both cases.}}
    \label{Fig:NoisyEstimates}
\end{figure}

\rev{
\section{Noisy estimates of the Integrand}
\label{Sec:NoisyEstrimates}
We conducted an experiment with a simple analytic function mimicking the setting in indirect lighting. We integrate function $f(x) = \int_0^1 \int_0^1 8xyz \diff y \diff z = 2x$ with its noisy one-sample estimates $\bar{f}(x) = 8xy_iz_i \approx f(x)$ for samples $(y_i, z_i)$. We use the same function scaled as a control variate $g(x) = 0.9 \cdot f(x)$, which is almost a perfect control variate, avoiding error due to insufficient approximation. Figure~\ref{Fig:NoisyEstimates} shows convergence graphs with noisy estimates and ground truth. While control variates with noisy estimates provide practically no improvement compared to vanilla Monte Carlo, the ground truth function values provide expected improvement, which is aligned with our light transport experiments. 
}

\bibliographystyle{eg-alpha}

\bibliography{main}

\end{document}